%% file: Espresso_WavelengthCalibration_arxiv.tex
\documentclass{aa}
\usepackage{graphicx}
\usepackage[font=small,labelfont=bf]{caption}
\usepackage{amsmath}
\usepackage{amssymb}
\usepackage{xfrac}
\usepackage{txfonts}	
\usepackage{lipsum}
\usepackage{stfloats}

\usepackage{color}
\definecolor{Navy}		{RGB}{  0,   0, 128}
\definecolor{MidnightBlue}	{RGB}{ 25,  25, 112}
\definecolor{yellow}   	{RGB}{255, 215,   0}
\definecolor{darkorange}{RGB}{255, 140,   0}
\definecolor{dodgerblue}{RGB}{ 30, 144, 255}
\definecolor{black}     {RGB}{  0,   0,   0}
\definecolor{dimgray}   {RGB}{105, 105, 105}
\definecolor{gray}   {RGB}{128, 128, 128}

\usepackage{hyperref}
\hypersetup{colorlinks=true, linkcolor=Navy, citecolor=Navy, urlcolor=Navy}

\newcommand{\Espresso}{\textsc{Espresso}}
\newcommand{\PtV}{PtV}

\renewcommand{\S}{Section~}

\DeclareRobustCommand{\ion}[2]{%
  \relax
  \ifmmode
    \ifx\testbx\f@series
      {\mathbf{#1\,\mathsc{#2}}}
    \else
      {\mathrm{#1\,\mathsc{#2}}}
    \fi
  \else
    \textup{#1\,{\mdseries\textsc{#2}}}%
  \fi
 }

\newcommand\blfootnote[1]{%
  \begingroup
  \renewcommand\thefootnote{}\footnote{#1}%
  \addtocounter{footnote}{-1}%
  \endgroup
}

\begin{document}
\graphicspath{{Figures/}}

\title{Fundamental physics with \Espresso{}: Towards an accurate wavelength calibration for a precision test of the fine-structure constant}
\titlerunning{Towards an Accurate \Espresso{} Wavelength Calibration}

\authorrunning{ T. M. Schmidt et al. }
          
\include{ListAuthors}

\include{ListAffiliations}

\date{Received September 4, 2020; accepted November 19, 2020}

\abstract{ 
Observations of metal absorption systems in the spectra of distant quasars allow to constrain a possible variation of the fine-structure constant throughout the history of the Universe. Such a test poses utmost demands on the wavelength accuracy and previous studies were limited by systematics in the spectrograph wavelength calibration. A substantial advance in the field is therefore expected from the new ultra-stable high-resolution spectrograph \Espresso{}, recently installed at the VLT.
In preparation of the fundamental physics related part of the \Espresso{} GTO program, we present a thorough assessment of the \Espresso{} wavelength accuracy and identify possible systematics at each of the different steps involved in the wavelength calibration process. 
Most importantly, we compare the default wavelength solution, based on the combination of Thorium-Argon arc lamp spectra and a Fabry-P\'erot interferometer, to the fully independent calibration obtained from a laser frequency comb. We find wavelength-dependent discrepancies of up to $24\,\mathrm{m/s}$. This substantially exceeds the photon noise and highlights the presence of different sources of systematics, which we characterize in detail as part of this study. 
Nevertheless, our study demonstrates the outstanding accuracy of \Espresso{} with respect to previously used spectrographs and we show that 
constraints of a relative change of the fine-structure constant at the $10^{-6}$ level can be obtained with \Espresso{} without being limited by wavelength calibration systematics.
}


\maketitle \blfootnote{Based on work by the fundamental constants working group of the ESPRESSO GTO Consortium.}

\section{Introduction}
The mathematical description of the phenomena of Nature, i.e., our laws of physics, requires a set of fundamental constants (e.g., $G$, $\hbar$, $c$, ...), which determine the scales of physical effects. The values of these constants cannot be predicted by theory but have to be determined experimentally.
In generalized field theories that try to relate physical constants to more fundamental concepts, these constants can in principle depend on time and space \citep[see e.g.,][]{Uzan2011}. Fortunately, such a possible change of fundamental constants can be tested, for instance, by ultra-precise laboratory experiments on Earth \citep[e.g.,][]{Rosenband2008}, but also with astronomical observations in the distant Universe. 

In the past, most attention has been directed towards the fine-structure constant $\alpha = \frac{e^2}{4\pi \epsilon_0 \, \hbar \, c} \approx \frac{1}{137}$, which defines the coupling strength of electromagnetic interactions and therefore affects the energy levels of atomic transitions. In practice, a change of $\alpha$ would shift the wavelengths of spectral lines and is therefore observable.
The strength of the effect depends on the electron configuration of an atom and different spectral lines can have a substantially different sensitivity to the value of the fine-structure constant \citep[e.g.,][]{Dzuba1999a}. By observing multiple transitions originating from the same absorption system, one can break the degeneracy between absorption redshift and fine-structure constant and therefore directly constrain the value of $\alpha$ at the time and place where the absorption happened. Thus, the constraint on $\alpha$ basically comes from an accurate measurement of relative wavelength differences.

\begin{figure}[th]
 \centering
 \includegraphics[width=\linewidth]{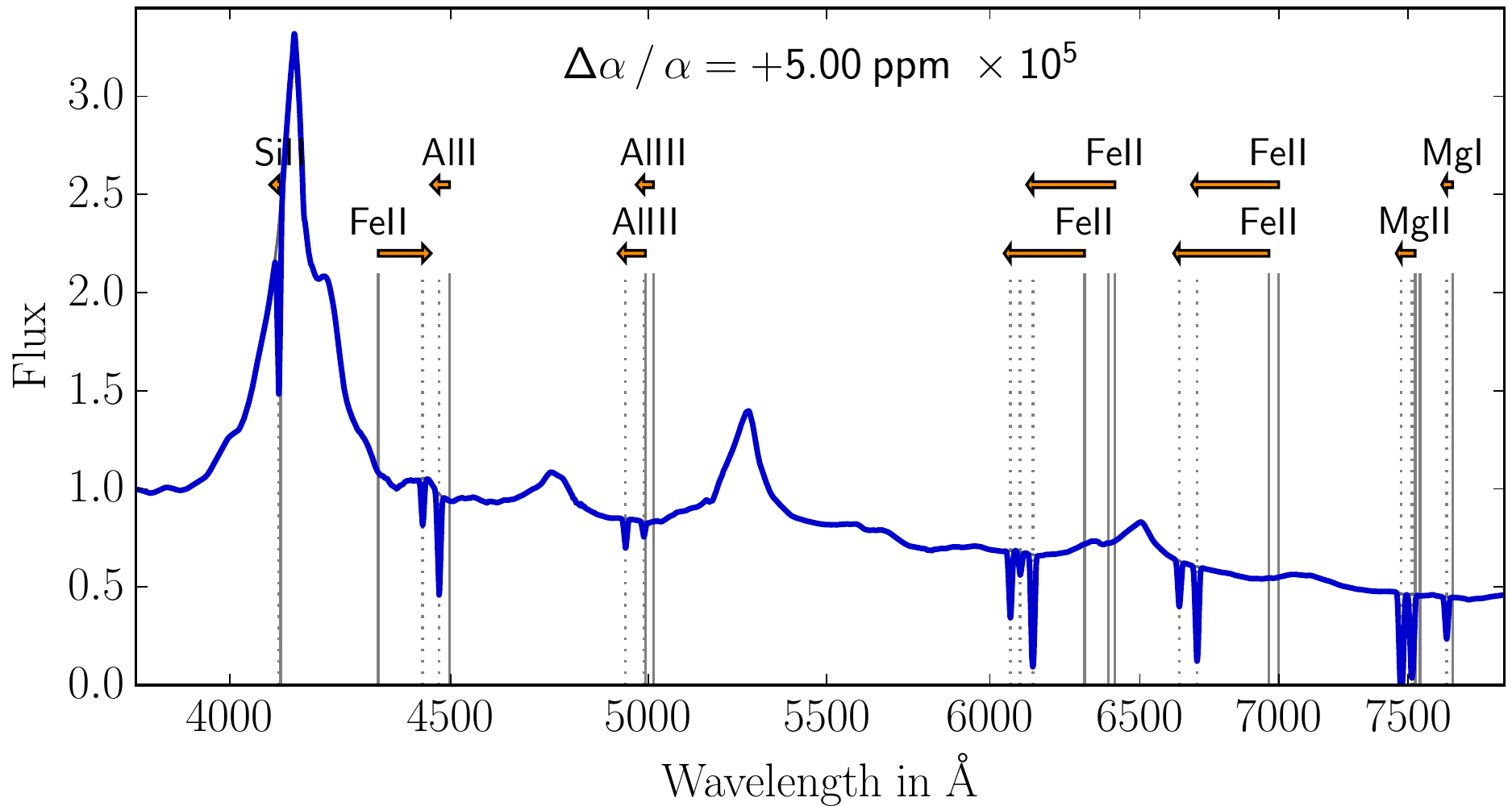} 
 \caption{
  Illustration of the general concept for measuring a variation of the fine-structure constant from a fiducial metal absorption systems at $z_\mathrm{abs}=1.7$ in a quasar spectrum. 
  The assumed change of $\Delta\alpha/\alpha$ by 5~ppm leads to differential line shifts, indicated by the arrows. For visibility, the magnitude of the shifts is exaggerated by $\times10^5$.
 }
 \label{Fig:FineStructureVisualization}
\end{figure}

The classical approach for such experiments is to study metal absorption lines in the spectra of distant quasars \citep[e.g.,][]{Wolfe1976,Webb1999}. Figure~\ref{Fig:FineStructureVisualization} illustrates the general concept of such an observation and indicates the amplitude of the wavelength shift for a few commonly used metal transitions.
Among the absorption lines with the strongest shift are the \ion{Fe}{ii} lines with rest~wavelengths between $2300\,\mathrm{\AA}$ and $2600\,\mathrm{\AA}$. They shift by about $20\,\mathrm{m/s}$ for a relative change of $\alpha$ of one part in a million (ppm) \citep[see e.g.,][]{Murphy2014}. This has to be compared to lines with low sensitivity to $\alpha$, often from light elements such as \ion{Mg}{ii} ($\simeq3\,\mathrm{m/s}$ shift per ppm) or \ion{Al}{iii} ($\simeq4\,\mathrm{m/s}$ per ppm).
A somehow special transition is the \ion{Fe}{ii} $\lambda=1608\,\mathrm{\AA}$ line, which has a relatively strong dependence on $\alpha$ but with opposite sign ($-12\,\mathrm{m/s}$ per ppm). It is therefore ideal to compare it with the other \ion{Fe}{ii} lines, but, unfortunately, its oscillator strength is relatively low and the line not always observed at sufficient strength and signal-to-noise ratio.

The current constraints reported on $\Delta\alpha/\alpha$, either from large ensembles of absorbers \citep[e.g.,][]{Murphy2003, Chand2004, King2012} or individual absorption systems \citep{Levshakov2007, Molaro2008b, Molaro2013a, Kotus2017}, have a claimed precision between one and a few ppm.
However, there is disagreement among these studies whether the fine-structure constant actually varies or not and to which degree the measurements are affected by (instrumental) systematics.
Since all these studies are working close to the instrument's wavelength accuracy, new measurements with more accurate spectrograps are clearly needed to settle the issue. These have to deliver a precision of at least 1~ppm to provide meaningful constraints, which poses utmost demands on the wavelength calibration of the spectrograph.
The line shifts stated above imply that the utilized spectrograph shall not exhibit peak-to-valley (\PtV) distortions of the wavelength scale that are substantially larger than $\simeq20\,\mathrm{m/s}$ to safely exclude wavelength calibration uncertainties. Achieving this goal is extremely challenging.

It turned out that the existing high-resolution echelle spectrographs Keck/HIRES \citep{Vogt1994} and VLT/UVES \citep{Dekker2000} are neither designed nor ideal for this task. For instance, \citet{Griest2010} and \citet{Whitmore2010} demonstrated by observations taken through iodine cells that the standard wavelength calibrations of HIRES and UVES, derived from exposures of thorium-argon (ThAr) hollow-cathode lamps (HCL), show intra-order distortions of up to $700\,\mathrm{m/s}$~\PtV{} and overall shifts between exposures up to $\pm1000\,\mathrm{m/s}$. This raised strong concerns if the $\simeq\,$1~ppm constraints on the fine-structure constant derived with these instruments are trustworthy.
More recently, attempts have been made to \textit{supercalibrate} quasar spectra \citep[][]{Molaro2008a, Rahmani2013, Evans2014, Whitmore2015, Murphy2017}, which intend to transfer the wavelength accuracy of Fourier-transform spectrometers at solar observatories \citep[e.g.,][]{Reiners2016} to the quasar observations by using spectra of solar light reflected off asteroids or \textit{solar twins}, i.e., stars very similar to the Sun. These studies confirm intra-order distortions of $300\,\mathrm{m/s}$~\PtV{} and global slopes up to $800\,\mathrm{m/s}$ over $1500\,\mathrm{\AA}$ for UVES and $600\,\mathrm{m/s}$ over $3000\,\mathrm{\AA}$ for HIRES \citep{Whitmore2015}.
The wavelength calibration of HARPS \citep{Mayor2003} is much more accurate \citep{Whitmore2015, Cersullo2019}, however, installed at the ESO 3.6\,m telescope, this spectrograph is simply not equipped with sufficient collecting area to observe any except the brightest quasar suitable for this experiment \citep{Kotus2017,Milakovic2020b}.

A substantial advance of this field is therefore expected from the new ultra-stable high-resolution Echelle SPectrograph for Rocky Exoplanets and Stable Spectroscopic Observations \citep[\Espresso,][]{Molaro2009, Pepe2010, Pepe2014, Megevand2014, Pepe2020}, recently installed at the VLT. 
\Espresso{} was designed for ultra-precise radial-velocity (RV) studies of exoplanets and for a precision test of the variability of fundamental constants. 

\Espresso{} is located at the incoherently-combined Coud\'e focus of the VLT and can in 1-UT~mode be fed by any of the four $8.2\,\mathrm{m}$ mirrors or in the 4-UT~mode with the (incoherently) combined light of all four VLT telescopes%
\footnote{However, in the 4-UT mode (\texttt{4MR}) only at \textit{medium} resolution of $R\approx72\,000$, approximately half of the resolution possible in the \textit{high-resolution} 1-UT mode (\texttt{1HR}), which offer $R\approx138\,000$.}.
Several design aspects are essential for its stability and accuracy. 
First, the spectrograph is located inside a thermally controlled ($\Delta{}T\simeq1\,\mathrm{mK}$) vacuum vessel 
and therefore not affected by environmental effects like air temperature and pressure, which influence the index of refraction and would therefore change the echellogram. 
Second, it has zero movable components within the spectrograph vessel and thus a fixed spectral format.
Third, \Espresso{} is fed by optical fibers%
\footnote{To facilitate high optical throughput, the light from the individual UTs is guided by classical optical trains composed of mirrors, prisms and lenses to the combined Coud\'e laboratory where the light is injected into relatively short optical fibers.},
facilitating a much more stable illumination of the spectrograph than possible with slit spectrographs. 
A high degree of immunity to changes of the fiber input illumination is achieved by the use of octagonal fibers, which have a high (static) mode-scrambling efficiency and provide a very homogeneous output profile \citep{Chazelas2012}. In addition, the near- and far-field of the fibers are exchanged between two sections of the fiber train, a design described as \textit{double-scrambler} \citep[e.g.,][]{Hunter1992, Bouchy2013}.
To minimize guiding errors and ensure an optimal injection of light into the optical fibers, active field and pupil stabilization by the means of guide cameras and piezo tip-tilt mirrors is incorporated in the spectrograph front-end \citep{Riva2014b, Calderone2016, Landoni2016}. In addition, atmospheric dispersion correctors (ADC) compensate the effect of differential refraction in Earth's atmosphere. 

These measures guarantee an extremely stable and homogeneous illumination of the spectrograph and therefore eliminate several issues that have plagued previous studies of a possible change of the fine-structure constant, namely slit and pupil illumination effects due to guiding errors, placement of the target on the slit, variable seeing, and atmospheric dispersion effects%
\footnote{For \Espresso{}, imperfect placement of the target image on the fiber, e.g., due to guiding errors or incomplete correction of differential atmospheric refraction, would lead to increased (differential) slit losses but not affect spectral features. Due to the $\lambda^{-\sfrac{1}{4}}$ dependence of atmospheric seeing, slit losses are anyway wavelength dependent. This might complicate spectrophotometric flux calibration but the sophisticated mode scrambling in the fiber feed ensures that the inferred wavelengths of spectral features remain unaffected.} %
 \citep[e.g.,][]{Murphy2001b}. 
In addition, the fiber feed also eliminates noncommon path aberrations between science target and wavelength calibration sources since in both cases the light passes through exactly the same optical path%
\footnote{The relevant procedure for fine-structure studies is that, before or after the science observation, light from the calibration source(s) is fed through the same fiber as the light from the science target. This is unrelated to the \textit{simultaneous reference} method used in RV studies.}.

To facilitate an accurate wavelength calibration, \Espresso{} is equipped with a comprehensive suite of calibration sources \citep{Megevand2014}.
This encompasses a classical ThAr hollow cathode lamp, which is complemented by a Fabry-P\'erot interferometer that produces an extremely large number of narrow (but marginally resolved), equally spaced lines \citep{Wildi2010, Wildi2011, Wildi2012, Bauer2015}. Although not providing absolute wavelength information, the Fabry-P\'erot interferometer allows in combination with the ThAr lamp a much more precise calibration, in particular on small scales, as would be possible with arc spectra alone \citep[demonstrated for HARPS by][]{Cersullo2019}.

A completely independent means of wavelength calibration%
\footnote{The LFC calibration requires an initial wavelength solution for line identification. However, the requirements for this are moderate and the final LFC solution is formally independent of the a-priori assumed wavelength solution. Details are given in \S\,\ref{Sec:LFC_Solution}.} 
comes from a laser frequency comb (LFC). The core element of a LFC is a passively mode-coupled femto-second laser that provides a dense train of extremely narrow and equally-spaced emission lines, which frequencies are directly stabilized against a (local or remote) atomic clock and therefore the fundamental SI time standard \citep[e.g.,][]{Murphy2007, Steinmetz2008, Wilken2010a, Wilken2010b, Wilken2012, Murphy2012, Probst2014}. With intrinsic accuracies reaching $10^{-12}$, LFCs promise to deliver unprecedented calibration accuracies for astronomical spectrographs. 
This novel calibration method was in the optical first applied to HARPS \citep{Wilken2010a, Coffinet2019, Cersullo2019}.
One application is for example the definition of a \textit{solar atlas} \citep{Molaro2013b}, obtained from LFC-calibrated asteroid spectra. A solar spectrum calibrated in this way can then, again via asteroid (or solar twin) observations, be used to calibrate quasar spectra taken at larger telescopes.
This is similar to the \textit{supercalibration} technique by \citet{Whitmore2015} but uses directly the absolute wavelength information provided by the LFC instead of relying on observations taken with solar Fourier-transform spectrometer, which by itself are calibrated by further secondary means. 
\Espresso{} should in the best case render these complex procedures obsolete. Since it is itself equipped with a LFC, no extra steps are required to achieve the same calibration accuracy. 
All these measures should make \Espresso{} the astronomical spectrograph that achieves the highest wavelength accuracy, only rivaled by Fourier-transform spectrometers at solar observatories or laboratories.

Despite the complex Coud\'e~train, the fiber feed, the extremely high resolution, and numerous extra steps taken to ensure stability and accuracy, \Espresso{} is designed for high efficiency and offers (arguably, and depending on the exact configuration and conditions) a similar throughput as UVES \citep{Pepe2020}, but at much higher resolution ($R=138\,000$ compared to $R\approx50\,000$ using a 0.8" slit) and with larger instantaneous wavelength coverage (from $3840\,\mathrm{\AA}$ to $7900\,\mathrm{\AA}$). It is therefore ideally suited for precision tests of fundamental physics and, logically, the \Espresso{} consortium has dedicated 10\% of the GTO time 
to fundamental physics related projects, in particular to a test for a possible variation of the fine-structure constant.

However, the claimed wavelength accuracy has to be demonstrated. The goal of this study is therefore a careful and thorough assessment of the \Espresso{} wavelength calibration%
%
. It has to be stressed that the calibration requirements for tests of a varying fine-structure constant are fundamentally different from those of RV studies aimed at detecting and characterizing exoplanets. 
Radial-velocity studies require extreme \textit{precision} and repeatability since the signal is the difference between observations taken with the same instrument at different times. Also, the aim is to measure a global RV shift for which all (or a selection of) absorption lines across the spectral range are combined. 
Distortions of the wavelength scale are therefore of little importance, at least if they remain stable.
This is different for a test of fundamental physics. Stability of the instrument is clearly highly convenient, but it is technically not \textit{essential} since the constraint comes from the wavelength difference between different absorption lines within the same observation.  Thus, it is crucial to have an \textit{accurate} wavelength scale that is free of distortions.
In fact, a global RV shift would be the only thing \textit{not} relevant for a constraint of the fine-structure constant since it is degenerate with the absorption redshift of the system and therefore unimportant. However, this is only true if the shift is constant in velocity space. A shift constant in wavelength, frequency or pixel position would cause a wavelength-dependent velocity shift and therefore not cancel out.

To assess the \Espresso{} wavelength accuracy,
we present here all steps required to obtain the wavelength solution of the spectrograph, starting from the basic data reduction (\S\,\ref{Sec:DataReduction}) and spectral extraction (\S\,\ref{Sec:Extraction}), fitting of the calibration source spectra (\S\,\ref{Sec:LineFitting}) and finally the computation of the actual wavelength solution (\S\,\ref{Sec:JointThArFPSolution} and \S\,\ref{Sec:LFC_Solution}). While going along, we identify issues that might lead to systematics for constraining $\alpha$ and wherever possible carry out consistency checks to test and demonstrate the accuracy of the derived results. The most informative test is the comparison of the wavelength solution obtained jointly from the ThAr arc lamp spectra and the Fabry-P\'erot interferometer to the one derived from the laser frequency comb (\S~\ref{Sec:WavelengthComparisons}). The comparison of these two fully independent solutions gives a clear picture of the wavelength accuracy and allows to predict the impact of systematic effects for a test of changing fine-structure constant (\S\,\ref{Sec:MockSystematics}).

\section{Basic Data Reduction}
\label{Sec:DataReduction}

The results presented in this study are based on a set of custom-developed routines for data reduction and wavelength calibration. We therefore do not make use of the \Espresso{} Data Reduction Software \citep[DRS,][]{Lovis2020}, which is the standard ESO instrument pipeline for this spectrograph%
\footnote{\url{https://www.eso.org/sci/software/pipelines/espresso/espresso-pipe-recipes.html}}.
The reasons for this approach are manifold. 
First, the DRS is designed as robust general-purpose instrument pipeline and optimized for exoplanet research, i.e., RV studies.
As described above, the requirements for a precision test of fundamental constants are different from those of RV studies. 
Our code therefore follows a design philosophy that is optimized for \textit{accuracy} instead of \textit{precision}. In particular high priority was given to the minimization of correlations of wavelength errors across the spectral range. 
The overall wavelength calibration strategy is of course governed by the instrument design and calibration scheme. Therefore, our code employs similar procedures than the DRS. Still, we try to optimize towards our science case, which results in the use of different algorithms for several tasks. For example, we try to wherever possible approximate functions locally by nonparametric methods instead of fitting global polynomials. These techniques should deliver better wavelength accuracy at the expense of precision, which is not the limiting factor 
for a test of fundamental constants.
Apart from this, a fully independent implementation allows for cross-check between the pipelines and allows to exclude systematics caused by the implementation itself (presented in Section~\ref{Sec:Comparison_DRS}). But most notably, our codebase is intended as a flexible testbed to experiment and develop improved algorithms, without the need to obey the strict requirements of official ESO instrument pipelines.

All data presented here were taken as part of the standard daily \Espresso{} calibration plan on August 31 2019, within a period of $\simeq2.5\,\mathrm{h}$. Instrumental drifts over this timespans are $\ll2\,\mathrm{m/s}$ and therefore negligible. Since the test of fundamental constants is in principle a \textit{single shot} experiment that does not rely on a monitoring campaign, the assessment of the stability or time evolution is beyond the scope of this paper. 
We focus on the 1-UT, high-resolution mode of \Espresso{} and if not stated otherwise, the detectors were read using a 1$\times$1 binning (\texttt{1HR1x1}). However, due to the faintness of the background quasars, all observations for the fundamental physics project will utilize the \texttt{1HR2x1} mode%
\footnote{Or in the future the \texttt{1HR4x2} heavy-binning mode.}, 
which employs a 2x binning in cross-dispersion direction. Therefore, we also carried out the full analysis for this instrument mode and wherever necessary also present the results for the \texttt{1HR2x1} mode and state this explicitly.


The basic data reduction of the raw frames follows a standard procedure and is very similar to the DRS.
Every frame is first processed in the following way: 
for each of the 16 readout amplifiers per detector, the bias value is determined from the overscan region and subtracted from the data. In the same step, the overscan region is cropped away and ADUs converted to $\mathrm{e}^{-}$ using the gain values listed in the fits headers ($1.099\,\mathrm{e^{-}/ADU}$ and $1.149\,\mathrm{e^{-}/ADU}$ for blue and red detector). This step also allows to determine the readout noise (RON), which is, depending on the amplifier,  in the \texttt{1HR1x1} readout mode between $8.7$ and $12.1\,\mathrm{e}^{-}$ for the blue detector and between $6.8$ and $14.2\,\mathrm{e}^{-}$ in the red detector. The corresponding values in \texttt{1HR2x1} mode are $3.0$ to $4.7\,\mathrm{e}^{-}$ and $2.2$ to $2.4\,\mathrm{e}^{-}$.
The RON is stored separately for each output amplifier and propagated throughout the full analysis.

Following this, the ten bias frames taken as part of the daily \Espresso{} calibration scheme are stacked and outliers identified to reject cosmic ray hits, which are present even in bias frames. The masked frames are then mean combined to form the masterbias, which encapsulates the spatial variation of the bias value. The structure in the masterbias is rather small ($\simeq \pm\,4\,\mathrm{e^{-}}$), except in the corners of the individual chip readout areas. Still, we subtract the masterbias from all subsequent frames to correct for this effect.
Since the dark current is low ($\simeq1\,\mathrm{e^{-}/h/pix}$) and the exposure times at maximum $40\,\mathrm{s}$, no dark current correction is applied.

According to the \Espresso{} calibration plan, every day two sets of ten spectral flatfield frames are taken in which always only one of the two fibers is illuminated. The two sets are treated independently, overscan corrected, converted to $\mathrm{e^{-}}$, the masterbias subtracted and then stacked including an outlier detection to reject cosmic ray hits.

For the wavelength calibration frames (Thorium-Argon arc lamp, Fabry-P\'erot Interferometer and laser frequency comb) only single frames are available per calibration session. These are just overscan and masterbias corrected.
No flatfielding, neither by chip flatfields nor spectral flatfields is applied. Pixel-to-pixel sensitivity variations will be accounted for as part of the spectral extraction process.
Also, no masking of bad pixels is applied since this is not needed within the context of this study. 

\section{Spectral Extraction}
\label{Sec:Extraction}

Following this basic reduction of the wavelength calibration frames, the next step is the spectral extraction, i.e., the gathering of  spectral information from the two-dimensional raw frames and their reduction to one-dimensional spectra. This is a crucial aspect of the data reduction procedure and therefore described in detail.

In general, extraction is done independently for each order. Since \Espresso{} is fed by two Fibers (A and B) and utilizes a pupil slicer design that splits the pupil image in half to create two images of the same fiber (Slice a and b) on the focal plane, four individual spectra per order are imaged onto the detectors. 


The extraction process should make use of \textit{optimal extraction}, a scheme initially described by \citet{Horne1986} and now commonly used.
As demonstrated in \citet{Horne1986}, the outlined scheme is \textit{flux complete} in the sense that it is able to deliver an estimate of the total received flux and therefore spectrophotometric results. In addition, it is \textit{optimal} in the sense that it minimizes the variance of the extracted spectrum.
However, it makes the fundamental assumption that there is a unique correspondence between wavelength and pixel position, neglecting the finite extent of the instrumental point-spread function (PSF) in dispersion direction.   
Under this assumption, the distribution of flux on the detector becomes a separable function of the spectral energy distribution of the source (SED) and the profile of the trace in spatial (or cross-dispersion) direction.
For a detailed discussion see e.g., \citet{Zechmeister2014}, in particular their Equations~2, 3, and 4.
This assumption drastically simplifies the extraction process, which is therefore mostly related to the description of the (spatial) trace profile, i.e., the relative intensity of different pixels corresponding to the same wavelength.

Because \Espresso{} is a fiber-fed echelle spectrograph, there is no spatial direction and the spectra projected on the detectors contain no scientific information in cross-dispersion direction. 
Since the echellogram is approximately aligned with the pixel grid and the individual traces never tilted by more than $10\,\mathrm{degrees}$ with respect to the detector coordinate, the spectral geometry is approximated by assuming that the wavelength direction coincides with the detector $Y$ direction and the cross-dispersion direction with the detector $X$ direction. 
Therefore, the trace profile $P_{yx}$ at pixel position $y$ is just a cut through the trace along the detector $X$ coordinate.

Formally, the extraction process can be described as follows:
For each detector position $(y|x)$, the bias and background subtracted raw electron counts $C_{yx}$ are divided by the normalized%
\footnote{We emphasize that in contrast to the formalism in \citet{Zechmeister2014}, $P_{yx}$ is normalized, i.e., $\sum_x P_{yx} = 1$. This is desired, so that $\bar{C}_{y}$ in Equation~\ref{Eq:OptimalExtraction} represents the physical number of detected photons and its error thus be $\simeq \sqrt{\bar{C}_{y}}$. In consequence, the blaze and spectral fluxing has to (or can) be determined separately.}
trace profile $P_{yx}$, which delivers for every pixel an independent estimate of the total detected number of photons in the given detector column. These can then be averaged in cross-dispersion direction using inverse-variance weighting ensuring that pixels in the center of the trace, which receive more flux and therefore show smaller relative errors, are weighted stronger than pixels in the wings of the trace. The estimate for the total number of photons received in a certain detector column $\bar{C}_y$ is therefore 
\begin{equation}
  \bar{C}_y = \frac{ \sum_x C_{yx} / P_{yx} \times W_{yx} }{ \sum_x W_{yx} }
  \label{Eq:OptimalExtraction}.
\end{equation}
By choosing weights $W_{yx}$ according to the inverse variance of the measurements $C_{yx}$ one obtains, as demonstrated in \cite{Horne1986}, an unbiased estimate of the true received flux with minimal varian    ce. The variance for the individual pixels can be computed as
\begin{equation}
 W_{yx} =  \frac{ P_{yx}^2 }{ var(C_{yx}) } = \frac{ P_{yx}^2 }{ C_{yx} + BG_{yx} + DARK_{yx} + RON_{yx}^2 }
 \label{Eq:InverseVariances},
\end{equation}
where $BG_{yx}$ and $DARK_{yx}$ are the contribution of scattered light and dark counts, while $RON_{yx}$ is the read-out noise.
The complete extraction procedure is therefore solely governed by the trace profile $P_{yx}$.

\subsection{Determining the Trace Profile}
\label{Sec:TraceProfile}

\begin{figure*}[tb]
 \centering
 \includegraphics[width=\linewidth]{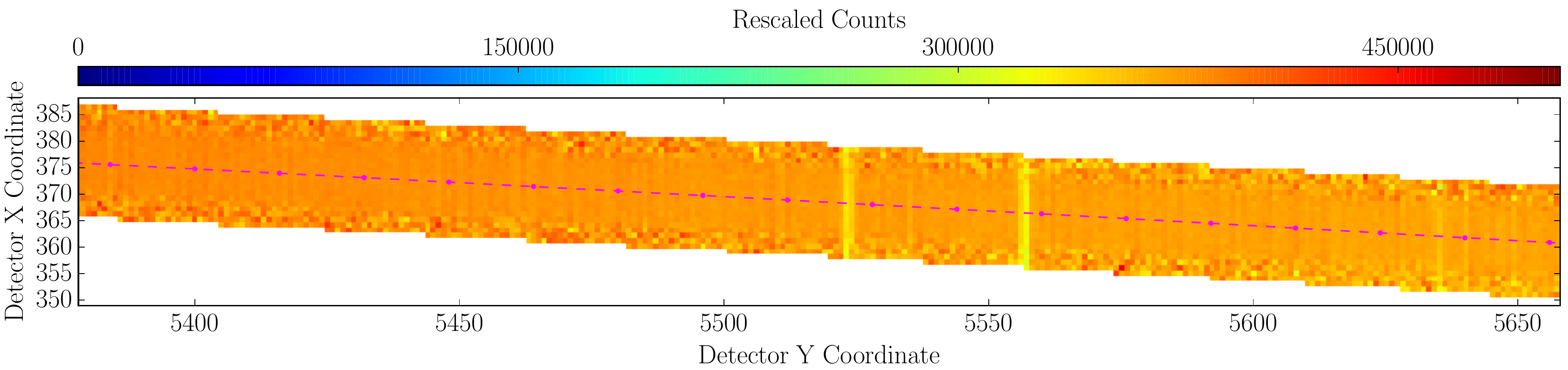}
 \caption{Small part of a single bias and scattered-light subtracted flatfield frame normalized by the trace profile. The pink dashed line indicates the estimated trace center. As expected, there is no structure in vertical direction. The only visible effect is increased noise towards the upper and lower edge of the trace. The vertical striping represents the individual pixel sensitivities but is mostly related to bad pixels.
 The striping, i.e., variations in dispersion direction,  is of no concern since it will be removed by the de-blazing and flux calibration procedures.}
 \label{Fig:OptialExtraction_2D_SHR_FLAT}
\end{figure*}

It is common practice to model the trace profile with analytic functions like a Gaussian and assume its properties (e.g., central position and width) to evolve slowly with detector position.
However, for \Espresso{} the trace profile is non-Gaussian and can not easily be described by analytic functions in an accurate and precise way. 
Fortunately, since \Espresso{} is fiber-fed and designed for extreme stability, the spectral format is entirely fixed and does not change with time. We therefore follow the approach described in \citet{Zechmeister2014}, which uses an empirical model of the trace profile that can be directly derived from the master flatfield frame. In this way, the extraction is applied completely blind and fully independent of the science spectrum, purely defined by the trace profile observed in the master flatfield.

To extract the trace profile from the master flatfield frame one has to know its location. This requires an initial fit to the trace profile, which, however, does not have to be particularly accurate. 
As described in Equation~\ref{Eq:OptimalExtraction}, the scaling and weighting of individual pixels is fully defined by the trace profile and thus the observed flux in the master flatfield frame. It is therefore completely independent of the initial trace center as long as the full extent of the trace is captured within an appropriately chosen window around the fitted trace center%
\footnote{One further requirement is that the empirical trace profile drops to zero far away from the trace center. Since the calibration frames contain a significant amount of scattered light, an adequate procedure to subtract this background is needed.}.
This substantially relaxes the requirement on the accuracy of the initial tracing and allows us to follow a rather simple approach.
One could fit the trace profile for each detector $Y$ position individually. However, given approx. 9000 pixels in detector~$Y$ direction and 340 traces, this poses a significant computational effort. To speed-up the computation, the data is binned by 16~pixels in detector $Y$ direction and then fitted by a simple Gaussian plus a constant background. Based on the central values obtained by these fits, an estimate of the trace center for every pixel in detector $Y$ direction $X_0(y)$ is obtained by cubic spline interpolation.

Following this, a global model of the scattered light is computed by first determining the median counts in $128\,\mathrm{pix}\times128\,\mathrm{pix}$ regions and then interpolating these measurements using  bivariate splines of third order. In an iterative sigma-clipping process, the traces themselves are masked and the background model refined using only the intra-order regions, which is then subtracted from the master flatfield.
To define the trace profile $P_{yx}$, a region of $\pm \Delta{}x^\mathrm{pr}$ pixels around the trace center%
\footnote{In practice, $\Delta{}x^\mathrm{pr} = 10.5\,\mathrm{pixels}$ is used in \texttt{1HR1x1} mode and $\Delta{}x^\mathrm{pr} = 5.5\,\mathrm{pixels}$ in \texttt{1HR2x1}}
is extracted from the master flatfield. 
As stated before, the exact extent of this window is of no particular importance, neither is the assumed center of the trace. The only requirement is that the trace is fully covered by the assumed region. 

As described in Equation~\ref{Eq:OptimalExtraction} and \ref{Eq:InverseVariances}, the trace profile $P_{yx}$ determines which pixels get extracted and therefore has to be zero far away from the trace center. 
Since the model is empirical and taken from the master flatfield, the profile will inevitably become noisy in the wings. This by itself should not be an issue since the noise in the to-be-extracted spectrum should always dominate over the noise in the stacked master flatfield. 
However, imperfections in the scattered light subtraction might prevent the trace profile to approach zero and therefore lead to biases. 
It was thus decided to artificially truncate the trace profile. This is done by introducing a window function $W^\mathrm{win}_{yx}$ that acts as additional weighting. Equation~\ref{Eq:OptimalExtraction} therefore becomes
\begin{equation}
  \bar{C}_y = \frac{ \sum_x C_{yx} / P_{yx} \: \times \: W^\mathrm{var}_{yx} \: \times \: W^\mathrm{win}_{yx}}{ \sum_x W^\mathrm{var}_{yx} \: \times \: W^\mathrm{win}_{yx} }
  \label{Eq:OptimalExtractionWindow},
\end{equation}
with $W^\mathrm{var}_{yx}$ as defined in Equation~\ref{Eq:InverseVariances}. For the window function $W^\mathrm{win}_{yx}$, \textit{fractional pixels} are extracted, i.e., the window function is nonbinary%
\footnote{If $W^\mathrm{win}_{yx}$ would be a binary mask, it could be combined with $P_{yx}$ in one function (as e.g., in \citealt{Zechmeister2014}). The need to account for fractional pixels requires to have two independent objects for the profile and the extraction window.} 
and for an individual pixel proportional to the area of the pixels that lies within a region of $\pm \Delta{}x^\mathrm{win}$ pixels around the trace center $X_0(y)$. Formally, this can be written as
\begin{equation}
 W^\mathrm{win}_{yx} = \int_{x_\mathrm{lo}}^{x_\mathrm{hi}} \; \Theta(\, x' - X_0(y) + \Delta{}x^\mathrm{win} \,) \; \Theta(\, -x' + X_0(y) + \Delta{}x^\mathrm{win} \,) \; dx',
\end{equation}
with $x_\mathrm{lo}$ and $x_\mathrm{hi}$ the lower and upper bounds of the pixel at position $y$ and $\Theta(x')$ the Heaviside step function. The size  $\Delta{}x^\mathrm{win}$ is chosen so that e.g., $\simeq95\%$ of the flux is extracted%
\footnote{Here, $\Delta{}x^\mathrm{win} = 5.0\,\mathrm{pixels}$ is used in \texttt{1HR1x1} and $\Delta{}x^\mathrm{win} = 2.5\,\mathrm{pixels}$ in \texttt{1HR2x1} mode.}.

Introducing the window function now makes the extraction formally dependent on the initial determination of the trace center. However, the detected counts normalized by the trace profile should be constant along the trace profile, i.e., the quantity $C_{yx} / P_{yx}$ should show no dependence on detector $X$ coordinate. Therefore, shifting or in other ways modifying the window might slightly change the variance of the extracted spectrum but the flux estimate $\bar{C}_y$ should remain unchanged. This is illustrated in Figure~\ref{Fig:OptialExtraction_2D_SHR_FLAT}, which shows a 2D representation of $C_{yx} / P_{yx}$ in detector coordinates.

\subsection{Ambiguity of the Trace Profile} 
\label{Sec:TraceProfileAmbiguity}

\begin{figure*}[hb]
 \centering
 \includegraphics[width=\linewidth]{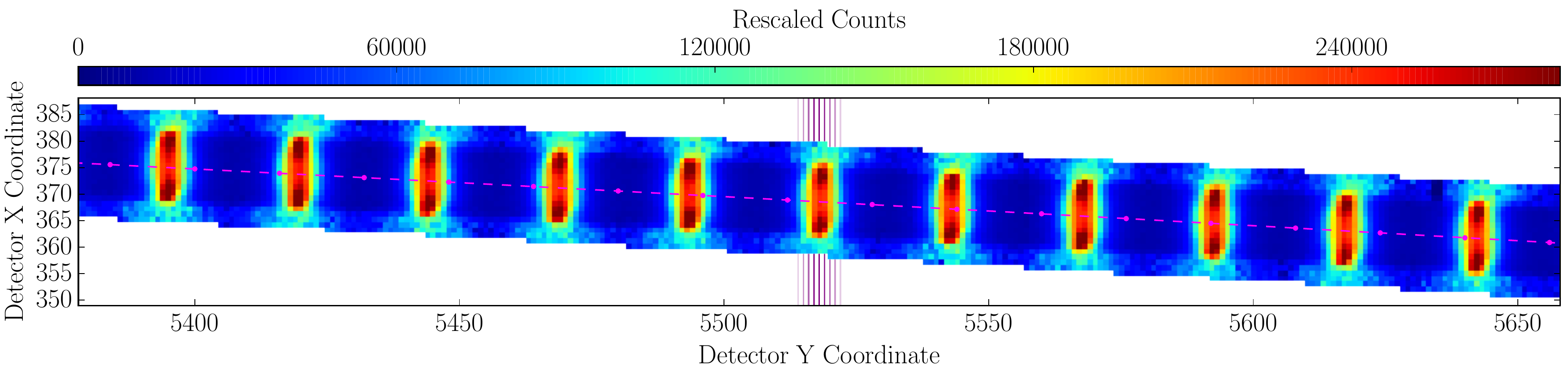}
 \caption{
  Extraction of a small part of a Fabry-P\'erot calibration frame. The plot shows the bias and scattered-light subtracted raw counts normalized by the trace profile. The region on the detector is identical to the one shown in Figure~\ref{Fig:OptialExtraction_2D_SHR_FLAT}.
  Clearly visible is the sequence of equally-spaced narrow emission lines. However, in contrast to the expectation, significant structure in cross-dispersion direction is present. The vertical purple lines correspond to the positions of the cuts shown in Figure~\ref{Fig:OptimalExtraction_1D_SHR}.
 }
 \label{Fig:OptialExtraction_2D_SHR_FP}
\end{figure*}

Many aspects of the extraction process, i.e., being optimal in the sense that it minimizes the variance of the extracted spectrum and the independence from the initial estimate of the trace center, are only valid as long as the assumed trace profile $P_{yx}$ is correct. As soon as the assumed profile deviates from the true one, this no longer holds and the extracted spectrum becomes imperfect. A slight increase of the variance in the output spectrum is acceptable to a certain degree, but the main concern within the context of a precision test of fundamental constants are biases in the extraction process that later introduce systematics in the wavelength solution. 

General shortcomings of the traditional \textit{optimal extraction} process have been described in detail by \citet{ABolton2010}. However, due to the complexity of the issue, we demonstrate in the following the particular effects for \Espresso.
For this, we focus on the extraction of a Fabry-P\'erot frame used for wavelength calibration. The Fabry-P\'erot interferometer (FP) produces a train of narrow (marginally resolved at $R\approx138\,000$), densely spaced ($\approx1.96\times10^{10}\,\mathrm{Hz}$ separation, corresponding to $0.1\,\mathrm{\AA}$ -- $0.4\,\mathrm{\AA}$) and equally bright lines and is therefore well-suited for wavelength calibration purposes.
The high number of equally spaced lines allows a very precise determination of the local wavelength solution, far better than possible with the sparse, unevenly distributed lines produced by the classical ThAr hollow cathode arc lamps.    

Figure~\ref{Fig:OptialExtraction_2D_SHR_FP} shows a small region of a FP spectrum in detector coordinates. Similar to Figure~\ref{Fig:OptialExtraction_2D_SHR_FLAT}, the raw counts are bias and scattered-light corrected and divided by the trace profile. Clearly visible is the series of narrow emission lines produced by the FP interferometer. However, in stark contrast to the flatfield frame, the FP frame exhibits very pronounced structure in cross-dispersion direction. Most importantly, the flux estimate for the lines is lower in the center of the trace compared to positions approximately five pixels above and below the center. In cross-dispersion direction, the FP lines basically resemble a \textit{double-horn} profile. 
In addition, the flux in the wings of the trace profile does not show the clear pattern of sharp emission lines but only some very limited modulation, staying far below the line flux measured in the core of the trace.
A similar but inverted behavior is seen in between the FP lines. Here, one finds lower flux in the trace center compared to its wings.
The data obviously does not match the expectations and clearly shows that the trace profile extracted from the flatfield frames is not capable of adequately describing the trace profile observed in FP frames. Since normalizing the observed counts by the trace profile does not lead to a constant flux estimate across the trace, the averaging process described in Equation~\ref{Eq:OptimalExtractionWindow} will not yield an unbiased estimate of the true flux and thus depends on the choice of the extraction window. 

The flatfield and FP frames were taken on the same day within less than one hour. Given the exquisite stability of \Espresso, instrumental drifts can be ruled out. The imperfect subtraction of scattered light could lead to a mismatch between the two trace profiles. However, we confirmed by not applying any scattered light subtraction at all that this is not the dominant effect driving the discrepancy seen in the extraction of the FP frame. 

\begin{figure*}[tb]
 \centering
 \includegraphics[width=\linewidth]{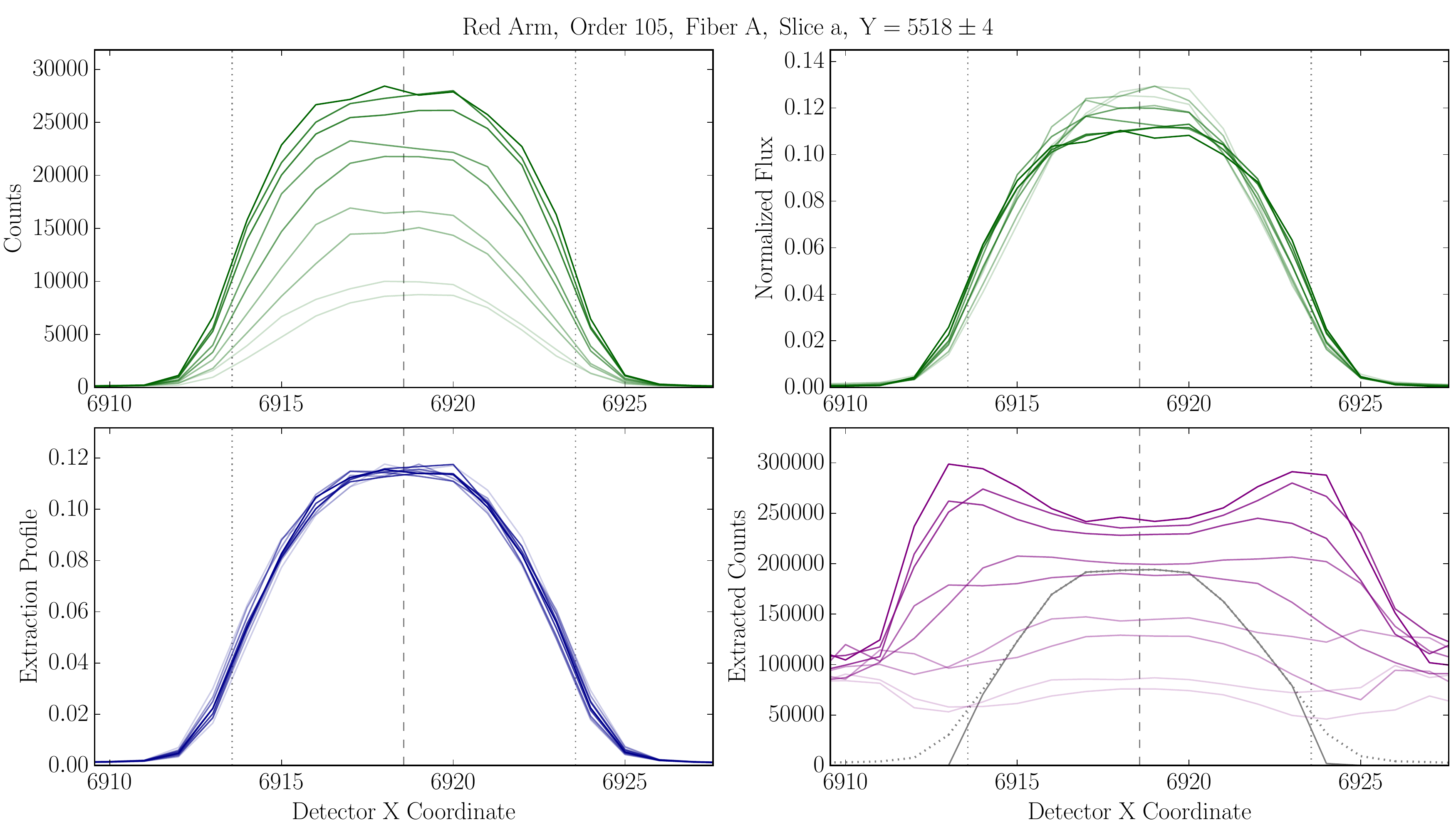}
 \caption{
  Illustration of various cuts through the trace in detector $X$ direction (cross-dispersion).
  \textit{Top left panel:} Nine trace profiles shown in terms of raw counts, i.e., $C_{yx}$. The central cut lies exactly at the peak of a FP~line. The others are offset in detector $Y$ (wavelength) direction by up to $\pm 4$~pixels. The saturation of the line color indicates the offset from the line.
  \textit{Top right panel:} Same as top-left panel but all profiles are normalized to unity flux, highlighting the different shapes.
  \textit{Bottom left panel:} Trace profiles extracted from a spectral flatfield frame at the same positions as the curves in the top panels, i.e $P_{yx}$. No significant dependence on the position is observed.
  \textit{Bottom right panel:} Raw counts from the FP frame normalized by the corresponding trace profiles, i.e., the quantity $C_{yx} / P_{yx}$. The gray dotted curve shows the inverse-variance weighting $W^\mathrm{var}_{yx}$ (arbitrary scale). The solid gray curve represents the full weighting function including the window function $W^\mathrm{win}_{yx}$, which limits the extraction to $\pm 5$~pixels (vertical dotted lines).   
  }
 \label{Fig:OptimalExtraction_1D_SHR}
\end{figure*}

Figure~\ref{Fig:OptialExtraction_2D_SHR_FP} also shows that the FP lines are slightly tilted with respect to the detector coordinates. This tilt is small and corresponds only to a fraction of a pixel difference between upper and lower part of the trace but still might be relevant.
The final wavelength solution is required to be \textit{accurate} to better than $\simeq20\,\mathrm{m/s}$, corresponding to only 4\% of a pixel. Thus, even very minute effects can impact the data products at a very significant level. Due to the tilt of the lines and the need of an explicit truncation of the trace profile, the initial fit to the trace center could have an impact on the final wavelength solution.

More detailed insights can be gained by inspecting cuts through the trace in detector $X$ direction (cross-dispersion direction) and Figure~\ref{Fig:OptimalExtraction_1D_SHR} shows a series of such cuts. The central profile (shown in the strongest color) coincides with the peak of an arbitrary chosen Fabrry-P\'erot line. In addition, eight additional profiles are shown, which are offset in detector $Y$ direction (wavelength direction) by up to $\pm 4$~pixels in 1~pixel steps.  

The top-left panel of Figure~\ref{Fig:OptimalExtraction_1D_SHR} illustrates these cuts in (bias and scattered-light corrected) raw counts. 
Obviously, the cut through the center of the FP line shows higher flux than the ones through the wings of the line. 
In addition, it is clear that the trace profiles are not Gaussian. Instead, they are flatter at the center and then drop off faster than a Gaussian at the flanks. However, this behavior is not identical for all nine profiles.

To highlight this difference in shapes, the top-right panel of Figure~\ref{Fig:OptimalExtraction_1D_SHR} shows the same cuts normalized to unity integrated flux. It becomes clear that the flattening of the profile at the center is more pronounced for the central profiles while the cuts through the wings of the line are actually more Gaussian.
This clearly indicates that the trace profile changes across the FP line. 

However, this is \textit{not} related to a change of the trace profile with detector position but instead intrinsic to the emission line. 
To demonstrate this, the bottom-left panel of Figure~\ref{Fig:OptimalExtraction_1D_SHR} shows cuts at the identical positions but extracted from the master flatfield. Clearly, they are all identical, showing that the trace profile for a white-light source does not vary, while for the FP line the trace profiles changes. This discrepancy is purely related to the spectral shape of the source, i.e., narrow emission line vs. broadband emission. The existence of \textit{one universal} trace profile that is independent of the SED of the source, which is the fundamental assumption of the \citet{Horne1986} optimal extraction scheme, is therefore not given. This demonstrates that the spectral shape of the source and the trace profile observed on the detector are coupled and not separable.

The observed behavior leads to undesired effects when averaging the flux estimates in cross-dispersion direction. The bottom-right panel of Figure~\ref{Fig:OptimalExtraction_1D_SHR} shows the bias and background subtracted counts normalized by the trace profile, i.e., $C_{yx} / P_{yx}$ as given in Equation~\ref{Eq:OptimalExtractionWindow}, for nine detector columns. All pixels of a given cut should represent the same unbiased estimate of the total received flux at that detector $Y$ position. Thus, the plotted lines should be flat with just increasing noise away from the center of the trace. However, one finds the already mentioned \textit{double-horn} profile with lower flux in the center of the trace compared to about $\pm 5$~pixels above or below. This is at least true for the cuts that go through the center of the FP line. Cuts through the wings of the line show the opposite shape, with the regions about $\pm6$~pixels away from the center showing less flux than the center itself. 
This will certainly impact the accuracy of the extracted spectra and possibly lead to biased estimates. In particular, the chosen width of the extraction window now directly influences the extracted flux and the resulting spectral line shape. It has to be stressed that the extraction process works perfectly fine for broadband spectra, but shows the illustrated systematics for spectra that are not flat. This will unavoidably have an impact on how spectral features will be recovered and therefore on the wavelength calibration and the measurement of quasar absorption lines. 

\begin{figure}
 \includegraphics[height=7.15cm]{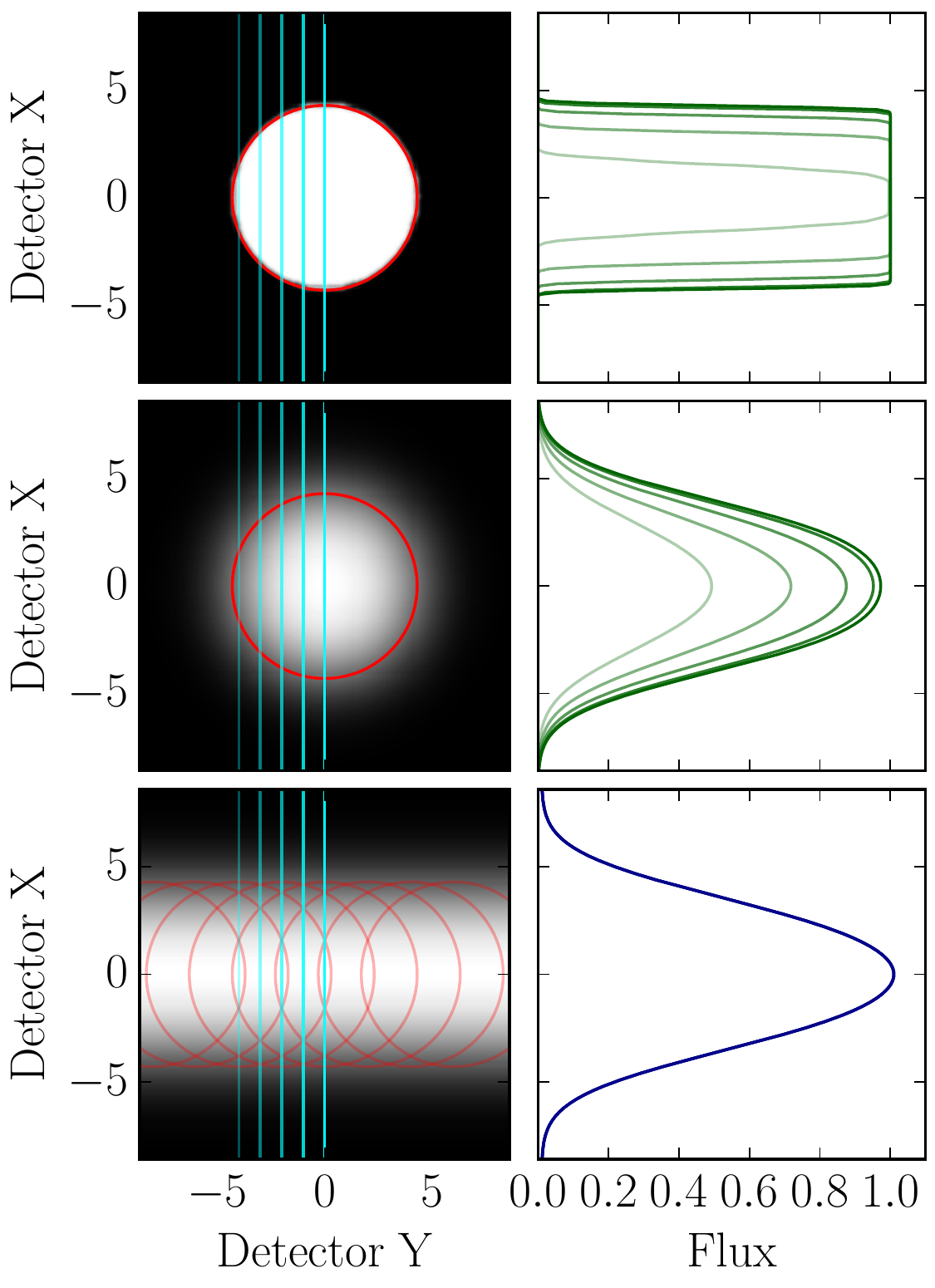} \hfill
 \includegraphics[height=7.15cm]{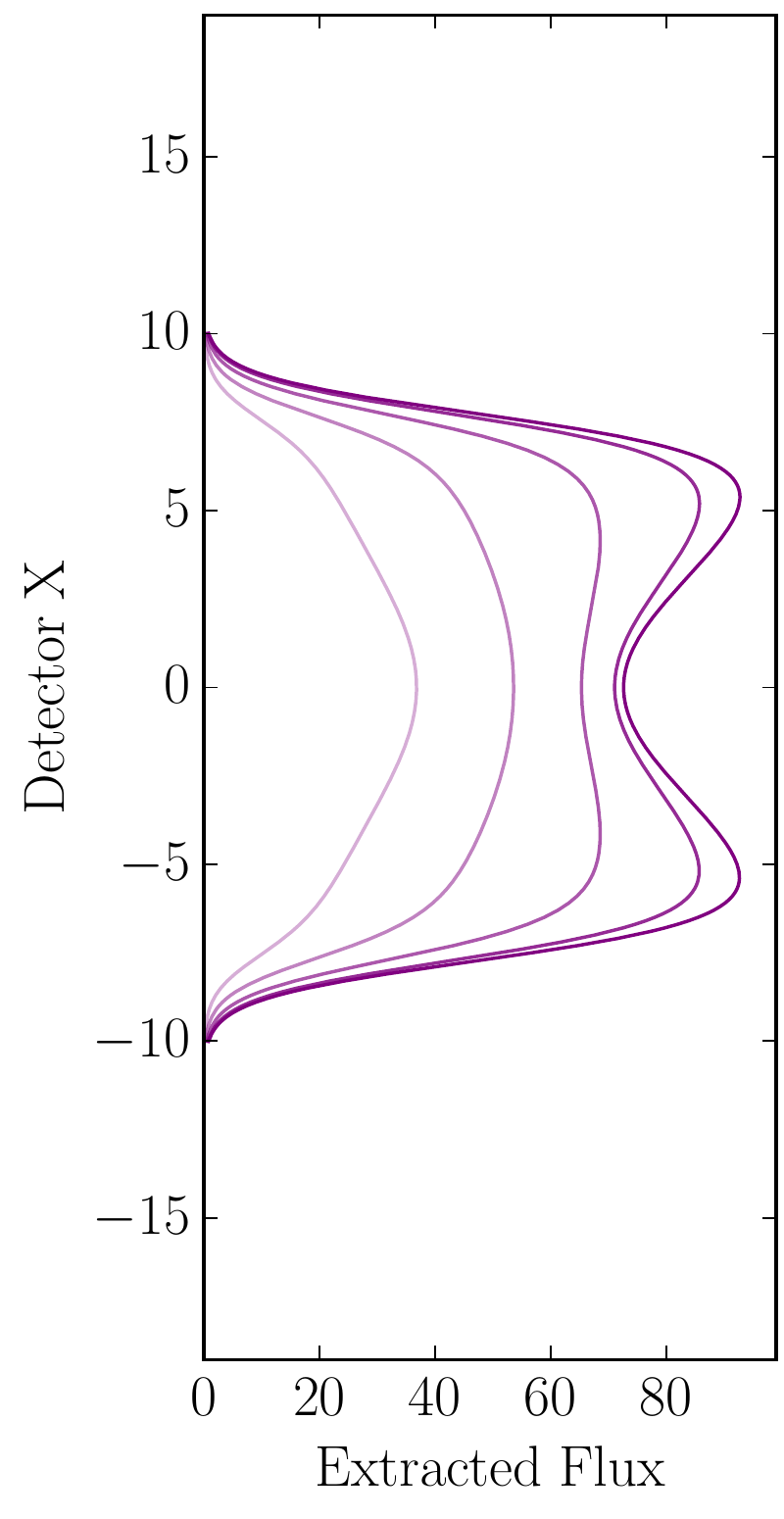}
 \caption{Simple model illustrating the mismatch of trace profiles between FP and flatfield spectra. The left column shows the flux distribution on the detector for a monochromatic source fed to an idealized fiber-spectrograph (top), the same monochromatic source affected by some instrumental blurring (center), and a spectrally-flat broadband source (bottom). The central column shows cuts through these flux distributions in cross-dispersion direction at different locations. The positions of the cuts are indicated in the left panels by vertical cyan lines. Decreasing color saturation corresponds to increasing offset from the center.
 The right panel displays the cross-sections from the central row divided by the profiles shown in the bottom row of panels. This corresponds to the normalization of FP spectra by trace profiles extracted from a flatfield spectrum. Clearly, there is a discrepancy between the trace profiles and the normalized flux does not exhibit the desired flat distribution along cross-dispersion direction.}
 \label{Fig:MockTraceProfile}
\end{figure}

These rather complicated properties of the trace profile can actually be understood considering the design of the spectrograph. 
We therefore show in Figure~\ref{Fig:MockTraceProfile} a toy model, illustrating the basic principles that lead to a mismatch in the trace profile between flatfield and FP spectra.

For an idealized spectrograph and a monochromatic light source, the flux distribution on the detector is a direct image of the (pseudo-)slit, in this case of the optical fiber feeding the spectrograph. This is shown in the top-left panel of Figure~\ref{Fig:MockTraceProfile} (assuming for simplicity a circular fiber instead of the octagonal ones utilized in \Espresso).
A cross-section in cross-dispersion direction through the flux distribution would resemble a top-hat profile.  
If the cross-section goes right through the center of this observed emission line, the width of the top-hat corresponds to the full diameter of the fiber, while taking a cut a few pixels offset from the flux center would result in a narrower top-hat.
A series of such cuts is displayed in the central-top panel of Figure~\ref{Fig:MockTraceProfile}.

The introduction of diffraction and instrumental aberrations (here approximated by simple Gaussian blurring) will soften the flux distribution on the detector. Still, the cross-section through the center of the emission line is wider and fatter compared to the off-center profiles, which are closer in shape to a Gaussian. This is illustrated in the central and central-left panel of Figure~\ref{Fig:MockTraceProfile}.

For a broadband source, the flux distribution on the detector  can be seen as a superposition of many point-spread functions corresponding to the individual wavelengths (indicated with multiple red circles).  In case of a spectrally flat source, all cross-sections will have an identical shape, independent of their position. This case is shown in the bottom row of Figure~\ref{Fig:MockTraceProfile} and corresponds to a spectral flatfield frame.

If now trace profiles determined from a flatfield frame are used to normalize the flux corresponding to an isolated emission line, e.g., from a FP or LFC spectrum, the resulting normalized flux will not show the desired flat shape. Instead, it will resemble a double-horn profile for the position right at the center of the line and a bell shape for its wings. This is illustrated in the right panel of Figure~\ref{Fig:MockTraceProfile}, which shows the profiles from the central row divided by the corresponding trace profiles from the bottom row, i.e., the quantity $C_{yx} / P_{yx}$ given Equation~\ref{Eq:OptimalExtraction}.

Although rather simple, this toy model shown in Figure~\ref{Fig:MockTraceProfile} reproduces at least qualitatively the effects found on real \Espresso{} data to a remarkably high degree (compare to Figure~\ref{Fig:OptimalExtraction_1D_SHR}). It demonstrates that for roundish fibers, a mismatch between trace profiles naturally and unavoidably occurs as soon as the spectral shapes deviate (spectrally flat broadband source vs. individual emission line) and also illustrates the fundamental limitation of the adopted extraction scheme based on \citet{Horne1986} and \citet{Zechmeister2014}.
The use of rectangular fibers might (similar to a slit) reduce this particular problem since in this case the trace profile should depend far less on the position at which it intersects the pseudo-slit image.
However, the only proper solution would be to model the flux distribution on the detector as superposition of individual 2D point-spread functions weighted by the spectral energy distribution of the source.

An algorithm that actually does this is \textit{Spectro-Perfectionism} described by \citet{ABolton2010}. 
However, this approach is conceptually and computationally extremely challenging and was only recently adopted by the Dark Energy Spectroscopic Instrument (DESI) consortium.
In addition, such an approach would, despite the tremendous requirements on computational resources, also pose very significant demands on the calibration procedure, which has to deliver a model of the 2D point-spread function for every wavelength of interest and therefore every location on the detector. It is unlikely that this can be done using the Fabry-P\'erot interferometer, since its lines are resolved by \Espresso. Also, the number of unblended and sufficiently bright ThAr lines produced by the hollow-cathode lamps used for wavelength calibration is extremely low and most-probably too sparse for a proper characterization of the PSF. In addition, also the ThAr lines  are marginally resolved. The laser frequency comb would in principle be suitable to characterize the \Espresso{} line-spread function since it delivers a plethora of extremely narrow ($\simeq100\,\mathrm{kHz}$) lines. However, it covers only 57\% of the \Espresso{} wavelength range and the individual lines might not be sufficiently separated ($\Delta{}\nu_\mathrm{LFC} = 18\,\mathrm{GHz}$) to fully disentangle them. 
For the time being, one therefore has to proceed with the classical extraction scheme based on \citet{Horne1986} and \citet{Zechmeister2014} described above but be aware that it is nonoptimal when it comes to details.

\section{Wavelength Calibration}
\label{Sec:WavelengthCalibration}

After extraction, a full \Espresso{} exposure is represented as 340 1D spectra (Order 161 to 117 from blue arm, Order 117 to 78 from red arm, $\times2$ fibers, $\times2$ slices), in the terminology of the DRS called the \texttt{S2D} format%
\footnote{Despite the misleading description, \texttt{S2D} does not describe 2D spectra but a collection of single-order, unmerged one-dimensional spectra.}.
Each individual spectrum has a length of 9232 pixels in detector $Y$ direction and the basic assumption of the extraction procedure is that these can be directly mapped to wavelengths.

To do so, the \Espresso{} calibration plan includes three types of wavelength calibration frames:
\begin{itemize}
 \item Exposures of a Thorium-Argon (ThAr) hollow cathode lamp provide absolute wavelength information by the means of \ion{Th}{i} emission lines. These quantum-mechanical transitions have been accurately measured to $\simeq5\,\mathrm{m/s}$ in laboratory experiments using Fourier-transform spectroscopy \citep[e.g.,][]{Redman2014}. However, many thorium lines are blended, contaminated by argon lines or simply not strong enough. The default line list therefore contains only 432 unique \ion{Th}{i} lines. In consequence, some \Espresso{} orders are covered by only two thorium calibration lines. This is by itself clearly not enough to derive an accurate and precise wavelength solution.  
 
 \item The Fabry-P\'erot interferometer (FP) therefore complements the information from the ThAr frames. It produces a dense series of rather narrow, nearly equally-spaced and equally-bright (apparent) emission lines across the full wavelength range \citep{Wildi2010, Wildi2011, Wildi2012}. 
 It was designed to produce lines with a separation of $\approx1.96\times10^{10}\,\mathrm{Hz}$, which corresponds to $0.1\,\mathrm{\AA}$ -- $0.4\,\mathrm{\AA}$. The lines are therefore marginally resolved and noticeably broader than the instrumental profile. The wavelengths are purely defined by the effective optical length of the Fabry-P\'erot cavity ($\simeq15.210\,\mathrm{mm}$). However this might change due to variations in temperature or pressure. 
 Without stabilization to any reference, the FP provides no absolute wavelength information. Instead, it has to be characterized by comparison to ThAr frames. Despite this, the large number of densely-spaced FP lines (about 300 per order) allows a very precise wavelength calibration on small scales, completely impossible with arc lamps alone. Therefore,  ThAr and FP provide highly complementary information and combining both allows to derive a precise and accurate wavelength solution over the full \Espresso{} wavelength range%
 \footnote{This of course requires that there is no significant instrumental drift between ThAr and FP exposures. However, this is no issue within the context of this study.}%
, which is in the following denoted as ThAr/FP solution.  
 
 \item A completely independent means of calibration is provided by the laser frequency comb (LFC), a passively mode-locked laser that emits a train of femtosecond pulses producing a set of very sharp emission lines \citep{Wilken2010a, Wilken2012, Probst2014, Probst2016}. The frequencies of the individual lines $\nu_k$ follow exactly the relation $\nu_k = \nu_0 + k \times \nu_\mathrm{FSR}$. The offset frequency $\nu_0$ and separation of the lines $\nu_\mathrm{FSR}$ are actively controlled and compared to a local (radio) frequency standard, which itself is stabilized against an atomic clock or GPS. Therefore, the accuracy of the fundamental time standard is transferred into the optical regime, in principle providing absolute calibration with and accuracy at the $10^{-12}$ level. However, this extreme accuracy of the laser frequency comb itself does not translate one-to-one into the accuracy of the final wavelength solution of the spectrograph and the details of this process are outlined in the following sections. 
 The LFC is operated with a mode-spacing of $\nu_\mathrm{FSR}=18.0\,\mathrm{GHz}$ and an offset frequency of $\nu_0=7.35\,\mathrm{GHz}$. The spacing of the LFC lines on the detector is therefore very similar to the ones of the FP. In  addition, the LFC lines have an extremely narrow width of $\simeq100\,\mathrm{kHz}$, corresponding to less than $\simeq \sfrac{1}{10\,000}$ of the spectrographs resolution in \texttt{1HR} mode. It is therefore the only calibration source that allows an accurate characterization of the instrumental line-spread function and spectral resolution.
 However, due to fundamental technical challenges and design choices, the \Espresso{} LFC covers only $\approx57\%$ of the wavelength range, with significant flux only from Order~132 ($4635\,\mathrm{\AA}$) to Order 85 ($7200\,\mathrm{\AA}$), and substantial uncovered regions at the blue and red end of the wavelength range%
 \footnote{We stress that the LFC flux levels are not stable and the usable wavelength range can change substantially from epoch to epoch.}
. 
 A concept based on two LFCs ($3800$\,--\,$5200\,\mathrm{\AA}$ and $5200$\,--\,$7600\,\mathrm{\AA}$) to cover the full spectral range of \Espresso{}, as initially envisioned \citep{Megevand2014}, is unfortunately not realized.
 In addition, the LFC is still rather unreliable with only $\simeq25\%$ availability since the start of regular observations in October 2018.
\end{itemize}

These wavelength calibrations, as well as flatfield and bias frames, are taken daily, usually in the morning, as part of the standard \Espresso{} calibration scheme. We make sure that all exposures processed together are from the same calibration session and therefore taken within less than 2.5\,hours. Instrumental drifts are therefore negligible.

\subsection{Line Fitting}
\label{Sec:LineFitting}

To process the wavelength calibration frames, the flux is extracted in the way described in Section~\ref{Sec:Extraction} and in addition de-blazed. This is essential to avoid a biasing of the determined line positions.
The blaze function is determined by optimal extraction from the master flatfield, identical to the way all other spectra are extracted. It therefore contains all sensitivity variations in detector $Y$ direction (wavelength direction), i.e., column-to-column and large-scale variations due to the blaze or transmission properties of the spectrograph, while the trace profile as outlined in Section~\ref{Sec:TraceProfile} captures sensitivity variations across the trace, i.e., in detector $X$ direction.
All extracted spectra are normalized by the blaze function determined in this way. Thus, the fluxes stated in the following are relative to the flux of the flatfield light source%
\footnote{An Energetiq laser-driven light source (LDLS) EQ-99X.}, 
which provides a rather featureless and flat spectrum.

After extraction and de-blazing, the individual emission lines are fitted. For the ThAr spectrum, this is done based on a line list with good initial guess positions. Lines are fitted with Gaussian functions plus a constant offset within a window of $\pm10\,\mathrm{km/s}$ 
around the initial guess position.
The used ThAr line list is identical with the one used by the \Espresso{} DRS \citep{Lovis2020} and the laboratory wavelengths and their uncertainties are taken from \citet{Redman2014}. 

\begin{figure}[h]
 \centering
 \includegraphics[width=\linewidth]{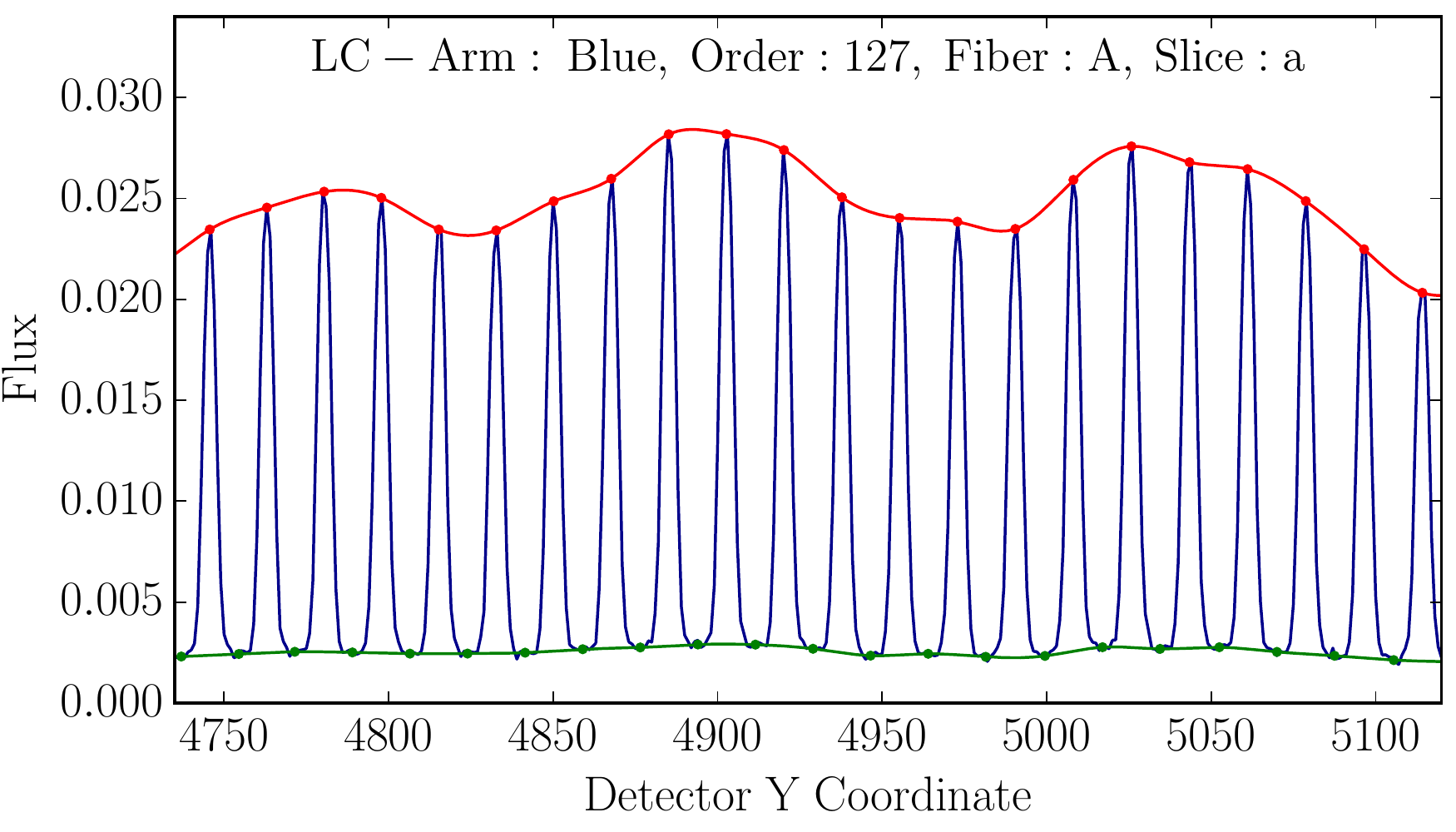} 
 \caption{
  Small part of a laser frequency comb spectrum, showing the dense forest of equally-spaced narrow emission lines. The flux is given relative to the flatfield. The LFC spectrum exhibits a significant background light contribution (green) and modulation of the line intensities (red), which are both modeled as part of the analysis with cubic splines.
  }
 \label{Fig:LFC_BackgroundEnvelope}
\end{figure}

Fitting of FP and LFC lines is done in an identical way. Starting from the center of an order, line peaks are identified and fitted, again with a Gaussian plus constant background model. Line fits are rejected if constraints on minimum or maximum width of the line or signal-to-noise ratio are not met. The photon-counting and RON errors from the raw frames are taken into account during the fitting process and fully propagated into the uncertainty of the line center estimate. Typical uncertainties on the position of individual, fully exposed spectra (reaching at the peak close to the full-well capacity of $\simeq60\,000\,\mathrm{ADUs}$) are about $1.5\,\mathrm{m/s}$ in \texttt{1HR1x1} binning mode and $2\,\mathrm{m/s}$ in \texttt{1HR2x1} mode.



\begin{figure*}
 \centering
 \includegraphics[width=\linewidth]{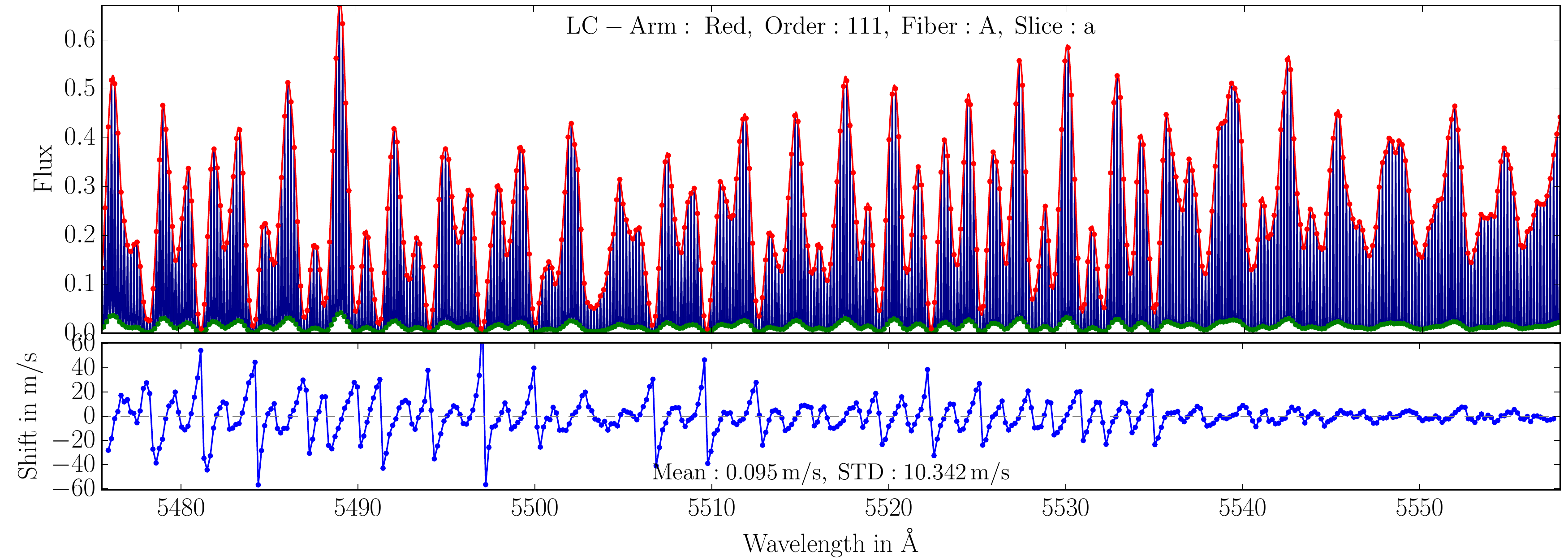}\\[10pt]
 \includegraphics[width=\linewidth]{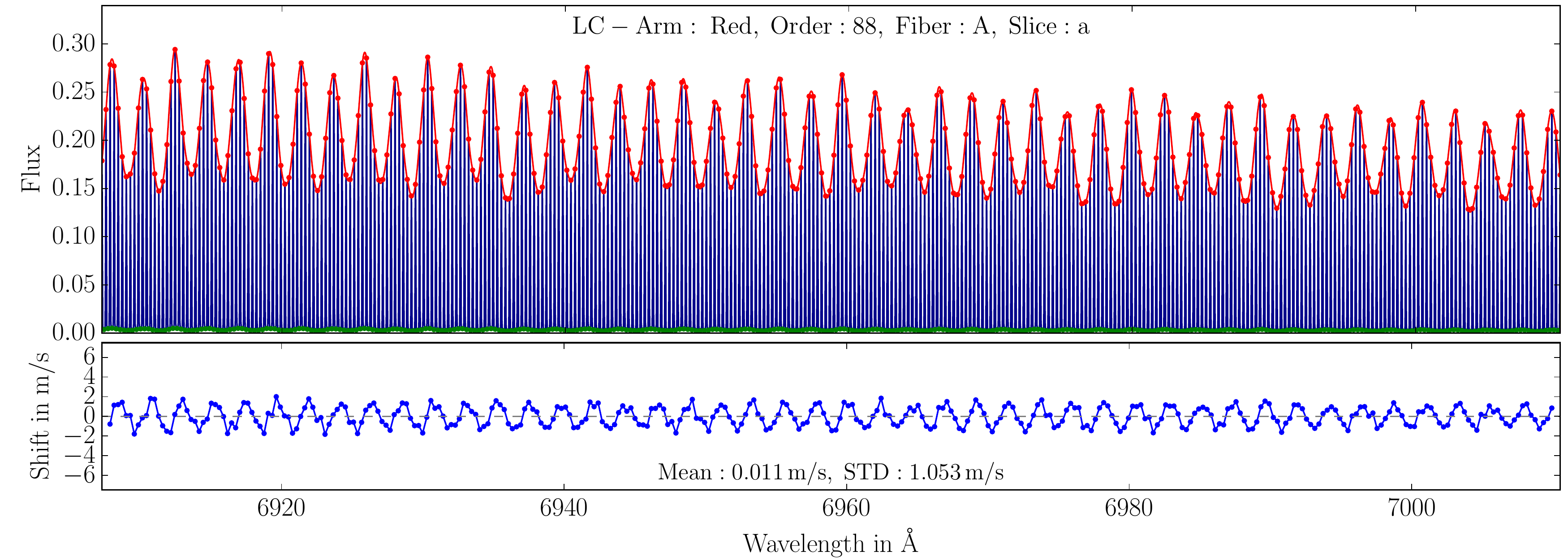}
 \caption{
  Laser frequency comb spectra of two selected spectral orders (top panels) together with the differences of the line positions obtained in fits with and without subtracting the background light model (bottom panels). Identical to Figure~\ref{Fig:LFC_BackgroundEnvelope}, models for the LFC flux envelope and background light are shown in red and green.
  The line shifts clearly correlate in amplitude and phase with the spectral shape of the background light.
  These two examples also highlight that the modulation patterns of flux envelope and LFC background light can be vastly different across the spectral range.
  }
 \label{Fig:LFC_BackgroundResiduals}
\end{figure*}

While this procedure provides usable line lists, it has to be noted that the laser frequency comb exhibits in addition to the individual emission lines a significant amount of background light that is likely related to amplified spontaneous emission in the optical Yb-fiber amplifiers of the LFC system \citep[see. e.g.,][]{Milakovic2020a}. 
Since the background light intensity is strongly modulated (see Figures~\ref{Fig:LFC_BackgroundEnvelope} and \ref{Fig:LFC_BackgroundResiduals}), it can lead to biases in the centroiding of lines, in particular if the local slope of the background is not modeled properly. A possible solution would be to fit for each line a higher order polynomial (at least of order one) in addition to the Gaussian and thereby accounting for the structure in the background. 
However, we follow a different approach by constructing a global background light model for each spectral trace. 
For this, the central 25\% between two neighboring emission peaks are median combined to form a spline point. All spline points are then connected by cubic spline interpolation to form a continuous model of the background light. This is illustrated in Figure~\ref{Fig:LFC_BackgroundEnvelope}. 
After subtracting the background model, the LFC lines are fitted a second time, resulting in improved estimates for the line centroids. 

Besides the background light, Figure~\ref{Fig:LFC_BackgroundEnvelope} shows that the LFC lines are not of equal intensity but modulated by a flux envelope function. This is most-likely caused by the Fabry-P\'erot cavities used in the LFC system to filter the spectrum. For technical reasons, the fundamental laser comb operates with a line separation (or equivalent repetition rate) of $250\,\mathrm{MHz}$. This is far too narrow to be resolved by \Espresso{}. The light from the fundamental comb is therefore filtered by three Fabry-P\'erot cavities, which remove the majority of the lines and let only one in 72 lines pass, thereby increasing the line separation to $18\,\mathrm{GHz}$ \citep[see e.g.,][]{Probst2016}. Similar to the treatment of the background light, we construct a model for the flux envelope by identifying the individual line peaks and connecting them by cubic spline interpolation (see Figure~\ref{Fig:LFC_BackgroundEnvelope}).
The modulation of flux envelope and background light are very similar (well visible also in Figure~\ref{Fig:LFC_BackgroundResiduals}), suggesting that the background light is likely produced by amplified spontaneous emission in the fiber amplifiers but its modulation related to the Fabry-P\'erot cavities.

This model for the flux envelope would in principle allow to fully normalize the LFC spectrum. However, it was decided \textit{not} to do so.
As stated before, the lines produced by the LFC are intrinsically extremely narrow ($\simeq100\,\mathrm{kHz}$). For each line, the modulation of the flux envelope is therefore only sampled over the extremely narrow range of the intrinsic line width. The observed line profile, however, is dominated by the instrumental line broadening.
Any intensity slope present within an individual LFC line is therefore \textit{diluted} by approximately the ratio between the width of the instrumental line profile and the intrinsic LFC line width. Since the instrumental profile dominates by more than a factor of $10\,000$, it was decided to not use the fitted flux envelope model for a normalization of the LFC spectrum. The behavior for the background light is different since it represents a broad-band component and in contrast to the emission lines is present at any wavelength.

To quantify the effect of the background light subtraction procedure, we compare the difference of the line positions obtained in the two fits. This is shown together with the LFC spectra in Figure~\ref{Fig:LFC_BackgroundResiduals} for two specific spectral orders.
As can be seen, there is a clear effect on the determined centroid of the individual lines that is closely correlated with the modulation of flux envelope and background light. The mean centroid shift per spectral order is always extremely small, on the order of a few $\mathrm{cm/s}$. The standard deviation of the line shifts depends on the amount of modulation. For the usual case of e.g., 20\% modulation, we find a few $\mathrm{m/s}$ (bottom spectrum shown in Figure~\ref{Fig:LFC_BackgroundResiduals}).
However, the amount and periodicity of the modulation varies strongly across the spectral range and the top panel of Figure~\ref{Fig:LFC_BackgroundResiduals} shows an extreme case with nearly 100\% modulation. Here, the standard-deviation of the shifts amount to $10\,\mathrm{m/s}$ and individual line positions differ by up to $60\,\mathrm{m/s}$.

\begin{figure*}
 \centering
 \includegraphics[width=\linewidth]{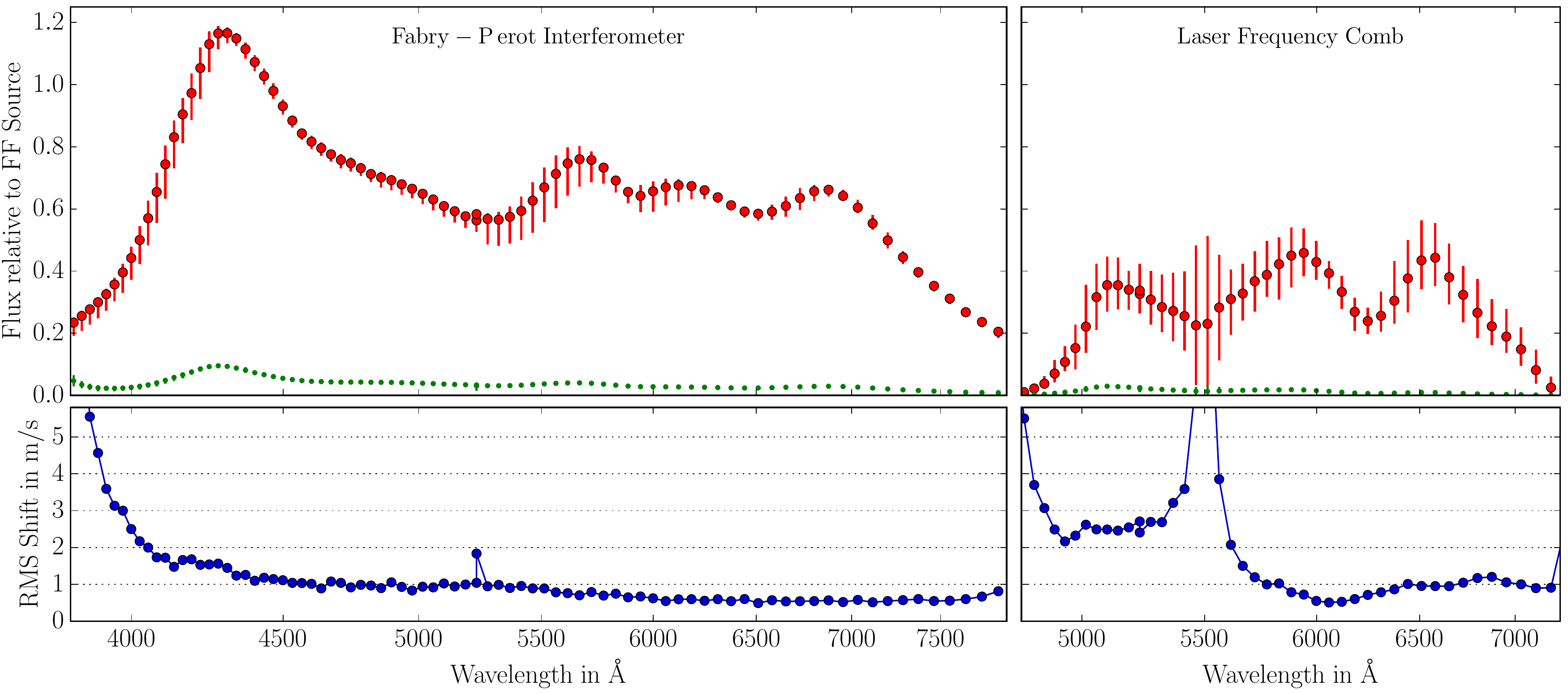}
 \caption{
  Flux envelope and background light characteristics of FP and LFC calibration light sources and the line centroid differences between the fits with and without background light subtraction.   
  The top panels show for each spectral order the median and 95\% range of flux envelope and background light, i.e., the statistical properties of flux and background light shown explicitly for two orders in Figure~\ref{Fig:LFC_BackgroundResiduals}. The bottom panels display per order the scatter of the centroid shifts, computed from the 16th--84th percentile range. For simplicity, only Slice~a of Fiber~A is shown. Note the reduced wavelength range but identical scaling for the LFC.
  }
 \label{Fig:LFC_BackgroundResidualsSummary}
\end{figure*}

Figure~\ref{Fig:LFC_BackgroundResidualsSummary} summarizes this behavior for the full wavelength range. It shows for each spectral order the median and 95\%~percentile range of flux envelope and background light.
Apparent is the fast drop-off in LFC intensity for wavelengths shorter than $5100\,\mathrm{\AA}$ and the gradual decrease redwards of $6500\,\mathrm{\AA}$. The most striking feature, however, is the extreme modulation of the LFC flux for wavelengths around $5500\,\mathrm{\AA}$, responsible for large centroid shifts. This highlights the importance of a proper modeling and subtraction of the LFC background light. 
However, the modulation pattern of the LFC flux envelope and background light is by no means stable and changes significantly with time, in particular after interventions to the LFC system. The data reduction and wavelength calibration routines therefore have to be able to adapt dynamically to changing conditions. In addition, the wavelength range for which a proper LFC wavelength calibration can be obtained might vary.  

As demonstrated in Figure~\ref{Fig:LFC_BackgroundResidualsSummary}, the spectrum of the Fabry-P\'erot interferometer shows far less modulation of the flux envelope and nearly constant background levels. In consequence, the typical differences in the line positions when subtracting the background model compared to not doing so are found to be  $\lesssim1\,\mathrm{m/s}$ over most of the wavelength range. 
Therefore, the detailed modeling and subtraction of the FP background light might for the scope of the fundamental physics project not be necessary.
In particular, the metal absorption systems used for constraining the fine-structure constant usually extend over a few hundred $\mathrm{km/s}$ and their wavelength determination does not rely on a single but many FP or LFC lines. Possible line shifts will therefore be averaged down and probably reduced to a negligible level. However, for the sake of consistency with the LFC spectra, we still perform the full procedure. In addition, proper modeling and removal of background light contamination is clearly required to reach photon-limited accuracy on single FP or LFC lines.
Figure~\ref{Fig:LFC_BackgroundResidualsSummary} also shows that a more homogeneous distribution of LFC flux envelope and background intensities would be desirable. Here, not the amplitude of the background light is the issue but its variation on small scales.

\subsection{Beat Pattern Noise}
\label{Sec:BeatPatternNoise}

The line fitting procedure described above delivers for each trace extracted from the FP or LFC spectra a list of lines with indices $i$ and line positions in pixel coordinates $y_i$, uncertainties, as well as widths and intensities. 
The frequencies of the lines for the FP are
$\nu^\mathrm{FP}_i = ( \, i^\mathrm{FP} + i_0^\mathrm{FP} \, ) \times \frac{c}{D_\mathrm{eff}(\lambda)}$
and
$\nu^\mathrm{LC}_i = \nu^\mathrm{LC}_0 + ( \, i^\mathrm{LC} + i_0^\mathrm{LC} \, ) \times \nu^\mathrm{LC}_\mathrm{FSR}$ for the LFC. 
While the LFC offset frequency $\nu^\mathrm{LC}_0$ and line separation $\nu^\mathrm{LC}_\mathrm{FSR}$ are actively controlled and known a-priori, the effective gap size of the Fabry-P\'erot interferometer $D_\mathrm{eff}(\lambda)$ still has to be characterized. The details of this are described later in Section~\ref{Sec:JointThArFPSolution}. However, $D_\mathrm{eff}(\lambda)$ varies only very slightly around $15.210\,\mathrm{mm}$.
Also, the index $i$ is still relative and set to zero for an arbitrary line at the center of the trace, hence the offset $i_0$. Still, the index $i$ is directly proportional to the frequency of the lines and the relation between index $i$ and position $y$ on the detector immediately reflects the wavelength solution of the given order. It is expected that this relation is monotonic and smooth.

The predicted smoothness can be tested. We therefore run a kernel smoothing filter over the relation $y(i)$. The smoothing algorithm was inspired by the \citet{SavitzkyGolay1964} filter and in a similar way locally approximates the data by a polynomial function. However, it is much more flexible, i.e., it does not require regularly sampled data, can handle missing data and allows for a weighting of the datapoints to take uncertainties into account. In addition, it is not restricted to a top-hat filter but can work with arbitrary filtering kernels, in particular a Gaussian. In this way, the algorithm is basically a low-pass filter, but by adopting a higher-order polynomial as local approximation it avoids the biasing introduced by classical running-mean or Gaussian-smoothing filters, which implicitly assume that the data is properly described by a constant relation.  

\begin{figure*}[tb]
 \centering
 \includegraphics[width=\linewidth]{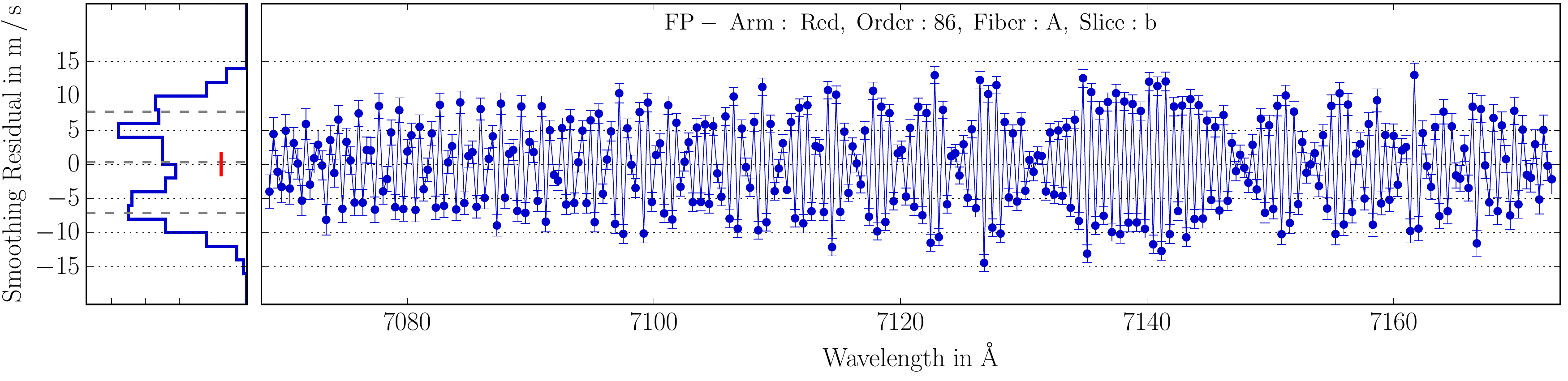} 
 \caption{Position difference between locations of individual FP lines and a smooth wavelength solution. The smooth wavelength solution was obtained in a nonparametric way by applying a kernel smoothing filter (basically a low-pass) to the FP line positions. Individual points are connected with lines to better visualize the quasi-periodic behavior.
 The left panel shows a histogram of these smoothing residuals with indications for the 16th, 50th, and 84th percentile of the distribution. The mean formal position error of the lines is visualized by a red bar.  
  }
 \label{Fig:PatternNoise_FP305}
\end{figure*}

We apply this filter using a third degree polynomial and a Gaussian kernel with $\sigma=35\,\mathrm{km/s}$, corresponding to a full width at half maximum (FWHM) of 5 to 11 Fabry-P\'erot or 6 to 12 LFC lines. Figure~\ref{Fig:PatternNoise_FP305} shows for a single FP order the difference between smoothed and unsmoothed line positions.
Since the kernel smoothing acts as a low-pass filter, the residuals of the filtering process shown in Figure~\ref{Fig:PatternNoise_FP305} represent the high-frequency component of the $y(i)$ or $y(\nu_i)$ relation. 

These smoothing residuals reveal numerous concerning effects.
First of all, the scatter is much larger than the photon noise (indicated with error bars for each individual FP line) and highly non-Gaussian. 
This is highlighted in the left panel of Figure~\ref{Fig:PatternNoise_FP305}, which shows a histogram of the smoothing residuals. 
The mean error of the individual line position measurements is indicated by a red bar. If the measurements were limited by photon noise, one would expect an approximately Gaussian-shaped distribution with a width close to the mean error ($\approx1.6\,\mathrm{m/s}$). Instead, the distribution nearly resembles a double-horn profile with a nonparametric standard deviation estimated from the 16th--84th percentile interval of about $8\,\mathrm{m/s}$. 
%
Thus, the smoothing residuals clearly exceed the photon noise. In addition, they are not the result of any other source of random stochastic scatter.
Instead, they exhibit a very peculiar, highly correlated pattern, indicative of systematic effects.
In particular, consecutive FP lines show with a striking regularity alternately positive and negative smoothing residuals. In addition, the amplitude of the smoothing residuals appears to be modulated along the order in a way that resembles a beat pattern and thus might be caused by some sort of interference. We therefore refer to these systematics as the \textit{beat pattern noise}.

It has to be stressed that the shown behavior does not depend on the details of the applied smoothing filter. With proper visualization of the data, these systematics can be picked up by eye without invoking any filtering. They are therefore intrinsic to the data and the determined line positions indeed suffer from a form of systematic, highly correlated noise.

The observed pattern of the smoothing residuals varies strongly from order-to-order. In general, its amplitude is slightly larger for Slice~b than Slice~a but very similar for the two Fibers. Also, it is present in Fabry-P\'erot and laser frequency comb spectra. The modulation of the pattern appears more pronounced and clearer in FP spectra compared to the LFC frames, but for the same Order, Fiber and Slice, the amplitude is slightly larger in the LFC frames, however, with very little similarity in the shape of the pattern itself.

\begin{figure*}[t]
 \centering
 \includegraphics[width=\textwidth]{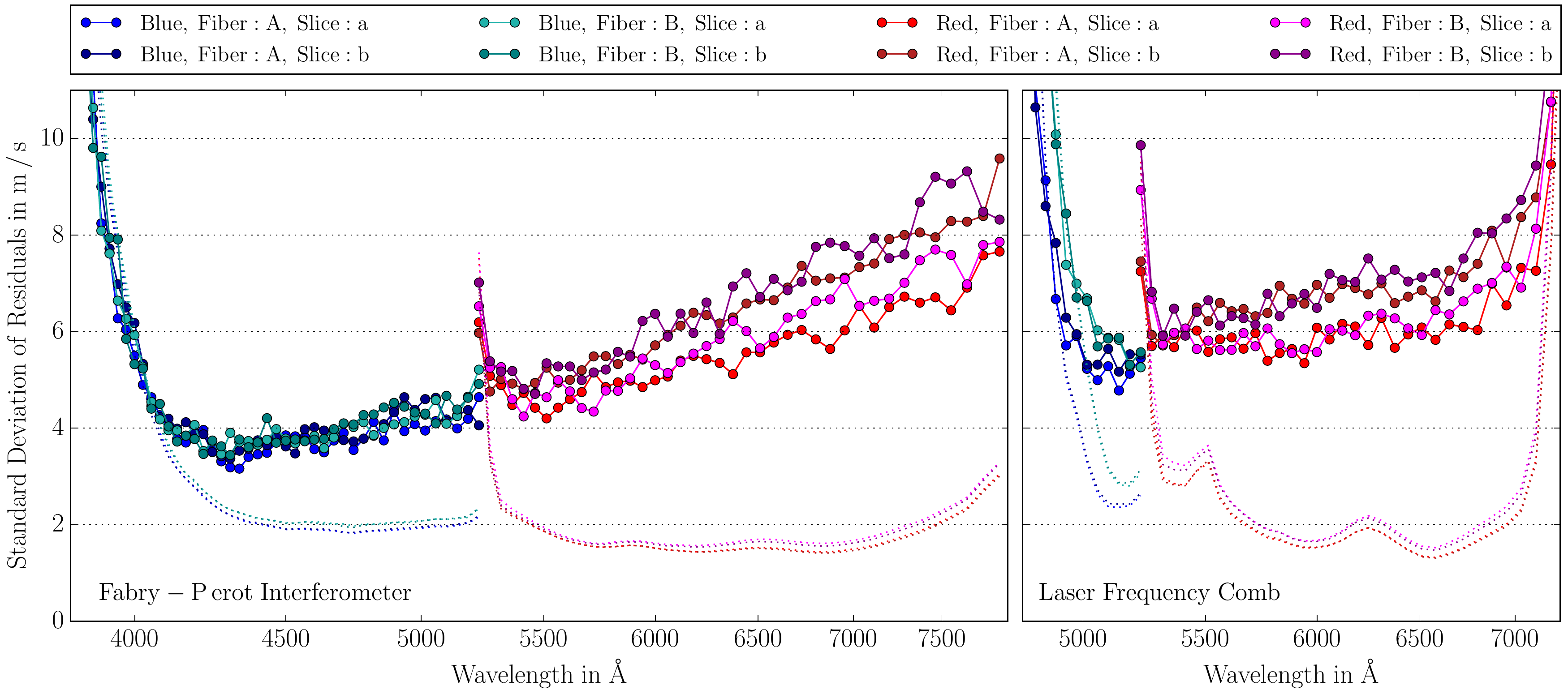}
 \caption{
  Summary of the difference between smoothed and non-smoothed FP and LFC line positions.
  Points show for each spectral order the standard deviation, derived from the 16th--84th percentile interval (compare to Figure~\ref{Fig:PatternNoise_FP305}).
  Dotted lines indicate the mean error of the individual line positions. 
  }
 \label{Fig:PatternNoise_Summary}
\end{figure*}

Figure~\ref{Fig:PatternNoise_Summary} illustrates this dependence in more detail. It shows the amplitude of the beat pattern noise for every spectral order across the spectral range, separated by Fiber and Slices and for the Fabry-P\'erot interferometer as well as for the laser frequency comb.
The amplitude of the smoothing residuals is determined from the 16th--84th percentile width of the distribution, identical to the width of the histogram in the left panel of Figure~\ref{Fig:PatternNoise_FP305}. In addition, Figure~\ref{Fig:PatternNoise_Summary} shows the mean photon-noise uncertainty of the line positions, again similar to the red bar in Figure~\ref{Fig:PatternNoise_FP305}.

Obviously, the beat pattern noise is present across the full wavelength range and for both calibration sources.
A detailed inspection of Figure~\ref{Fig:PatternNoise_Summary} leads to the following conclusions:

\begin{itemize}
 \item The beat pattern noise has an amplitude between 4 and $10\,\mathrm{m/s}$ and increases steadily with wavelength. A larger standard deviation of the residuals is observed in regions where the scatter is dominated by photon noise.
 \item The photon-limited precision for individual line positions is about or below $2\,\mathrm{m/s}$. Over most parts of the spectral range, the amplitude of the beat pattern noise is between $2\times$ and $5\times$ as large. Therefore, photon-limited precision is never reached, except in regions where the calibration sources deliver very little flux and the precision is anyway low.
 \item The beat pattern noise is about 15\% stronger (larger rms) in Slice~b compared to Slice~a, but close to identical for the two Fibers. This effect is consistent across the whole wavelength range and for both calibration sources.
 \item No discontinuity at the transition from blue to red arm is observed. The effect therefore has to be unrelated to components behind the dichroic, i.e., cross-dispersers, cameras, and detectors. This also excludes fringing effects as possible source for the beat pattern noise. The two detectors have different thicknesses, which would lead to different fringing patterns and most-probably to a different amplitude of the systematics in the two arms.
 \item The amplitude of the effect is very similar for the FP and LFC spectra. However, both sources are fundamentally different, which makes it extremely unlikely that the observed effect is related to the calibration sources themselves. For instance, the LFC produces coherent light and internally operates with mono-mode fibers. This could in principle lead to laser speckles, which might cause the observed interference pattern. The Fabry-P\'erot interferometer, however, uses as light source a LDLS, basically a glowing plasma ball, and therefore emits incoherent light, which is transported by multimode fibers. It thus seems unreasonable to assume that e.g., possible laser speckles or other effects specific to one of the calibration sources might be responsible for the beat pattern noise.
 \item Figure~\ref{Fig:PatternNoise_Summary} shows a slightly larger amplitude of the pattern noise for the laser frequency comb than for the Fabry-P\'erot interferometer. This might at least partially be related to the different line shapes of the two calibration sources. The FP lines are marginally resolved and noticeably wider than the intrinsically extremely narrow LFC lines. 
 One might also speculate that subtle differences in the line shape could explain the higher amplitude in Slice b compared to Slice a. However, the line-spread functions of the two fibers are as well not fully identical and still the amplitude of the pattern noise seems to be equal.
\end{itemize}

In addition, we checked if there is any correlation between the displacements of individual lines and their pixel phase in detector $Y$ direction. However, no correlation was found. Also, it was verified that the effect appears for \texttt{1HR1x1} and \texttt{1HR2x1} binning mode at approximately comparable amplitude.
Therefore, the root cause for the beat pattern noise is so far not understood. From the considerations above, certain sources can be excluded, but the true origin remains elusive.

It has to be noted that the instrumental line-spread function is non-Gaussian and different for the two Fibers and Slices. Modeling it with a Gaussian therefore leads to a formally poor fit. However, this should within the context of this problem not matter.
While a mismatch between the actual line-spread function and the used Gaussian model might have significant implications for the accuracy of the wavelength solution, it is hard to imagine that this could be the sole cause of the observed beat pattern noise. The described systematics can already be identified when inspecting only a few consecutive FP or LFC lines of a single trace. Over such small scales, the instrumental profile does not change significantly and a possible error in the line centroids due to an inaccurate line model would be constant for all lines. However, what is observed is an alternating offset between consecutive lines (see Figure~\ref{Fig:PatternNoise_FP305}).   
Differences in the amplitude of the effect between slices and the two calibration sources might indeed be related to the slightly different line-spread functions in the two slices and the different intrinsic line width of the two sources. However, this can not be the root cause for the issue.

One aspect that is currently suspected to possibly be related to the observed systematics is the spectral extraction procedure. As described in Section~\ref{Sec:Extraction}, the adopted scheme is in certain ways nonoptimal and several aspects that could lead to inaccuracies in the extracted flux were outlined. The observed beat pattern hints towards some sort of interference or aliasing effect. The pixelization of the detector grid would at least in principle provide one such periodic component.

As a cross-check, a simplified extraction scheme was tested in addition to the \textit{optimal extraction} described in Section~\ref{Sec:Extraction}. The simplified scheme does not weight the observed flux according to a trace profile. Instead, the raw counts are simply summed in detector $X$ direction (cross-dispersion) without any further assumptions. Choosing a wide extraction window ensures that all flux is captured and that the location of the window has very little influence on the extracted flux. 
The amplitude of the beat pattern noise was found to be slightly larger but not substantially different compared to the optimal extraction scheme.
This behavior was expected. In cases where the shot-noise from the source dominates over other noise contributions (like RON, Bias and scattered light), optimal extraction becomes formally equivalent to a simple summation of raw counts \citep{Horne1986}.   
Still, it shows that both extraction schemes work within their limitations, i.e., the assumption that the 
flux distribution on the detector is a separable function of trace profile and source SED, correctly.
However, this rather strong assumption is\,--\,as demonstrated in Section~\ref{Sec:TraceProfileAmbiguity}\,--\,not fully satisfied. 

Preliminary tests indicate that indeed a 2D modeling of the lines in detector space might be able to substantially reduce the beat pattern noise. 
However, a quantitative assessment is highly nontrivial since it basically requires the development of a far more sophisticated spectral extraction algorithm that fully forward-models the flux on the detector by a superposition of 2D instrumental point-spread functions, similar to the \textit{Spectro-Perfectionism} algorithm described by \citet{ABolton2010}.

The beat pattern noise therefore remains an unsolved problem and its impact on the final wavelength accuracy is not entirely clear. 
On one hand, it is a small-scale effect and averages down quickly. Applying the kernel smoothing filter to the determined line positions should therefore remove the beat pattern noise to a large degree by simply \textit{smoothing it away}. In addition, the metal absorption systems used to constrain a possible variation of the fine-structure constant have typical extents of a few $100\,\mathrm{km/s}$. This as well should mitigate the effect to some degree. However, the complexity of the involved effects does not allow to make clear predictions for this.

On the other hand, systematic effects that are present in the data and not fully understood are always concerning. If the effect is, as outlined above, not intrinsic to the either FP or LFC, one has to assume that it is present in every spectrum taken. This would mean that the observed systematics act as an additional correlated noise term on all spectral measurements, i.e., also on science spectra and ThAr calibration frames.
The presence of the beat pattern noise in the ThAr line measurements could be particular severe since they are used as anchors and provides all the absolute wavelength information for the ThAr/FP solution (described in detail in the following Section~\ref{Sec:JointThArFPSolution}). In contrast to FP and LFC, the ThAr spectra provide only relatively few, sparse and unevenly distributed lines. It is therefore not possible to directly check for the presence of the beat pattern noise in the ThAr spectra and the options to mitigate the effect by averaging several lines are extremely limited. Therefore, the beat pattern noise, which in principle is a small-scale effect, could induce undesirable systematic effects in the ThAr/FP wavelength solution on intermediate and large scales, which in the end might compromise the full \Espresso{} wavelength calibration.

\subsection{Joint ThAr/FP solution}
\label{Sec:JointThArFPSolution}

Despite the unsolved issue regarding the beat pattern noise, we proceed to derive the joint ThAr/FP wavelength solution, based on the ThAr and Fabry-P\'erot line positions determined in Section~\ref{Sec:LineFitting}.
Since we consider the observed systematic displacement of the FP lines a source of noise, which is not supposed to be there, we adopt in the following the smoothed line positions, obtained by applying the kernel smoothing filter using third degree polynomials and a Gaussian kernel with $\sigma=35\,\mathrm{km/s}$ to the raw line positions (see Section~\ref{Sec:BeatPatternNoise}). 

As described above, the Fabry-P\'erot interferometer delivers a dense train of lines equally spaced in frequency space. The high density of lines ($\simeq300$ per order, separation $2\times10^{10}\,\mathrm{Hz}$ corresponding to $0.1\,\mathrm{\AA}$ to $0.4\,\mathrm{\AA}$) and therefore exquisite sampling on small scales will later allow to define a very precise nonparametric wavelength solution. First, however, the wavelengths of the individual FP lines have to be determined. Formally, these are are given by
\begin{equation}
 \lambda^\mathrm{FP}_k = D_\mathrm{eff}(\lambda) \; \frac{1}{k^\mathrm{FP}}
 \label{Eq:FabryPerot},
\end{equation}
where $k^\mathrm{FP}$ describes the index of an FP line and $D_\mathrm{eff}(\lambda)$ the effective gap size of the Fabry-P\'erot cavity%
\footnote{We define with $D_\mathrm{eff}(\lambda)$ the effective optical length of the FP cavity, which is twice as long as the physical separation of the two mirrors since the light travels back and forth in the resonator.}.
The FP interferometer is therefore fully characterized by the separation of two mirrors forming the resonator. However, this quantity is not entirely constant but shows a slight but highly significant dependence on wavelength, which is probably related to the dielectric coatings on the mirror surfaces. In addition, the \Espresso{} FP interferometer is not locked to any absolute wavelength reference. The device is built out of Zerodur, a material with very low thermal expansion, and placed inside a thermally controlled vacuum vessel \citep[similar to][]{Wildi2012}. Still, residual changes in temperature or pressure can lead to a drift of the FP line pattern. It is therefore necessary to characterize the Fabry-P\'erot interferometer, i.e., determine $D_\mathrm{eff}(\lambda)$, based on reference spectra providing absolute wavelength information. For this, spectra of a Thorium-Argon hollow cathode lamp (ThAr) are used. In addition, also the line indices $k^\mathrm{FP}$ have to be properly identified.

The first step in connecting the wavelength information from FP and ThAr spectra is to assign to each ThAr line an effective, non-integer FP line index $i^\mathrm{ThAr}_j$, which describes the position of a ThAr line on the detector relative to the (neighboring) FP lines. This is done independently for each trace.
At this stage, the absolute FP line indices $k^\mathrm{FP}$ are not known yet. Instead, only the relative indices $i^\mathrm{FP}$ with respect to an arbitrary chosen reference line approximately in the center of the trace is available.
To determine for each ThAr line with index $j^\mathrm{ThAr}$ the effective FP line index $i^\mathrm{ThAr}_j$, the pixel positions $y^\mathrm{FP}_i$ and line indices $i^\mathrm{FP}$ of the FP lines are interpolated using a cubic spline. This is then evaluated at the pixel positions of the ThAr lines $y^\mathrm{ThAr}_j$ to deliver the effective FP line indices $i^\mathrm{ThAr}_j$.
Assuming (for now) that the effective gap size of the FP is constant over a single spectral order, these then follow according to Equation~\ref{Eq:FabryPerot} the relation
%
\begin{equation}
 \nu^\mathrm{ThAr}_j = \frac{\mathrm{c}}{D_\mathrm{eff}} \; i^\mathrm{ThAr}_j + \nu_0^\mathrm{FP}
 \label{Eq:FabryPerot_ThAr}.
\end{equation}
Since the laboratory frequencies of the ThAr lines $\nu^\mathrm{ThAr}_j$ are known, i.e., taken from the \citet{Redman2014} catalog, and the effective FP line indices $i^\mathrm{ThAr}_j$ have been determined above, one can fit for the effective gap size $D_\mathrm{eff}$ and the offset frequency $\nu_0^\mathrm{FP}$ corresponding to the FP line with $i^\mathrm{FP}=0$. Obviously, this requires that each spectral order contains at least two properly measured ThAr lines.
The determined parameters $D_\mathrm{eff}$ and $\nu_o^\mathrm{FP}$ now define for each order an approximate characterization of the Fabry-P\,erot interferometer and by evaluating Equation~\ref{Eq:FabryPerot_ThAr} at the line indices $i^\mathrm{FP}$, initial wavelengths can be assigned to all FP lines.

The second step is to combine the still independent FP wavelength solutions of all spectral orders to one common relation. 
For each pair of consecutive extracted traces, which always have some overlap, wavelengths and line indices are compared and an average integer-valued index offset between the two traces is determined. The initial FP line frequencies obtained above are precise enough so that this is always possible without ambiguity.   
By applying the determined offsets to the line indices $i^\mathrm{FP}$ (and similarly the $i^\mathrm{ThAr}_j$), all FP (and ThAr) lines across all orders are brought to the same scale.
In addition, a global index offset $i_0^\mathrm{FP}$ has to be determined. Therefore, the ensemble of ThAr lines is fitted a second time, in a way very similar as described in Equation~\ref{Eq:FabryPerot_ThAr}, but now over the full spectral range, as
\begin{equation}
 \lambda^\mathrm{ThAr}_j = D_\mathrm{eff} \; \frac{1}{i^\mathrm{ThAr}_j + i_0^\mathrm{FP}}
 \label{Eq:FabryPerot_ThAr_Global}\;.
\end{equation}
Here, $D_\mathrm{eff}$ is again assumed to be constant, which is across the full spectral range clearly not the case but can be understood as an \textit{average} effective gap size%
\footnote{
For \Espresso, $D_\mathrm{eff}(\lambda)$ varies by more than a full wavelength across the spectral range, which leads to some ambiguity in how to define the average effective gap size and in consequence $i_0^\mathrm{FP}$. However, this has no physical relevance. 
It only has to be noted that the exact numerical value of the index is (within $\pm1$) a choice and that $D_\mathrm{eff}(\lambda)$ is degenerate with $i_0^\mathrm{FP}$. Therefore, a different choice for the indexing will lead to a different $D_\mathrm{eff}(\lambda)$, but this will according to Equation~\ref{Eq:FabryPerot} still result in the same wavelengths for the individual FP lines.
}.
With the determination of $i_0^\mathrm{FP}$, the FP line indices can be expressed in terms of $k^\mathrm{FP} = i^\mathrm{FP} + i_0^\mathrm{FP}$, which directly describes the physical mode of the Fabry-P\'erot interferometer. For the \Espresso{} FP, the average gap size is $D_\mathrm{eff} \simeq 15.210\,\mathrm{mm}$ and the observed modes are between $19\,300 \lesssim k^\mathrm{FP} \lesssim 40\,200$.

While most of the procedure described above was primarily required to determine the proper line indices $k^\mathrm{FP}$, the final step is to precisely determine $D_\mathrm{eff}(\lambda)$. 
From Equation~\ref{Eq:FabryPerot} follows, that, after the relative effective FP indices $i^\mathrm{ThAr}_j$ have been converted to absolute ones $k^\mathrm{ThAr}_j$, each individual ThAr line can be understood as an independent measurement of the FP effective gap size 
\begin{equation}
 D_\mathrm{eff}^{j}( \lambda^\mathrm{ThAr}_j ) = k^\mathrm{ThAr}_j \times \lambda^\mathrm{ThAr}_j
 \label{Eq:Deff} .
\end{equation}
This is a sparsely sampled and noisy representation of $D_\mathrm{eff}(\lambda)$. Determining  the FP effective gap for every wavelength therefore becomes an interpolation problem. Unfortunately, the uneven and sparse distribution of the ThAr lines and the requirement to determine $D_\mathrm{eff}(\lambda)$ to a relative accuracy of few~$\times{}10^{-8}$ make this an extremely challenging task. 
In addition, every ThAr is observed at least four times (two fibers, two slices) or even eight times in the region of order overlap. Hence, there are always multiple estimates of $D_\mathrm{eff}(\lambda)$ at the same wavelength%
\footnote{It has to be stressed that all ThAr measurements, i.e from the different arms, orders, fibers and slices, have to be described by the same model. There is only one Fabry-P\'erot device and hence there can only be a single and unique $D_\mathrm{eff}(\lambda)$ function that describes it. The instrumental effects for the two fibers or slices might be different, but if necessary, these shall be modeled explicitly somewhere else and not be confused with the FP effective gap.}.
This renders e.g., spline interpolation completely unsuitable. A fit with polynomials, however, would introduce unwanted long-range correlations into the wavelength solution and require an extremely high order to accurately describe the data, which then would lead to instabilities.
We therefore use for the interpolation and modeling of the FP effective gap size the nonparametric kernel smoothing filter described already in Section~\ref{Sec:BeatPatternNoise}. The algorithm, inspired by the \citet{SavitzkyGolay1964} filter, was purposely developed for this task and offers all the capabilities required, e.g., it is able to accept multiple measurements at the same wavelength, properly weights them by their given uncertainty, can handle unevenly sampled and heteroscedastic data, is\,--\,if desired\,--\,rather flexible, does not assume the data to follow any specific functional form and in particular allows full control over the correlations introduced by the smoothing procedure.
In addition, our algorithm allows to evaluate the smoothed function at any arbitrary positions instead of only at the sample positions and thereby provides the needed interpolation functionality.

Since Version~1.5.1, the same approach based on a kernel smoothing filter is also adopted by the \Espresso~DRS to describe $D_\mathrm{eff}(\lambda)$. This proved to be more accurate than the high order ($\simeq24$) polynomial used before and at the same time allows control over the introduced correlations \citep{Lovis2020}. 

\begin{figure*}
 \centering
 \includegraphics[width=\linewidth]{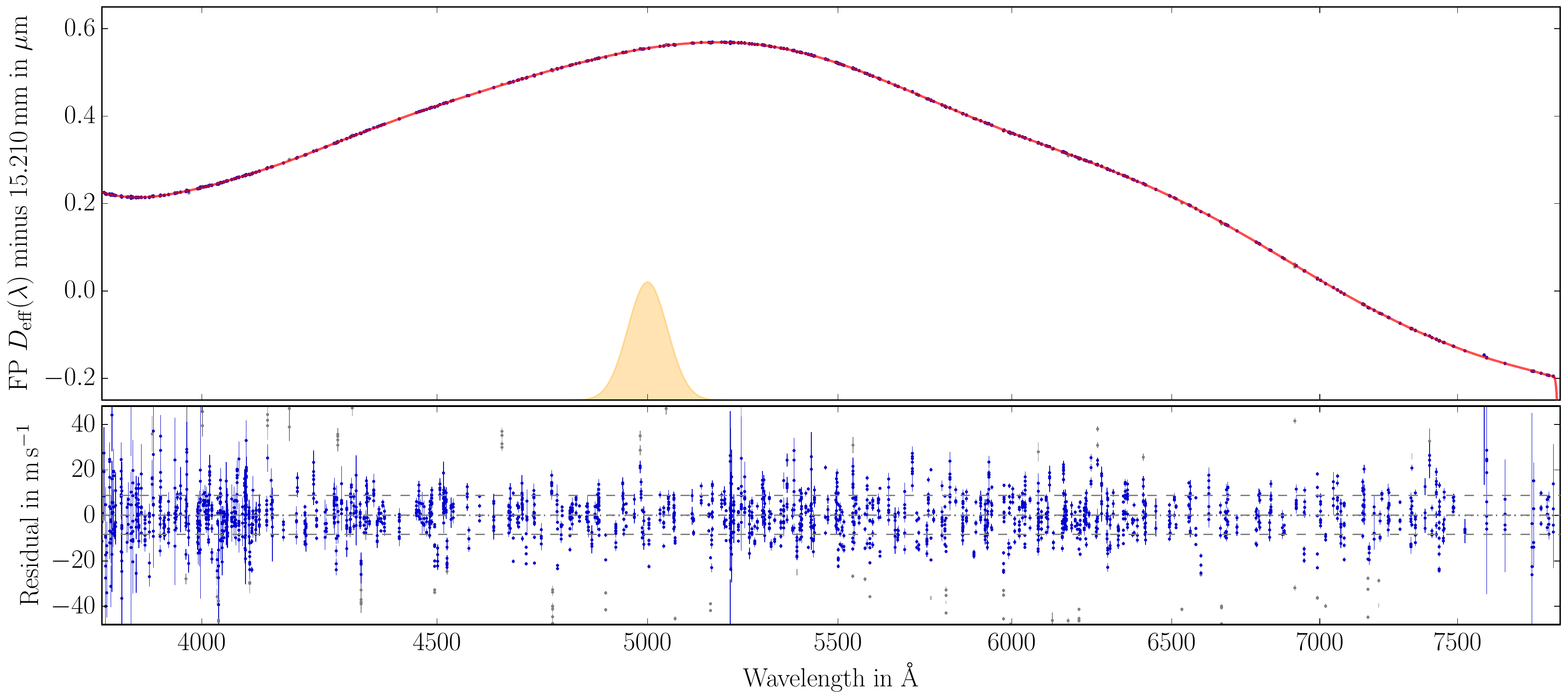}
 \caption{
  Visualization of the $D_\mathrm{eff}(\lambda)$ determination.
  The top panel shows the individual measurements of the FP effective gap size obtained from the ThAr lines (blue) and the interpolated smooth function (red). For better visualization, a value of $15.210\,\mathrm{mm}$, approximately the mean $D_\mathrm{eff}$, is subtracted from measurements and $D_\mathrm{eff}(\lambda)$ function. 
  The bottom panel shows the corresponding residuals.
  Uncertainties of the $D_\mathrm{eff}^{j}(\lambda^\mathrm{ThAr}_j)$ measurements are based on the combined laboratory wavelength and line fitting error. Points rejected by the sigma-clipping procedure are shown in gray.  A representation of the smoothing kernel used by the interpolation scheme is displayed in orange.
  }
 \label{Fig:Deff}
\end{figure*}

The kernel smoothing filter is also used to detect outliers. It is run multiple times in an iterative process and uses a sigma-clipping scheme to reject outliers.
The rejection criterion is based on the uncertainty of the individual measurements. However, as outlined below, the dispersion around the smoothed solution is inconsistent with the uncertainty of the individual data points. 
Since the source of the additional scatter is not understood, we adopt a simple sigma-clipping scheme where points are rejected if 
\begin{equation}
 \frac{ \: \lvert \: D_\mathrm{eff}^{j}(\lambda^\mathrm{ThAr}_j) \:-\: D_\mathrm{eff}(\lambda) \: \rvert \: }{ \sqrt{ {\sigma_{D_\mathrm{eff}^{j}(\lambda^\mathrm{ThAr}_j)}}^2 +  {\sigma_{D_\mathrm{eff}}}^2 } } > N_\sigma 
 \label{Eq:Deff_SigmaClip}.
\end{equation}
Here $D_\mathrm{eff}^{j}(\lambda^\mathrm{ThAr}_j)$ and $\sigma_{D_\mathrm{eff}^{j}(\lambda^\mathrm{ThAr}_j)}$ are the individual measurements from the ThAr lines and their uncertainties, $D_\mathrm{eff}(\lambda)$ the smoothed solution for the FP effective gap and $\sigma_{D_\mathrm{eff}}$ the dispersion of the datapoints around the smoothed solution based on the 16th--84th percentile width of the distribution. The rejection threshold is controlled by setting $N_\sigma=2.8$.  

Figure~\ref{Fig:Deff} shows the full ensemble of 2513 $D_\mathrm{eff}^{j}( \lambda^\mathrm{ThAr}_j )$ measurements obtained from 432 unique \ion{Th}{I} lines as well as the final $D_\mathrm{eff}(\lambda)$ model. The Fabry-P\'erot effective gap size is about $15.210\,\mathrm{mm}$, but varies over the full spectral range of \Espresso{} by about $\pm0.4\,\mathrm{\mu{}m}$ in a nontrivial way. The $D_\mathrm{eff}(\lambda)$ model was obtained using a large Gaussian smoothing kernel with $\sigma=3000\,\mathrm{km/s}$, truncated at $\pm9000\,\mathrm{km/s}$ and a rather high polynomial degree of 5. As can be seen from Figure~\ref{Fig:Deff}, the adopted scheme based on the kernel smoothing filter is able to describe the FP effective gap with high accuracy, high precision and little remaining systematics in the residuals.

The uncertainties of the $D_\mathrm{eff}^{j}(\lambda^\mathrm{ThAr}_j)$ measurements are based on the combined uncertainty of the laboratory wavelength from the \citet{Redman2014} catalog and the line fitting error. The line list is composed of particularly strong \ion{Th}{i} lines with typical uncertainties of the used Ritz wavelengths between $0.6$ and $3.0\,\mathrm{m/s}$, much less than the average uncertainty of $11\,\mathrm{m/s}$ in the \citealt{Redman2014} catalog. This results, together with photon-counting errors between $0.7$ and $3.2\,\mathrm{m/s}$, in combined uncertainties between $1.0$ and $4.6\,\mathrm{m/s}$. All values are given for \texttt{1HR1x1} binning and correspond to the 16th to 84th percentile interval.

For the weighting in the kernel smoothing filter, the computed scatter around the $D_\mathrm{eff}(\lambda)$ solution is quadratically added to the combined photon and wavelength calibration error, i.e., $\sigma = \sqrt{ {\sigma_{D_\mathrm{eff}}}^2  + {\sigma_{D_\mathrm{eff}^{j}(\lambda^\mathrm{ThAr}_j)}}^2 }$, identical to Equation~\ref{Eq:Deff_SigmaClip}. The motivation for this is that the latter two contributions are obviously not the dominant source of scatter and that including the empirical noise estimate in the weighting scheme results in a more uniform solution.

The bottom of Figure~\ref{Fig:Deff} visualizes the residuals between individual ThAr measurements and $D_\mathrm{eff}(\lambda)$ model. Obviously, the scatter of the ThAr measurements is significantly larger than the combined laboratory and shot-noise uncertainty. The standard deviation (obtained from the 16th--84th percentile width) is about $\pm10\,\mathrm{m/s}$ and therefore nearly $4\times$ as large as the mean error of the measurements.
Obviously, there are additional sources of noise.

Following the discussion in Section~\ref{Sec:BeatPatternNoise}, the beat pattern noise is an obvious suspect for the excess scatter since it has to be assumed to also affects the ThAr line positions. However, for FP and LFC spectra, the observed amplitude of the beat pattern noise shows a clear wavelength dependence and even for the longest wavelengths barely reaches $10\,\mathrm{m/s}$. Figure~\ref{Fig:Deff} does not show such a trend and in general larger residuals. It is therefore unclear if the observed scatter can be (fully) attributed to the beat pattern noise.
Another potential source of scatter might come from undetected blending of the ThAr lines. However, the \ion{Th}{i} lines are carefully selected 
and residual blending effects should at most affect a limited number of lines. These might even be rejected by the outlier detection. Even if not, blending of a line should cause the identical discrepancy in all four traces of a spectral order. The distribution of the residuals (shown in the bottom panel of Figure~\ref{Fig:Deff}) does not support such a hypothesis. Instead, it appears as if a large fraction of ThAr lines suffers from very significant scatter.

\begin{figure}[htb]
 \centering
 \includegraphics[width=\linewidth]{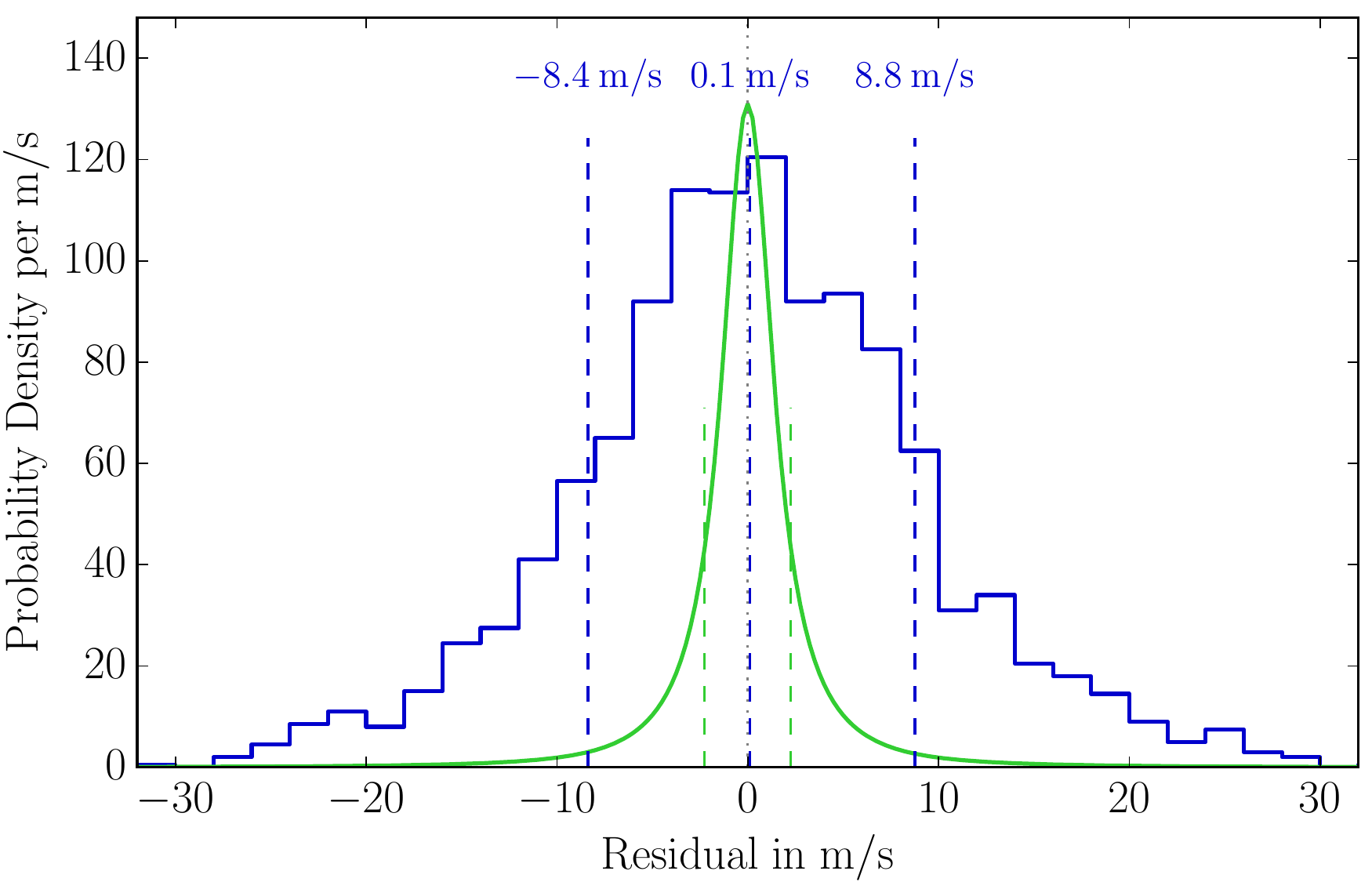} 
 \caption{
  Histogram of the $D_\mathrm{eff}(\lambda)$ residuals shown in the bottom panel of Figure~\ref{Fig:Deff}. The 16th, 50th, and 84th percentiles are labeled in the plot. 
  The expected distribution based on the formal uncertainties of the datapoints (laboratory wavelength and fitting error) is shown in green, for visualization purpose with a Y-scale reduced to $\sfrac{1}{4}$. The dispersion of the expected distribution is $2.3\,\mathrm{m/s}$. 
  }
 \label{Fig:Deff_Hist}
\end{figure}

This is better visualized in Figure~\ref{Fig:Deff_Hist}, which shows a histogram of the $D_\mathrm{eff}(\lambda)$ residuals. 
Clearly, the distribution is rather wide and exhibits extended wings. The formal standard deviation is $10.4\,\mathrm{m/s}$, while the 16th and 84th percentiles are located at $-8.4\,\mathrm{m/s}$ and $+8.8\,\mathrm{m/s}$. 
These values can be compared to the width of the distribution expected from the uncertainties of the individual $D_\mathrm{eff}^{j}(\lambda^\mathrm{ThAr}_j)$ measurements alone, which has a dispersion of less than $2.3\,\mathrm{m/s}$ (also shown in Figure\ref{Fig:Deff_Hist}). Therefore, the observed scatter is $3.8\times$ as large as the photon shot-noise and laboratory wavelength uncertainties combined.

Clearly, this excess scatter negatively impacts the determination of the FP effective gap size and undoubtedly propagates into the ThAr/FP wavelength solution.
It would therefore be highly desirable to identify the cause of this noise and remove it. However, for the time being, it is only possible to choose a wide smoothing kernel and thereby mitigate the issue to some degree, at the expense of enhanced correlations. 
In general, the size of the smoothing kernel has to be chosen with care. On one hand, a small kernel is clearly desirable to reduce correlations along the wavelength solution and makes the model more flexible, which increases the accuracy to which $D_\mathrm{eff}(\lambda)$ can be described.
On the other hand, a large kernel will average over more $D_\mathrm{eff}^{j^\mathrm{ThAr}}( \lambda^\mathrm{ThAr}_j )$ measurements, reducing the noise and thereby creating a smoother and more precise description of the FP effective gap. This is in particular important given the large scatter in the individual ThAr measurements. So far, the kernel size was chosen by eye and seems to give good results. However, a formal optimization of this meta-parameter might be done in the future.

Despite the excess scatter, the characterization of the Fabry-P\'erot interferometer works in general quite well and the adopted scheme still holds room for further tuning and optimization. Successfully determining $D_\mathrm{eff}(\lambda)$ therefore completes the characterization of the Fabry-P\'erot interferometer and together with the line indices $k^\mathrm{FP}$, precise wavelengths (or frequencies) can be assigned to every FP line (see Equation~\ref{Eq:FabryPerot}).
The concluding step in deriving the joint ThAr/FP solution is therefore to compute wavelengths for every pixel along extracted traces by interpolating between FP lines. Due to the high density of FP lines, this is a rather simple task for which we adopt a straight-forward cubic spline interpolation.  
In contrast to describing the wavelength solutions with polynomial functions, the adopted nonparametric approach does not add additional long-range correlations and is therefore more suitable in the context of accurate wavelength calibration.

\subsection{LFC solution}
\label{Sec:LFC_Solution}

Calculating the LFC wavelength solution is much simpler. As stated before, the frequencies of the individual LFC lines are given by 
\begin{equation}
 \nu^\mathrm{LC}_k = \nu^\mathrm{LC}_0 + k^\mathrm{LC} \times \nu^\mathrm{LC}_\mathrm{FSR}.
 \label{Eq:LFC}
\end{equation}
The offset frequency $\nu^\mathrm{LC}_0$ and pulse repetition rate, equivalent to the line spacing $\nu^\mathrm{LC}_\mathrm{FSR}$, are actively controlled with respect to a local $10\,\mathrm{MHz}$ reference frequency by the means of phase-locked loops (PLL). 
The accuracy of these frequencies should therefore be as good as the local oscillator frequency, which is provided by a Meinberg Lantime M600/GPS device with an OCXO~DHQ crystal oscillator (Rafael Probst, personal communication). The specified relative accuracy is about $10^{-12}$ and therefore far in excess of what is needed%
\footnote{\url{https://www.meinbergglobal.com/english/specs/gpsopt.htm}}.
The setpoints for the two defining frequencies are given in the fits header to exactly $\nu^\mathrm{LC}_\mathrm{FSR} = 18.0\,\mathrm{GHz}$ and $\nu^\mathrm{LC}_0 = 7.35\,\mathrm{GHz}$.

The $\nu^\mathrm{LC}_0$ offset frequency is stabilized as part of the fundamental comb and hence before the filtering by the Fabry-P\'erot cavities. This in principle leads to a $250\,\mathrm{MHz}$ ambiguity in the final  offset frequency but this can be resolved by the use of the built-in wavemeter. The wavemeter readings are currently not propagated to the fits headers but a manual check confirmed that the offset frequency of $\nu^\mathrm{LC}_0 = 7.35\,\mathrm{GHz}$ is indeed correct (Tilo Steinmetz, personal communication). 

Given this, only the line index $k$ in Equation~\ref{Eq:LFC} has to be determined. This is done by referring to an a-priori wavelength solution, in this case the ThAr/FP solution derived in Section~\ref{Sec:JointThArFPSolution}.
However, for an unambiguous identification of a line index, the a-priori wavelength solution only needs to be good enough to half the spacing between LFC lines, i.e., better than $\simeq3.5\,\mathrm{km/s}$. This requirement is trivially fulfilled and consistency checks confirmed no misidentifications.
Since the a-priori wavelength solution is only required for identification and possible uncertainties do not propagate into the final LFC wavelength solution, the ThAr/FP and LFC wavelength solution can be considered fully independent.   

Given that the positions of all LFC lines are fitted as described in Section~\ref{Sec:LineFitting} and the line indices $k$ are successfully identified, one immediately obtains the frequencies of all LFC lines with negligible uncertainty (see Equation~\ref{Eq:LFC}).  The final LFC wavelength solution can then be constructed by simply interpolating between the LFC lines. For this, a cubic spline is used, taking into account the uncertainties of the individual lines, providing the wavelength solution $\nu^\mathrm{LFC}(y)$ or equivalently $\lambda^\mathrm{LFC}(y)$. Although the value of a spline at a given position formally depends on all spline points, the dependence in practice decreases rapidly with increasing distance from the position of evaluation. Therefore, the spline-interpolated wavelength solution shows very little long-range correlations and the wavelength solution at any given point only depends on the LFC lines within a limited region around this position.

An additional source of correlation comes from the fact that we use the smoothed LFC line positions to mitigate the impact of the beat pattern noise (see Section~\ref{Sec:BeatPatternNoise}). However, this way of smoothing and averaging allows perfect control over the introduced correlations since the smoothing kernel is explicitly defined. In particular, all points separated by more than the extent of the kernel remain fully uncorrelated. 

The described approach ensures that correlations across the wavelength range, which later might induce systematics in a constraint on the fine-structure constant are kept to a bare minimum and limited to small scales. 
The LFC solution can be considered fully independent and uncorrelated for all pairs of wavelengths separated by more than $\simeq100\,\mathrm{km/s}$%
\footnote{Based on a smoothing using a Gaussian kernel with $\sigma=35\,\mathrm{km/s}$, corresponding to a FWHM of 6 to 12 LFC lines}.
In addition, all traces are treated completely independent. Although there is of course some commonality, the different fibers and slices can therefore in some aspects be considered spectra of the same source taken with different spectrographs. This later allows to perform viable internal consistency checks. 

\section{Comparison of Wavelength Solutions}
\label{Sec:WavelengthComparisons}

In general, it is difficult to demonstrate the accuracy of a wavelength calibration. The best way is of course to observe some external calibration standard.
However, at the accuracy required for a precision test of fundamental constants, no such standard exists. One therefore has to rely on internal tests.
Fortunately, \Espresso{} offers two fully independent wavelength calibrations. While they can not give a definite answer about the absolute accuracy, they can at least be used to check for consistency of the wavelength solutions. Given that the calibration sources are fundamentally different, we are confident that possible systematics would affect the two solutions differently, making them detectable.
All undetected effects that also affect the science spectrum will cancel out anyway. The comparison of ThAr/FP and LFC wavelength calibration therefore gives valuable insights into the accuracy of the \Espresso{} wavelength calibration.

\subsection{Comparison of ThAr Lines to LFC Wavelength Solution}
\label{Sec:Comparison_ThAr-LFC}

As outlined in Section~\ref{Sec:JointThArFPSolution}, the determination of the joint ThAr/FP solution is rather complicated and involves many steps. Directly comparing it to the LFC solution might therefore at first not yield that many insights.
A far more direct check is to simply compare the ThAr lines to the LFC wavelength solution. 

The ThAr laboratory wavelengths are taken from the \citet{Redman2014} catalog and the line positions are fitted in a straight-forward way as described in Section~\ref{Sec:LineFitting}.
Also the LFC calibration is rather simple. Since LFC offset frequency and pulse repetition rate are actively stabilized and individual modes identified without problems, the frequencies of all LFC lines are a-priori known to extreme accuracy. Therefore, it is only required to fit the LFC line positions and to interpolate between them to derive the full wavelength solution $\lambda^\mathrm{LFC}(y)$. Due to the high density of the LFC lines, this interpolation is simple and done using cubic splines (see Section~\ref{Sec:LFC_Solution}).

The only modest complication arises from the subtraction of the LFC background light. However, the details of this do not matter. Figure~\ref{Fig:LFC_BackgroundResidualsSummary} shows that, at least for regions without excessive modulation of the background light, the fitted positions of the LFC lines do only shift by about $1\,\mathrm{m/s}$ to $3\,\mathrm{m/s}$, even if no background subtraction is applied at all. The beat pattern noise remains a problem, but for the LFC solution we mitigate this to a large degree by applying the kernel smoothing filter to the measured line positions. Still, the LFC wavelength solution remains fully local with no correlations on scales larger $\gtrsim100\,\mathrm{km/s}$ and no correlation at all between different traces (see Section~\ref{Sec:LFC_Solution}).

\begin{figure}[th]
 \centering
 \includegraphics[width=\linewidth]{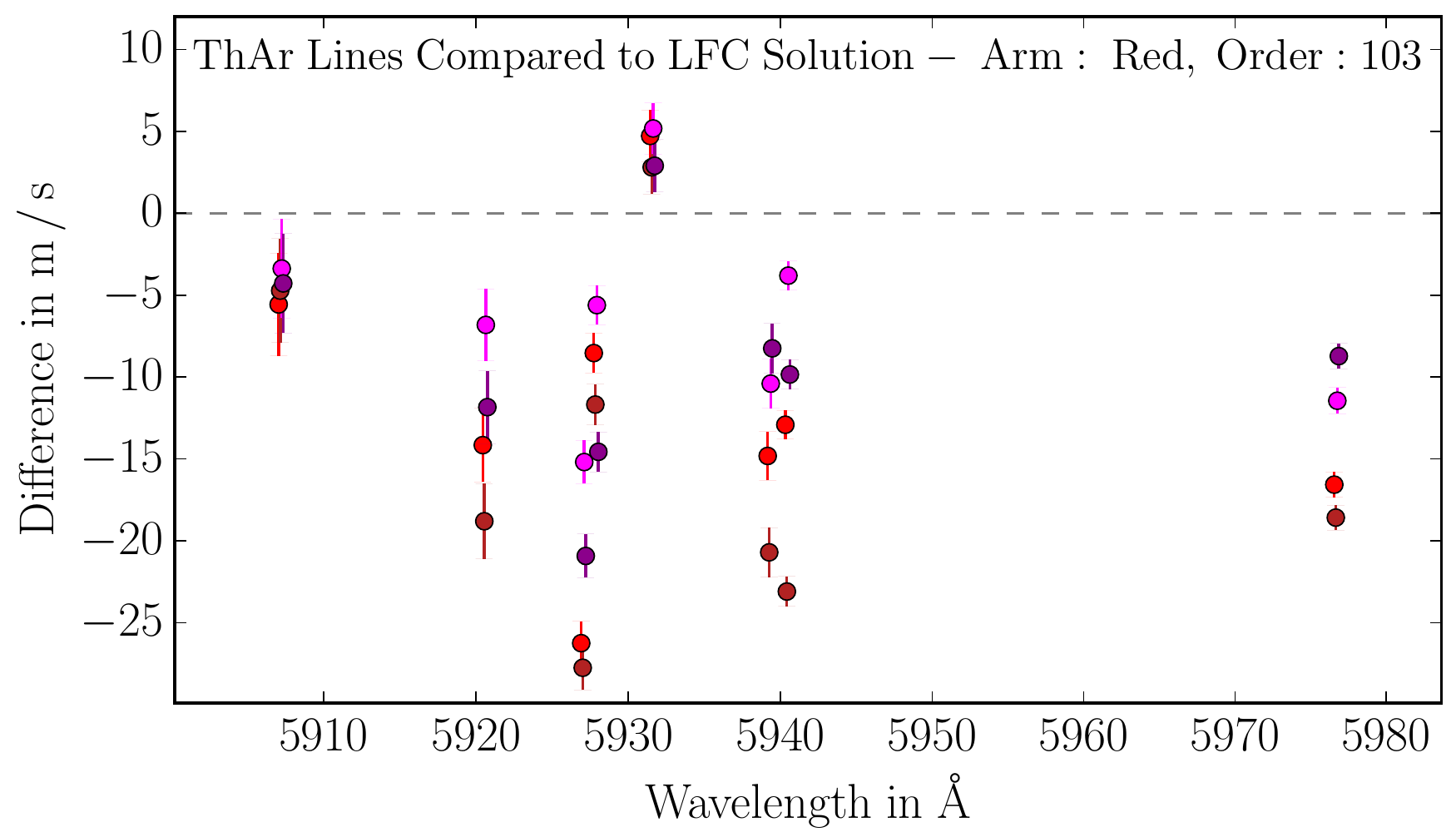} 
 \caption{
  Difference between ThAr laboratory wavelengths and LFC-calibrated measured line positions for one specific spectral order.  
  Colors indicate different fibers and slices and are chosen identical to e.g., Figure~\ref{Fig:PatternNoise_Summary}.
  }
 \label{Fig:ThAr_Comparison_103R}
\end{figure}

This allows a direct comparison of the ThAr laboratory wavelengths provided by \citet{Redman2014} and the ones derived from our own LFC wavelength solution. For this, the LFC wavelength solution is evaluated at the measured pixel position of the ThAr lines and the obtained wavelength compared to the ThAr laboratory data. The obtained differences are shown for one specific order in Figure~\ref{Fig:ThAr_Comparison_103R}.

Obviously, the ThAr lines exhibit a large scatter and for barely any line the two wavelength measurements are formally consistent. The indicated errors represent the combined laboratory and line fitting uncertainty of the ThAr lines. No uncertainties of the LFC wavelength solution are included, but these should be small, given that the formal error of individual LFC lines is $\approx2\,\mathrm{m/s}$ and the smoothing and interpolation procedure further reduce them.

\begin{figure*}
 \centering
 \includegraphics[width=\linewidth]{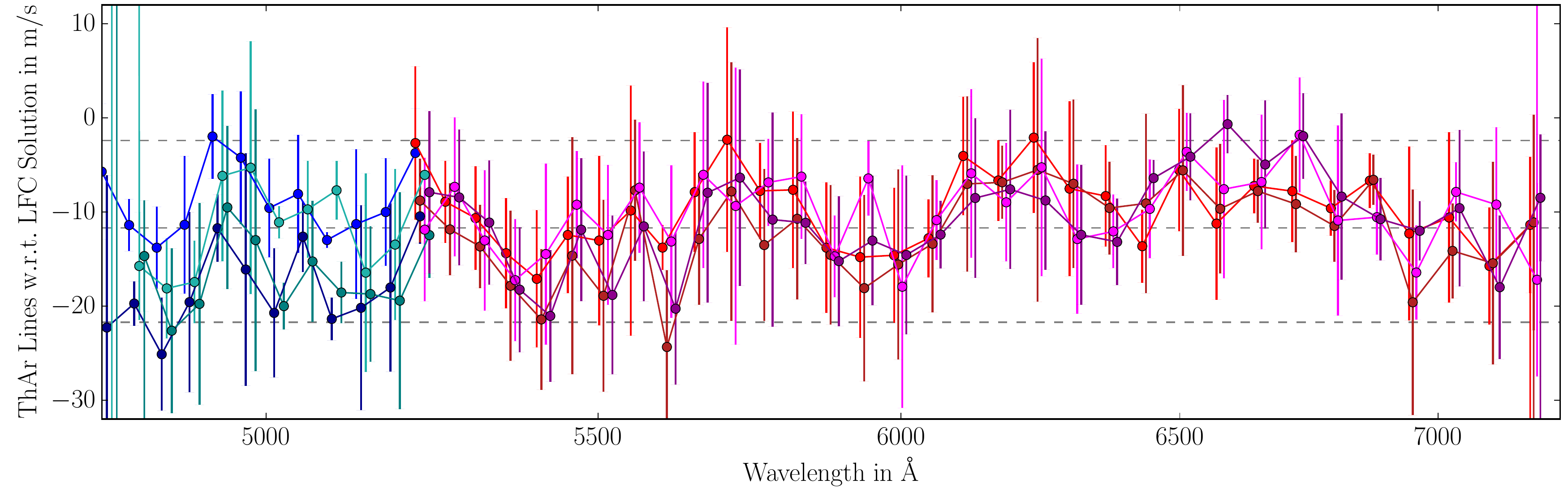}
 \caption{
  Differences between ThAr laboratory wavelengths and LFC-calibrated measurements summarized by spectral order.  Since the formal uncertainties of the individual ThAr lines are small (indicated in Figure~\ref{Fig:ThAr_Comparison_103R}), the vertical error bars represent the scatter within each order. Colors indicate different fibers and slices (similar to e.g., Figure~\ref{Fig:PatternNoise_Summary}) and points are slightly displaced in wavelength direction for clarity. 
  Indicated with horizontal lines are the 16th, 50th, and 84th percentiles of all ThAr wavelength measurements.
  }
 \label{Fig:ThAr_Comparison}
\end{figure*}

Figure~\ref{Fig:ThAr_Comparison_103R} shows that for most ThAr lines the measurements obtained from the different fibers and slices are not consistent with each other. 
Possible errors in the laboratory wavelength or line-blending effects would affect all four traces in the same way.
One possible reason for the discrepancy is therefore again the beat pattern noise affecting the ThAr line positions.
In addition, Figure~\ref{Fig:ThAr_Comparison_103R} suggests a systematic offset between ThAr laboratory wavelengths and the LFC calibrated measurements. This can be seen more clearly when referring to all ThAr lines for which a LFC solution is available. Figure~\ref{Fig:ThAr_Comparison} therefore shows a summary of the wavelength differences grouped by spectral orders.

This clearly reveals a global offset between laboratory wavelength and LFC measurement of $\approx-12\,\mathrm{m/s}$. It also shows that the scatter obtained from the 16th--84th percentile interval is about $10\,\mathrm{m/s}$. This is similar to the scatter observed during the characterization of the Fabry-P\'erot interferometer (see Figure~\ref{Fig:Deff_Hist}). Although the current test is completely independent of the $D_\mathrm{eff}(\lambda)$ measurement, it suggests a common source of the scatter intrinsic to the ThAr line measurements.

In addition, Figure~\ref{Fig:ThAr_Comparison} shows some structure in the residuals. However, the large-scale distortions ($\gtrsim500\,\mathrm{\AA}$) are less pronounced than the scatter within each order. For the red arm, the four independent traces behave rather consistently with only a few $\mathrm{m/s}$ difference between each other. In the blue arm, however, both traces of Slice~b seem to deviate from Slice~a and exhibit a difference between laboratory wavelength and LFC calibrated measurement, which is about $5\,\mathrm{m/s}$ more negative than for Slice~a.

The source of the structure in the ThAr residuals is so far not understood.
The \citet{Redman2014} catalog is (mostly) based on measurements obtained with the 2m Fourier-Transform Spectrometer (FTS) at the National Institute of Standards and Technology (NIST). The absolute calibration of these FTS measurements is provided by 30 \ion{Th}{i} lines for which precise laser-based opto-galvanic absolute wavelengths were determined by \citet{Sansonetti1984}, \citet{DeGraffenreid2002} and \citet{DeGraffenreid2012}. These authors state for the precision of their wavenumbers $2\sigma$~errors of $0.0004\,\mathrm{cm^{-1}}$, $0.0002\,\mathrm{cm^{-1}}$ and $0.0003\,\mathrm{cm^{-1}}$, corresponding to $1\sigma$ uncertainties between $1.9\,\mathrm{m/s}$ and $4.1\,\mathrm{m/s}$. 
Although there seem to be some indications for systematic distortions in the FTS wavelength calibration \citep[see e.g.,][ Figure~2]{Redman2014}, these should be limited to an amplitude of $\simeq5\,\mathrm{m/s}$ and therefore smaller than the distortions shown in Figure~\ref{Fig:ThAr_Comparison}. This makes it unlikely that the discrepancy between ThAr laboratory wavelengths and LFC calibrated measurements originates from the FTS calibration.
In particular, the discrepancy between the slices in the blue arm can not be related to the laboratory wavelengths. 

Another aspect might be a possible aging of the ThAr hollow-cathode lamp or in general different operation conditions. With time, argon carrier gas gets trapped by the sputtered cathode material, which leads to a lower gas pressure in the lamp. To compensate, the voltage has to be increased to keep the current constant. Also, since FTS spectrometers are inefficient, ThAr lamps for laboratory measurements are usually operated at much higher currents than for calibration of astronomical spectrographs. This clearly has an effect on the line intensities of argon and thorium and also impacts the wavelengths of the argon transitions. However, the thorium lines are unaffected by this \citep{Nave2018}. So as long as exclusively unblended \ion{Th}{i} lines are used for wavelength calibration, these aspects should not matter.

On the other hand, the LFC frequencies are known to extreme accuracy and, in particular due to the fundamental principle of operation, distortion-free. The relative frequency accuracy of the fundamental comb should be around $10^{-12}$ and possible distortions introduced by the mode filtering in the Fabry-P\'erot cavities kept below $1\,\mathrm{cm/s}$ (Rafael Probst, personal communication).  Also the value of the offset frequency of $\nu_0=7.35\,\mathrm{GHz}$ was explicitly checked and found to be correct (Tilo Steinmetz, personal communication). One therefore has to assume that the discrepancy between NIST and LFC calibrated wavelengths arises within the spectrograph or as part of the data treatment.

One might speculate that different intrinsic line shapes of ThAr and LFC lines could lead to the observed discrepancy. In particular an asymmetric line shape of one of the sources could introduce a systematic offset. However, in \texttt{1HR} mode, the observed width of the ThAr lines is only about 1.5\% larger than the ones of the fully unresolved LFC lines. Also, \ion{Th}{i} has no isotopic substructure and the width of the lines is dominated by Doppler-broadening, which is symmetric by nature. It is therefore not clear if such a small difference in intrinsic line width can actually be responsible for the $12\,\mathrm{m/s}$ difference.

Furthermore, one has to consider whether the data reduction and calibration algorithms could cause the observed discrepancy between ThAr and LFC wavelengths. For example, in Section~\ref{Sec:TraceProfileAmbiguity} it was outlined that the utilized extraction procedure is in the end nonoptimal. However, this applies to both, ThAr and LFC spectra, and it is therefore hard to imagine how this could introduce systematic offsets between ThAr and LFC lines, given that e.g., the exactly identical extraction profiles are used.
The cause for the discrepancy between the two wavelength measurements therefore remains elusive.

\subsection{Comparison of ThAr/FP to LFC Wavelength Solution}
\label{Sec:Comparison_ThArFP-LFC}

\begin{figure*}[t]
 \centering
 \includegraphics[width=\linewidth]{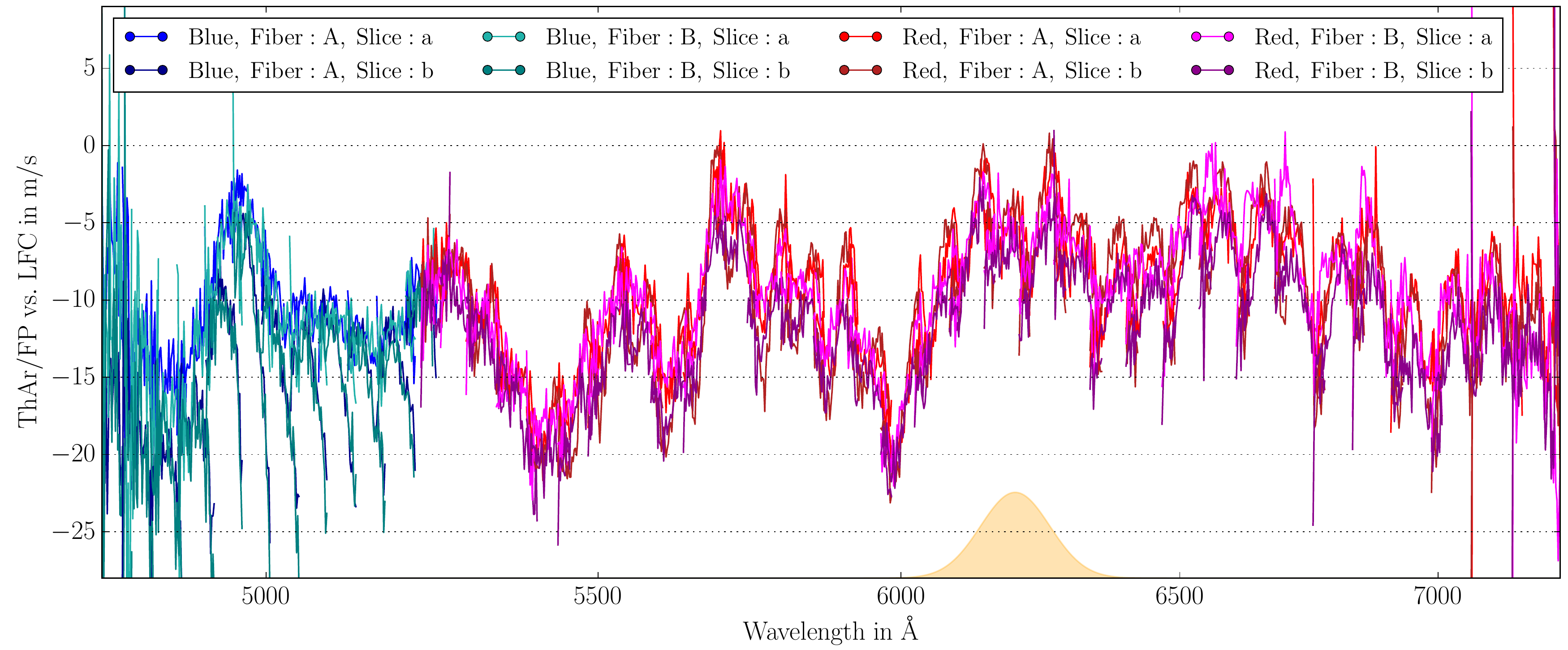}
 \caption{
  Difference of joint ThAr/FP and LFC wavelength solution. Colors indicate different fibers and slices.
  For better visualization, the sampling of the wavelength solutions is reduced to $100\,\mathrm{km/s}$. On these scales, the wavelength solutions should anyway be coherent due to the applied smoothing filter. Parts of the extracted traces where the blaze function drops below 25\% of the peak throughput are excluded. This also reduces the overlap of neighboring spectral orders. The smoothing kernel utilized to determine $D_\mathrm{eff}(\lambda)$ is visualized in orange.
  }
 \label{Fig:ThArFP-LFC_Comparison_1x1}
\end{figure*}

Following the comparison between ThAr laboratory wavelength and LFC calibrated measurements, one can proceed to compare the full ThAr/FP wavelength solution to the one obtained from the laser frequency comb. This is shown on a pixel-by-pixel basis in Figure~\ref{Fig:ThArFP-LFC_Comparison_1x1}.

As one can see, the difference between the two solutions shows a complex, nontrivial pattern.  
Overall, there is a global offset between the two solutions of approximately $-10\,\mathrm{m/s}$. Such an offset is of course unsatisfactory but in the end not essential for a precision test of fundamental constants. Instead, the crucial aspect for deriving constraints on a possible variation of the fine-structure constant are, as visualized in Figure~\ref{Fig:FineStructureVisualization}, spurious velocity shifts between transitions at different wavelengths, caused by distortions of the wavelength scale. For these, Figure~\ref{Fig:ThArFP-LFC_Comparison_1x1} shows deviations between ThAr/FP and LFC calibration of slightly more than $20\,\mathrm{m/s}$ peak-to-valley, if one excludes some more extreme outliers in the blue arm. 

These large-scale properties, i.e., a global offset of $-10\,\mathrm{m/s}$ and fluctuations between $-22\,\mathrm{m/s}$ and zero, are clearly inherited from the ThAr lines, since only these provide absolute wavelength information and define the large-scale structure of the ThAr/FP solution (see determination of $D_\mathrm{eff}(\lambda)$ in Section~\ref{Sec:JointThArFPSolution} for details). 
Therefore, it is not surprising that the overall shape of the difference between ThAr/FP and LFC looks rather similar and has the same statistical properties as the difference between ThAr lines and LFC solution (Figure~\ref{Fig:ThAr_Comparison}).
Thus, all aspects and possible causes for the discrepancy between ThAr laboratory wavelengths and LFC-calibrated measurements discussed in the previous Section~\ref{Sec:Comparison_ThAr-LFC} apply here as well.

However, using also the information from the Fabry-P\'erot interferometer allows in addition a comparison on much smaller scales and with far less noise.
Here, several additional effects can be seen. First, there is basically no discontinuity in the arm overlap region. Therefore, it will be no problem to stitch together the spectra of the two arms. 
Second, the differences between ThAr/FP and LFC solution are quite consistent for the spectra extracted from different fibers and slices. This is an indicator of the statistical uncertainty of the wavelength solutions, which has to be $\lesssim1\,\mathrm{m/s}$.

However, several orders in the blue arm show a very substantial deviation for both Fibers in Slice~b. A similar effect was already visible in Figure~\ref{Fig:ThAr_Comparison}. However, Figure~\ref{Fig:ThArFP-LFC_Comparison_1x1} reveals that this is not a constant offset between the slices but instead an effect that evolves strongly along each spectral orders and reaches up to $15\,\mathrm{m/s}$ discrepancy between Slices~a and b at the red end of each order.
It is of course unclear if this issue is related to the ThAr/FP or the LFC wavelength solution since Figure~\ref{Fig:ThArFP-LFC_Comparison_1x1} only shows the difference between both. The limited fidelity of the ThAr-LFC comparison does not allow definitive conclusions, but the fact that the discrepancy is observed in the ThAr-LFC and ThAr/FP-LFC comparison at broadly consistent magnitude indicates that this particular intra-order effect might be related to the LFC calibration. The $D_\mathrm{eff}(\lambda)$ determination is based on all four traces of each spectral order and the smoothing kernel (indicated in Figure~\ref{Fig:ThArFP-LFC_Comparison_1x1}) extends over more than one order. If the discrepancy between the slices would only affect the ThAr lines, it would appear as enhanced scatter in the $D_\mathrm{eff}(\lambda)$ determination (compare e.g., to Figure~\ref{Fig:Deff}) but would not propagate in this way into the ThAr/FP solution. There either has to be a systematic difference between the slices that evolves along the orders and affects ThAr and FP lines (more or less independently) at similar magnitude or only the LFC. So probably, this different behavior for the two slices originates in the LFC calibration.
One might relate this discrepancy between the two slices to a difference in the line-spread functions.
Indeed, the instrumental line-spread function differs slightly for the two fibers and slices and also along individual orders. However, this affects LFC and ThAr spectra in the same way. One would have to postulate a very subtle effect in which the slightly different change of the instrumental line-spread function along the spectral orders for the two slices causes due to the wider intrinsic line width of the ThAr and FP lines a differential shift with respect to the LFC lines. 
Without detailed investigation and simulation, such a hypothesis can neither be confirmed nor ruled out.

Further intra-order patterns can be seen redwards of $\gtrsim6000\,\mathrm{\AA}$. Here, the four traces per order are consistent with each other but they show for basically every spectral order a clear pattern in which the difference between ThAr/FP and LFC solution is minimal in the center of the spectral order and increases towards both ends. This causes in Figure~\ref{Fig:ThArFP-LFC_Comparison_1x1} a pronounced modulation with the periodicity of the orders.
Also here, one might suspect line-spread function effects as possible cause. But again, a line-spread function that evolves along the trace affects FP and LFC lines. A varying discrepancy between ThAr/FP and LFC solution can only occur if this affects the extended (35\% wider) FP lines in a different way than the unresolved LFC lines. 

The different properties of the intra-order pattern for the two arms can possibly be explained by the slightly different optical design or maybe even just by the optical alignment. Also the spectral format differs between the arms. 
While on the red arm the orders overfill the detector and are truncated, the traces in the blue arm never reach the edge of the detector but instead the blaze function drops to zero flux well before the detector edge, leaving about 1000~pixels unused on each side. 

\begin{figure*}[t]
 \centering
 \includegraphics[width=\linewidth]{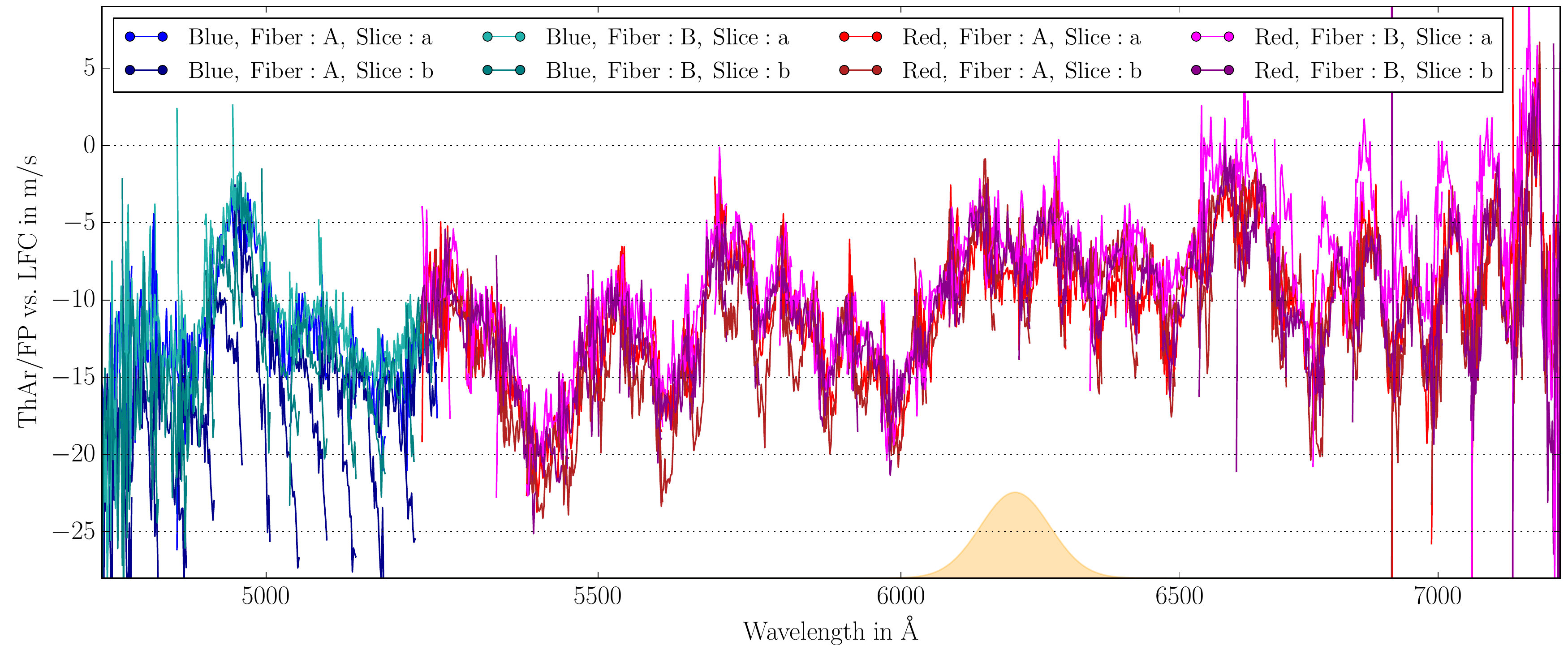}
 \caption{Same comparison of ThAr/FP and LFC wavelength solution as shown in Figure~\ref{Fig:ThArFP-LFC_Comparison_1x1} but for data obtained in \texttt{1HR2x1} readout mode instead of \texttt{1HR1x1} binning. Both datasets are from the same day, taken within 2h, fully independent and processed in an identical way. 
  }
 \label{Fig:ThArFP-LFC_Comparison_2x1}
\end{figure*}

Figure~\ref{Fig:ThArFP-LFC_Comparison_2x1} shows the same comparison between ThAr/FP and LFC wavelength solution as Figure~\ref{Fig:ThArFP-LFC_Comparison_1x1} but for calibrations taken with the \texttt{1HR2x1} readout mode instead of the \texttt{1HR1x1} binning. Both datasets are fully independent but taken on the same day within less than two hours and processed in an identical way. In general, Figure~\ref{Fig:ThArFP-LFC_Comparison_2x1} shows a bit more scatter, which can at least to some degree be explained by the $\sqrt{2}$ larger photon noise in the \texttt{1HR2x1} binning mode.
Apart from this, both figures show very similar patterns in particular on small and medium scales. For instance, one can identify peaks around $4900\,\mathrm{\AA}$, $5300\,\mathrm{\AA}$, $5700\,\mathrm{\AA}$ and valleys near $5400\,\mathrm{\AA}$ and $6000\,\mathrm{\AA}$. The same structure can actually be seen in just the ThAr-LFC comparison (Figure~\ref{Fig:ThAr_Comparison}), underlining again that most of this structure is inherited from the ThAr lines to the ThAr/FP solution.

One difference that can be identified between the two binning modes is the behavior towards the long-wavelength end of the spectral range. For the \texttt{1HR1x1} data, the difference between ThAr/FP and LFC calibration remains approximately constant around $-12\,\mathrm{m/s}$ or drops slightly redwards of $6700\,\mathrm{\AA}$.  In the \texttt{1HR2x1} mode, the wavelength difference seems to increase (the absolute value of the difference decreases) in this region, reaching up to $-5\,\mathrm{m/s}$, however, with a strong modulation on intra-order scales. It has to be noted that this spectral range is particularly poorly sampled with ThAr lines (see Figure~\ref{Fig:Deff}) and suffers from heavy blooming caused by extremely saturated \ion{Ar}{i} lines. It might therefore be worth to reinvestigate the $D_\mathrm{eff}(\lambda)$ determination in this region and possibly improve the ThAr/FP solution.

In addition, some deviations of individual traces can be seen in Figure~\ref{Fig:ThArFP-LFC_Comparison_2x1}. This for instance affects Slice~b of Fiber~A in the range $5500<\lambda<6000$ and Slice~a of Fiber~B for $6500<\lambda$. These inconsistencies are not present in the \texttt{1HR1x1} mode (compare to Figure~\ref{Fig:ThArFP-LFC_Comparison_1x1}). However, they amount to only $3\,\mathrm{m/s}$ and are therefore not considered a major issue. 

One can therefore conclude that apart from mostly minor deviations, considering the overall scatter and distortions in the ThAr/FP and LFC wavelength solutions,  
the used binning mode has no substantial effect on the wavelength calibration. In both cases, there is a global offset between both wavelength solutions of $-10\,\mathrm{m/s}$ and distortions on various scales which amount to differences of about $20\,\mathrm{m/s}$ to $25\,\mathrm{m/s}$ peak-to-valley.

\section{Discussion}

In the following we discuss the results presented above. In particular, we compare the wavelength calibration obtained with our code to the one delivered by the \Espresso{} DRS. Also, we compare the wavelength accuracy of \Espresso{} to spectrographs previously used for fine-structure constant observations and we assess the impact the discovered wavelength distortions might have for a precision test of fundamental constants.

\subsection{Comparison to the \Espresso{} DRS}
\label{Sec:Comparison_DRS}

\begin{figure*}[t]
 \centering
 \includegraphics[width=\linewidth]{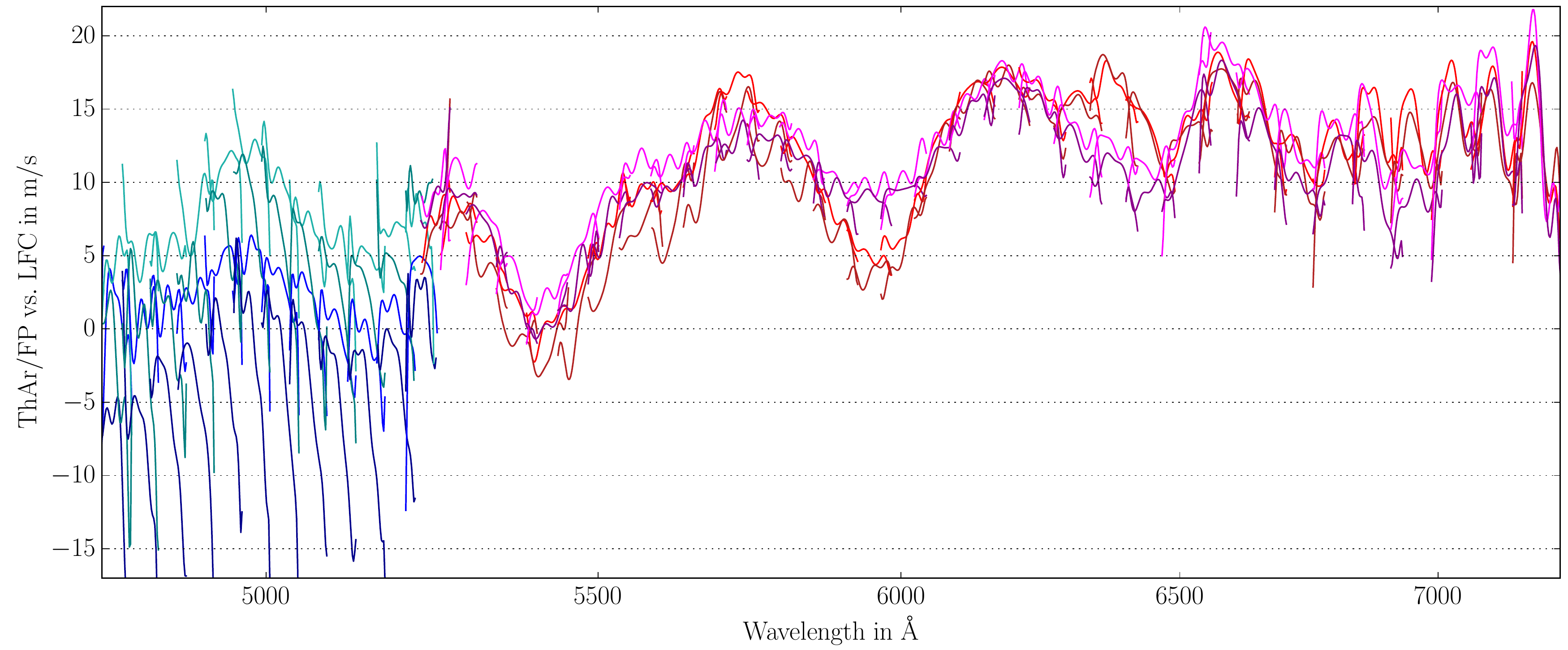}
 \caption{Comparison of ThAr/FP and LFC wavelength solution for the same wavelength calibration frames as shown in Figure~\ref{Fig:ThArFP-LFC_Comparison_2x1}, however, for wavelength solutions computed by the standard \Espresso{} DRS data reduction pipeline.
  }
 \label{Fig:ThArFP-LFC_DRS_Comparison_2x1}
\end{figure*}

As outlined in Section~\ref{Sec:DataReduction}, the results presented in this study are based on our own purpose-developed reduction and calibration code. It is therefore worth to compare our wavelength solutions to the ones delivered by the official ESO pipeline \citep{Lovis2020}. 
Figure~\ref{Fig:ThArFP-LFC_DRS_Comparison_2x1} therefore shows the difference between ThAr/FP and LFC wavelength solution as computed by the \Espresso{} DRS version 2.1.1. The input data are again the calibration frames taken on August 31, 2019 in \texttt{1HR2x1} binning mode. The wavelength solutions shown in Figure~\ref{Fig:ThArFP-LFC_DRS_Comparison_2x1} are thus based on the same data as the ones shown in Figure~\ref{Fig:ThArFP-LFC_Comparison_2x1} and both plots are directly comparable. In particular, both figures show only parts of the extracted traces where the blaze function is above 25\% of the peak throughput.

The \Espresso{} DRS follows in some aspects different philosophies than our code. Most notably, the wavelength solution for each order is described by a 9th order polynomial instead of the nonparametric cubic spline interpolation utilized by our code (see Section~\ref{Sec:JointThArFPSolution}).
Therefore, the wavelength differences shown in Figure~\ref{Fig:ThArFP-LFC_DRS_Comparison_2x1} appear free of statistical noise, but also clearly show for each order a 9th order polynomial shape and have the tendency to diverge towards the edges of the orders.

Despite these fundamental differences, Figure~\ref{Fig:ThArFP-LFC_Comparison_2x1} and \ref{Fig:ThArFP-LFC_DRS_Comparison_2x1} show overall rather similar differences between the two wavelength solutions.
In particular the shapes of the distortions on the largest scales are quite similar and are obviously inherited from the individual ThAr measurements (see Figure~\ref{Fig:ThAr_Comparison}).
The maximum peak-to-valley difference between the ThAr/FP and LFC solutions is about $35\,\mathrm{m/s}$ for the DRS and therefore $\approx5\,\mathrm{m/s}$ larger than for our solution presented in Figure~\ref{Fig:ThArFP-LFC_Comparison_2x1}. However, this is in both cases driven by strong deviations at the edges of the orders. 

While the magnitude of the peak-to-valley differences between ThAr/FP and LFC solution are similar for both pipelines, the global offsets are substantially different. For our code, we find that the mean difference between ThAr/FP and LFC solution is $-11\,\mathrm{m/s}$, i.e., the ThAr/FP solution is blueshifted with respect to the LFC solution. The DRS shows the opposite behavior. Here, the mean difference between ThAr/FP and LFC solution is $+10\,\mathrm{m/s}$. Fortunately, tests for a variation of the fine-structure constant are insensitive to such a global velocity offset since it is fully degenerate with the absorber redshift.

Furthermore, Figure~\ref{Fig:ThArFP-LFC_DRS_Comparison_2x1} shows for the red arm intra-order distortions of $\lesssim 8\,\mathrm{m/s}$, which is comparable to Figure~\ref{Fig:ThArFP-LFC_Comparison_2x1}. Also very similar to our results, the pattern in the red arm is dominated by large-scale structures bluewards of $\approx6400\,\mathrm{\AA}$ while at longer wavelengths intra-order distortions become the dominant feature.
Apart from this, there are substantial coherent deviations ($\approx5\,\mathrm{m/s}$) between Fiber~A and B over nearly $100\,\mathrm{\AA}$ around $6000\,\mathrm{\AA}$ and $6400\,\mathrm{\AA}$ not seen in our calibration.

Substantially more discrepancies are apparent for the blue arm. While for the red arm all four traces show a (mostly) consistent difference between ThAr/FP and LFC solution, which is directly inherited from the ThAr vs. LFC discrepancy (Figure~\ref{Fig:ThAr_Comparison}), the different fibers and slices become inconsistent with each other in the blue arm. 
For our reduction, this was already seen in Figure~\ref{Fig:ThArFP-LFC_Comparison_1x1} (\texttt{1HR1x1}) where Slice~b of both fibers starts to deviate from Slice~a towards the red end of basically all orders of the blue arm. For the \texttt{1HR2x1} mode, this effect is actually less pronounced for Slice~b of Fiber~B and thus only for Slice~b of Fiber~A the difference between ThAr/FP and LFC solution deviates strongly by up to $15\,\mathrm{m/s}$ from the difference found in the other traces (Figure~\ref{Fig:ThArFP-LFC_Comparison_2x1}).

The same effect, i.e., a strong increase (more negative) in the difference between ThAr/FP and LFC wavelength solution towards the red end of the orders is also found in the DRS wavelength solutions (Figure~\ref{Fig:ThArFP-LFC_DRS_Comparison_2x1}). Again, the deviation is strongest for Slice~b of Fiber~A ($\lesssim20\,\mathrm{m/s}$) and smallest for Slice~a ($\approx5\,\mathrm{m/s}$). 
In addition, there seems to be an overall offset between the four traces of up to $10\,\mathrm{m/s}$ (Fiber~A, Slice~b vs. Fiber~B, Slice~a) in the DRS wavelength solution that is not present in our calibration. All together, the wavelength calibration for the blue arm is far less accurate than the red arm with discrepancies of up to $30\,\mathrm{m/s}$~\PtV{} over small scales (one or a few orders). The reason for this is so far not clear. The main difference between the arms is that in the blue arm the traces underfill the detector and are therefore recorded to their full blaze-limited extent. 
For the red arm, however, the orders overfill the detector and are truncated. 
One might therefore speculate that there are in general issues at the outer red and blue end of all traces which are just not seen in the red arm since they fall outside the detector.

Overall, it is noteworthy that two completely independent pipelines, using in certain key aspects different approaches, deliver rather similar results. This clearly indicates that the observed discrepancies are real and not an artifact of the individual implementation. In particular the dominating large-scale structure in the ThAr/FP vs. LFC comparison is extremely similar in both reductions.

\subsection{Comparison to HIRES, UVES and HARPS}
\label{Sec:Comparison_HIRES_UVES_HARPS}

As described in detail in Section~\ref{Sec:WavelengthComparisons}, we find various kinds of systematics in the calibrations that are significantly larger than the photon noise and discrepancies between ThAr/FP and LFC solution of up to $20\,\mathrm{m/s}$. This is clearly not fully satisfactory and demands further efforts to improve on this. However, it is worth to compare \Espresso{} to other spectrographs that were used to derive constraints on fundamental physics and for which a detailed characterization of the wavelength accuracy is available.  

Previously, \citet{Griest2010} and \citet{Whitmore2010} used iodine cells placed in the spectrographs' lights path to assess the wavelength accuracy of Keck/HIRES and VLT/UVES. The iodine reference spectrum was taken with the FTS at the Kitt Peak solar observatory (KPNO). This allowed to compare the wavelength calibration derived from ThAr arc lamps to the (supposedly) much more accurate FTS spectrum%
\footnote{Even solar spectra obtained with FTS spectrometers can exhibit wavelength distortions up to $300\,\mathrm{m/s}$. See \citet{Reiners2016} for a comparison.}.
\citet{Griest2010} found for HIRES very significant intra-order distortions of up to $500\,\mathrm{m/s}$~\PtV{} and global offsets up to $1000\,\mathrm{m/s}$, but also state a large variability between nights. \citet{Whitmore2010} conducted a very similar study for UVES and found  smaller, but still very significant intra-order and global distortions up to $250\,\mathrm{m/s}$~\PtV.

A more detailed analysis was carried out by \citet{Whitmore2015} who used solar twins, i.e., stars with a spectrum very similar to the solar one, and compared these to the KPNO solar spectrum by	 \citet{Chance2010}.
The wavelength distortions they find are composed of two dominating components: intra-order distortions and global, approximately linear slopes.  
Both spectrographs exhibit intra-order distortions of approximately $300\,\mathrm{m/s}$. In addition, \citet{Whitmore2015} report slopes of $800\,\mathrm{m/s}$ over $1500\,\mathrm{\AA}$ for UVES (similar for both arms) and  $600\,\mathrm{m/s}$ over $3000\,\mathrm{\AA}$ for HIRES.
Combined, these result in total wavelength distortions of up to $1000\,\mathrm{m/s}$~\PtV{}, with substantial variations between epochs.

These findings can be compared with our current results for \Espresso. As shown in Figure~\ref{Fig:ThArFP-LFC_Comparison_1x1}, the maximum discrepancy between ThAr/FP and LFC solution is only $22\,\mathrm{m/s}$~\PtV{} and therefore nearly a factor 40$\times$ smaller than for UVES or HIRES. Similarly, we see intra-order distortions of $\approx5\,\mathrm{m/s}$~\PtV{} (except for Slice~b of the blue arm with up to $\approx15\,\mathrm{m/s}$~\PtV), which is as well only about $\sfrac{1}{60}$ of the distortions seen in UVES or HIRES.
A preliminary check also confirmed that the pattern for \Espresso{} does not show large variations with time. The same comparison as shown in Figure~\ref{Fig:ThArFP-LFC_Comparison_2x1} for August 2019 looks at least similar for calibrations taken in November 2018 and February 2020.
Given that HIRES and UVES were in the past the workhorses for fine-structure constant measurements, it is highly encouraging that \Espresso{} performs in terms of wavelength calibration accuracy between one and two orders of magnitude better than the previously used spectrographs.

A much closer comparison can be done with HARPS, which is in many ways (e.g., spectrograph design, fiber feed and calibration strategy) the precursor of \Espresso. Also, it was the first ESO instrument to be equipped with a laser frequency comb \citep{Wilken2010a}.
\citet{Cersullo2019}  compared ThAr/FP to LFC wavelength solution for HARPS, in a way very similar to our test shown in Figure~\ref{Fig:ThArFP-LFC_Comparison_1x1} and \ref{Fig:ThArFP-LFC_Comparison_2x1}.
They find discrepancies up to $40\,\mathrm{m/s}$~\PtV.
So even compared to HARPS, \Espresso{} performs twice as good.

This also highlights that with current methods only limited improvement can be expected from \textit{supercalibration} techniques because very few spectra are available for comparison that have a higher accuracy than provided by \Espresso{} itself. 
For example, even the IAG solar flux atlas \citep{Reiners2016, Baker2020} only states a 'precision and accuracy of $\pm10\,\mathrm{m/s}$'. However, systematics in the wavelength calibration of Fourier-transform spectrometer are likely vastly different from the ones of grating spectrographs.
Therefore, comparisons to asteroids or solar twins will clearly provide highly valuable cross-checks to further strengthen the confidence in the \Espresso{} wavelength solution and be able to identify or exclude systematics that might have not be found by our ThAr/FP vs. LFC comparison \citep[e.g.,][]{Murphy2020}.
Additional tests, in particular with respect to a global wavelength shift, might be possible using the laser guide stars of the VLT/4LGS facility \citep[see e.g.,][]{Vogt2019}.

\subsection{Impact on $\Delta\alpha/\alpha$ Measurement}
\label{Sec:MockSystematics}

In Section~\ref{Sec:WavelengthComparisons} we compared ThAr/FP and LFC wavelength solutions and found significant deviations. 
To get at least a rough estimate of the impact these wavelength calibration uncertainties might have on a $\Delta{}\alpha/\alpha$ measurement, we perform a simple simulation.
Based on a list of atomic transitions containing laboratory wavelengths $\lambda_i^0$ and sensitivity coefficients $q_i$, we create for an absorption system at a given redshift mock wavelength measurements $\lambda_i^\mathrm{obs}$ of the form
\begin{equation}
 \lambda_i^\mathrm{obs} = \frac{ 1 + z_\mathrm{abs} }{ 1/\lambda_i^0 + 2 \: q_i \: \frac{\Delta{}\alpha}{\alpha} } \; \times \; \left( 1 + \frac{\delta{}v_i}{\mathrm{c}} \right)
 \label{Eg:MockWavelengths},
\end{equation}
to which velocity offsets $\delta{}v_i$ representative for the systematics in the wavelength calibration are added.
We chose an absorption redshift of $z_\mathrm{abs}=1.7$, a deviation of the fine-structure constant of $\Delta{}\alpha/\alpha = 1\,\mathrm{ppm}$%
\footnote{
The assumed redshift and $\alpha$-value of the absorption system have no impact on the accuracy of the $\Delta\alpha/\alpha$ measurement and are only picked for folkloristic reasons.
}
and include the transitions \ion{Al}{ii}~$1670\,\mathrm{\AA}$, \ion{Fe}{ii}~$2382\,\mathrm{\AA}$ and $2600\,\mathrm{\AA}$, as well as \ion{Mg}{ii}~$2803\,\mathrm{\AA}$%
\footnote{
No \ion{Mg}{i} and \ion{Al}{iii} lines were used since they stem from different ionization states. 
\ion{Mg}{ii}~$2796\,\mathrm{\AA}$ was excluded since the relevant long-range distortions will at least in the ThAr/FP solution be correlated with the nearby \ion{Mg}{ii}~$2803\,\mathrm{\AA}$ line (note the size of the smoothing kernel in Figure~\ref{Fig:ThArFP-LFC_Comparison_2x1}).
Similarly, the \ion{Fe}{ii} $2586\,\mathrm{\AA}$, $2344\,\mathrm{\AA}$ and $2374\,\mathrm{\AA}$ lines were rejected due to likely correlation with \ion{Fe}{ii}~$2382\,\mathrm{\AA}$ and $2600\,\mathrm{\AA}$. The \ion{Fe}{ii}~$1608\,\mathrm{\AA}$ and \ion{Si}{II}~$1526\,\mathrm{\AA}$ lines were rejected since they are often rather weak.
}.
These choices are of course arbitrary but make the assumed system similar to the $z_\mathrm{abs}=1.6916$ absorber in the HE\,2217-2818 sightline \citep[e.g.,][]{Molaro2013a} and also corresponds to the illustration shown in Figure~\ref{Fig:FineStructureVisualization}.
Line shifts, described by the $q_i$~coefficients, are taken from \citet{Murphy2014} and are about $+2\,\mathrm{m/s}$ per ppm for the low-mass ions and $+21\,\mathrm{m/s}$ for the \ion{Fe}{ii} lines.

Since the origin of the discrepancy presented in Figures~\ref{Fig:ThArFP-LFC_Comparison_1x1} and \ref{Fig:ThArFP-LFC_Comparison_2x1} is not understood and it is without external information impossible to discern from which of the two wavelength solutions this discrepancy arises, we conservatively assume that both wavelength solutions could at all wavelengths deviate from the true value by the most extreme values seen in Figure~\ref{Fig:ThArFP-LFC_Comparison_2x1}. We also apply this assumption to the wavelengths, for which due to the limited spectral range of the LFC, no comparison could be performed. 
Thus, we draw for each transition velocity offsets $\delta{}v_i$ from a flat distribution between $-22\,\mathrm{m/s}$ and $+2\,\mathrm{m/s}$ and add these perturbations to the model wavelengths%
\footnote{The nonvanishing mean is of no relevance since this is degenerate with the absorber redshift. We therefore simply take upper and lower bound from Figure~\ref{Fig:ThArFP-LFC_Comparison_2x1}.}.
The mock wavelengths created in this way are then fitted to obtain estimates for ${z_\mathrm{abs}}$ and ${\Delta{}\alpha/\alpha}$ including the wavelength calibration offsets. Since we are only interested in the systematic effects, no uncertainties are assumed and all four transitions receive equal weights in the fit. In reality, this would depend on the strength of the individual transitions, the velocity structure, saturation effects and the data quality.   

\begin{figure}[htb]
 \centering
 \includegraphics[width=\linewidth]{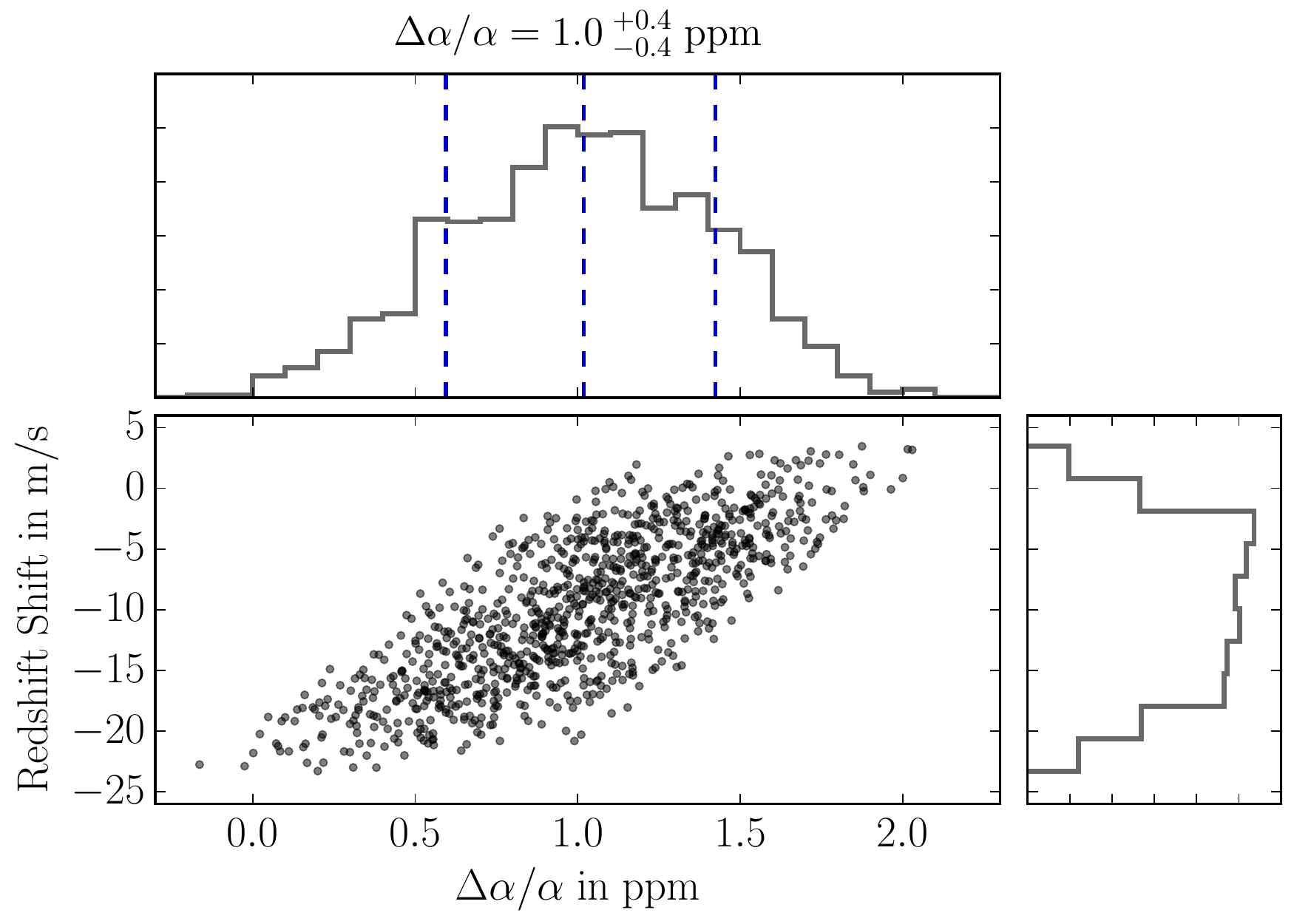} 
 \caption{
 Effect of wavelength calibration uncertainties on the precision of fine-structure constraints. 
 For 1000 mock realizations, wavelength offsets between $-22$ and $+2\,\mathrm{m/s}$ were added to the data before fitting for $\Delta\alpha/\alpha$ and the absorber redshift. The latter one is displayed relative to the initially assumed value.
 }
 \label{Fig:AlphaPrecision}
\end{figure}

We repeat this process for 1000 random realizations of the velocity offsets and show the inferred $\Delta\alpha/\alpha$ values and the difference between apparent and true absorption redshift (which is a pure nuisance parameter) in Figure~\ref{Fig:AlphaPrecision}.   
It has to be stressed that in reality there would be no stochasticity in the process.  Distortions of the wavelength scale are not random but purely systematic. Our Monte-Carlo approach to pick velocity offsets at random only hides the fact that we have no better description of the calibration systematics. Figure~\ref{Fig:AlphaPrecision} therefore rather shows a range of possibilities than true probabilities.

As one can see, the assumed wavelength calibration errors can lead to deviations in $\Delta\alpha/\alpha$ of up to 1.2~ppm, which follows directly from the assumed $24\,\mathrm{m/s}$~\PtV{} calibration systematic and the largest line shift difference of $\approx19\,\mathrm{m/s}$. The concentration of the points in Figure~\ref{Fig:AlphaPrecision}, however, depends on the number of transitions selected. 
Using just \ion{Fe}{ii}~$2600\,\mathrm{\AA}$ and \ion{Mg}{ii}~$2803\,\mathrm{\AA}$ would result in a constant density of scatter points within the parallelogram-shaped envelope. Including a larger number of transitions causes a more concentrated distribution. 

A clear goal is to deliver constraints on the fine-structure constant at the 1~ppm level. Whether the accuracy estimated in Figure~\ref{Fig:AlphaPrecision} is sufficient for this depends on the adopted criteria.
%
%
Following concepts used for stochastic uncertainties, the shown $\pm0.4\,\mathrm{ppm}$ uncertainty would qualify for a 2\,ppm constraint at $5\,\sigma$ significance.
However, the uncertainties due to imperfect wavelength calibration are not statistical but systematic and the treatment in terms of standard deviations not fully applicable. Therefore, the limit on the achievable accuracy is better described by the maximum deviation of 1.2~ppm, close to the 1\,ppm goal.

\subsection{Further Remarks and Outlook}

There are some aspects that have not been addressed in this study but might be worth a detailed investigation in the future. In particular, all aspects regarding the stability of the spectrograph were beyond the scope of this paper and are in general not particularly relevant for a precision test of fundamental physics. Still, some insights into the origin of the discovered systematics might be gained by studying their time evolution. 

In addition, it might be useful to expand the ThAr line list. As described in Section~\ref{Sec:JointThArFPSolution}, determining the Fabry-P\'erot effective gap from the individual ThAr measurements is a rather challenging interpolation problem. Currently, we use the ThAr line list provided by the \Espresso{} DRS, which includes strong and well-selected but also relatively few \ion{Th}{i} lines. Such an approach is totally sufficient for RV studies and the \Espresso{} DRS anyway uses a static description of $D_\mathrm{eff}(\lambda)$ that is only adjusted for global RV drifts \citep{Lovis2020}. For fundamental physics related studies, it might however be helpful to use a substantially extended ThAr line list. This would make the interpolation problem (Figure~\ref{Fig:Deff}) easier and might deliver a more accurate description of the FP effective gap and therefore a better ThAr/FP wavelength solution.

Another aspect that will require further attention in the future is the line-spread function. As described previously, all emission lines are described within this study by Gaussian profiles. However, the true instrumental point-spread function in 2D as well as the line-spread function in extracted 1D spectra is non-Gaussian. Also, it can be slightly asymmetric, vary along an individual order or across different orders and is slightly different for the two slices and fibers. None of these effects are large, but given the extreme accuracy we require, they might be relevant and some of the systematics we found could at least partially be related to line-spread function effects, in particular intra-order distortions and the discrepancy between Slice a and b (Figure~\ref{Fig:ThArFP-LFC_Comparison_1x1}). Also for the global offset of $\approx12\,\mathrm{m/s}$ between ThAr and LFC lines, which persists throughout our study, one might wonder if it could be related to the different line shapes, given that the ThAr lines appear not fully unresolved but (only) 1.5\% wider than the LFC lines.
The complexity of a detailed investigation of these aspects exceeds the scope of this paper but might be presented somewhere else.  

It also has to be stressed that all tests presented within this study are based on spectra composed of (apparent) narrow \textit{emission} lines. All constraints on the fine-structure constant are, however, derived from metal \textit{absorption} lines in quasar spectra.
If indeed the observed discrepancies in the wavelength solutions are related to line-spread function effects and the fact that ThAr and FP lines are intrinsically slightly wider than LFC lines, one might also expect a yet different behavior for absorption-line spectra. 
However, without thorough investigation and modeling, it is quite unclear what this would mean in detail. 

In general, a precise measurement of the instrumental line-spread function can only be obtained from the LFC lines since only these are truly unresolved. Unfortunately, the LFC covers only 57\% of the \Espresso{} spectral range. So even if a proper modeling of the line-spread function would resolve some of the discovered discrepancies, it might only be available for a limited wavelength range. The transitions used to constrain the fine-structure constant are, however, distributed over a wide range in wavelengths. The exact location of the lines depends of course on the absorber redshift, but restricting the range to the region covered by the LFC (approximately $5000\,\mathrm{\AA}$ to $7000\,\mathrm{\AA}$, see Figure~\ref{Fig:LFC_BackgroundResidualsSummary}) would render many well-studied absorption systems basically unobservable with \Espresso{}.
For example, for the $z_\mathrm{abs}=1.69$ systems along the HE\,2217-1818 sightline \citep{Molaro2013a}, the \ion{Mg}{ii} lines would not be covered anymore, leaving only the \ion{Fe}{ii} lines around $\simeq2500\,\mathrm{\AA}$ for a constraint of the  fine-structure constant (compare to Figure~\ref{Fig:FineStructureVisualization}). However, these five lines have a limited spread in sensitivity with respect to $\alpha$.
In addition, relying only on the LFC wavelength range  will make it at any redshift impossible to observe the \ion{Fe}{ii}~$1608\,\mathrm{\AA}$ line together with the other \ion{Fe}{ii} lines at $\lambda_\mathrm{rest} > 2300\,\mathrm{\AA}$. 
Usually, this is a very valuable combination of transitions, since the \ion{Fe}{ii}~$1608\,\mathrm{\AA}$ line shifts in opposite direction compared to the other \ion{Fe}{ii} lines and using only transitions originating from the same ion avoids possible systematics related to different velocity structures and isotopic abundances.
For many absorption systems, it is thus crucial to use the full spectral range of \Espresso{} and a restriction to the limited range covered by the currently installed laser frequency comb would substantially diminish the scientific return. Therefore, it is important that future improvements of the \Espresso{} wavelength calibration are applicable to the full spectral range of \Espresso{}.

\section{Summary}

The observation of metal absorption systems in spectra of distant quasars allows to constrain the value of fundamental physical constants throughout the history of the Universe, in particular of the fine-structure constant $\alpha$. 
However, previous studies provided inconclusive results whether the fine-structure constant was indeed different in the past. 
Since the constraint on $\alpha$ comes from a very accurate wavelength measurement of the absorption lines, possible calibration inaccuracies of the existing spectrographs (e.g., HIRES and UVES) were a serious limitation preventing progress in this field.
A significant advancement is therefore expected from the novel ultra-stable high-resolution spectrograph \Espresso{} recently installed at the VLT, which is specially designed for precise radial-velocity studies and tests of fundamental physics.
Crucially, \Espresso{} is equipped with advanced wavelength calibration sources, namely a Fabry-P\'erot interferometer and a laser frequency comb.

In preparation of the test for a possible variation of the fine-structure constant carried out as part of the \Espresso{} GTO program, we conduct a thorough assessment of the  spectrograph's wavelength calibration. 
To do so, we develop our own data reduction and calibration software, which is fully independent from the \Espresso{} Data Reduction Software
and acts as a flexible test bed for experiments and the development of advanced calibration techniques. We use it to carry out a careful investigation of all aspects relevant for the accuracy of the wavelength solution.

In Section~\ref{Sec:Extraction}, we focus on the crucial aspect of spectral extraction. We show that the fundamental assumption underlying the utilized \textit{flat spectral extraction} algorithm \citep[outlined in][]{Zechmeister2014}, i.e., that there is a unique correspondence between wavelength and pixel position, is not fully satisfied for \Espresso{}. In Figures~\ref{Fig:OptialExtraction_2D_SHR_FLAT}, \ref{Fig:OptialExtraction_2D_SHR_FP}, and \ref{Fig:OptimalExtraction_1D_SHR}, we demonstrate by comparing spectral flatfields to FP spectra that the trace profile in cross-dispersion direction does indeed depend on the spectral structure of the observed source. This effect, already described by \citet{ABolton2010}, will unavoidably cause imperfections in the spectral extraction process. However, without implementing a far more sophisticated spectral extraction algorithm that fully models the 2D instrumental point-spread function, we are unable to quantify the resulting impact on the extracted spectra.

In Section~\ref{Sec:LineFitting}, we investigate the properties of the LFC, in particular regarding the delivered flux levels, and show that a careful modeling of the strongly modulated LFC background light (Figure~\ref{Fig:LFC_BackgroundEnvelope} and \ref{Fig:LFC_BackgroundResiduals}) is necessary to achieve accurate centroiding of the lines. Not doing so can introduce systematic displacements with a standard deviation of up to $3\,\mathrm{m/s}$ and extreme shifts for individual lines of up to $60\,\mathrm{m/s}$ (Figure~\ref{Fig:LFC_BackgroundResidualsSummary}). The Fabry-P\'erot interferometer, however, is far less affected by this effect since its background light component is substantially less structured.

Furthermore, we discover in the positions of FP and LFC lines a form of highly correlated noise that causes displacements of individual lines by $\simeq${}$10\,\mathrm{m/s}$. The displacements of consecutive lines compared to a smooth wavelength solution tend to have opposite signs  and the amplitude of the pattern is strongly modulated along orders (Figure~\ref{Fig:PatternNoise_FP305}). This \textit{beat pattern noise} is present at similar magnitude in FP and LFC spectra, in both arms, fibers and slices of the spectrograph. Its amplitude varies between 4 and $\mathrm{10\,m/s}$, increasing with wavelength, and is therefore up to 4$\times$ larger than the photon-limited precision (Figure~\ref{Fig:PatternNoise_Summary}). Despite detailed investigation (see Section~\ref{Sec:BeatPatternNoise}) we cannot determine the origin of this issue and can only speculate that it could be related to the spectral extraction.

The derivation of the joint ThAr/FP wavelength solution requires a careful characterization of the Fabry-P\'erot interferometer.
In particular, the FP \textit{effective gap} size $D_\mathrm{eff}(\lambda)$ has to be determined based on the observed ThAr arc lines.
To model the $D_\mathrm{eff}(\lambda)$ function, we follow a nonparametric approach based on a purpose-developed kernel smoothing filter (Section~\ref{Sec:JointThArFPSolution}), which performs much better than the classically used high-order polynomials and allows full control over the introduced correlations, crucial for a precision test of fundamental physics. Since Version~1.5, a similar approach is utilized by the \Espresso{}~DRS.
Despite this, we find large residuals with a dispersion of $8.5\,\mathrm{m/s}$ in the $D_\mathrm{eff}(\lambda)$ determination, far in excess of the photon-noise limited precision of $2.3\,\mathrm{m/s}$ (Figure~\ref{Fig:Deff} and \ref{Fig:Deff_Hist}). 
Most notably, the measurements originating from the same \ion{Th}{i} line but observed in different fibers and slices usually show significant disagreement (see Figure~\ref{Fig:ThAr_Comparison_103R}). This discrepancy cannot be caused by external reasons (like e.g., the  ThAr laboratory line list) but clearly has to be related to the spectrograph or the data processing itself.  

A key feature of \Espresso{} is that, with the joint ThAr/FP and laser frequency comb solutions, it offers two fully independent and extremely precise high-fidelity wavelength calibrations. To assess their accuracy, we therefore perform (over the limited wavelength range where this is possible) a detailed comparison of the two wavelength solutions (Section~\ref{Sec:WavelengthComparisons}). We find significant discrepancies up to $22\,\mathrm{m/s}$~\PtV{} in \texttt{1HR1x1} mode ($28\,\mathrm{m/s}$~\PtV{} for \texttt{1HR2x1}), which are dominated by complex large-scale distortions (Figure~\ref{Fig:ThArFP-LFC_Comparison_1x1} and \ref{Fig:ThArFP-LFC_Comparison_2x1}) and includes a global offset of $\approx10\,\mathrm{m/s}$ between ThAr/FP and LFC solution.
Most of the large-scale distortions can already be seen in a much simpler comparison just between the ThAr lines and the LFC solution (Figure~\ref{Fig:ThAr_Comparison}) and are therefore unrelated to the Fabry-P\'erot interferometer but inherited into the ThAr/FP solution. 

A cross-check between our results and the same comparison between ThAr/FP and LFC wavelength solution delivered by the \Espresso{} DRS shows that both codes, despite significant differences in the modeling of the wavelength solutions, yield overall quite similar results (Section~\ref{Sec:Comparison_DRS}, Figure~\ref{Fig:ThArFP-LFC_DRS_Comparison_2x1}). The discrepancies between the two wavelength solutions are slightly smaller for our computation, but the majority of the features appears in the outputs of both codes, highlighting that these issues are indeed inherent to the spectrograph or the general calibration approach and not related to the implementation. 

Despite the numerous systematics discovered within the course of our study and the obvious discrepancies between the two wavelength solutions, the \Espresso{} wavelength calibration still has to be considered excellent.
This is particularly true when comparing with HIRES or UVES, for which wavelength distortions up to $800\,\mathrm{m/s}$~\PtV{} were reported (Section~\ref{Sec:Comparison_HIRES_UVES_HARPS}). 
Even with respect to HARPS, basically the predecessor of \Espresso{}, we find in this study substantially smaller differences between ThAr/FP and LFC wavelength solutions. \Espresso{} therefore has to be considered the astronomical spectrograph with the highest demonstrated wavelength accuracy, only rivaled by Fourier-transform spectrometers at Solar observatories. 

Following this, we demonstrate in Section~\ref{Sec:MockSystematics} that the discovered wavelength calibration uncertainties, conservatively taking the $24\,\mathrm{m/s}$~\PtV{} discrepancy between ThAr/FP and LFC solution as general calibration uncertainty at any wavelength, correspond (for a representative absorption system) to errors in the fine-structure constant of less than $\pm1.2\,\mathrm{ppm}$. This is of the same order as the expected statistical errors achievable in long (20 to 40\,h) integration of the brightest and most-suitable quasar sightlines.
Pushing to even tighter limits possibly requires significant improvements to the \Espresso{} wavelength calibration.
This has to be achieved by better understanding the causes for the observed inconsistencies.
The identification and precise characterization of the relevant systematics in this study is clearly the first step towards future improvements. A fundamental limitation could, however, be the restricted wavelength range of the laser frequency comb.

\begin{acknowledgements}

The authors thank the anonymous referee for helpful comments.

TMS would like to thank Rafael Probst and Tilo Steinmetz for their support regarding the \Espresso{} LFC system.

This research has made use of Astropy, a community-developed core Python package for Astronomy \citep{Astropy2013,Astropy2018}, and Matplotlib \citep{Hunter2007}.

TMS  acknowledges the support by INAF/FRONTIERA through the "Progetti Premiali" funding scheme of the Italian Ministry of Education, University, and Research.

The INAF authors acknowledge financial support of the Italian Ministry of Education, University, and Research with PRIN 201278X4FL and the "Progetti Premiali" funding scheme. 

MTM acknowledges the support of the Australian Research Council through \textsl{Future Fellowship} grant FT180100194 and the Institute for Fundamental Physics of the Universe for hosting a collaborative visit in June 2019.

This work has been carried out within the framework of the National Centre of Competence in Research PlanetS supported by the Swiss National Science Foundation. 
RA acknowledge the financial support of the SNSF.

This work was supported by FCT - Funda\c{c}\~ao para a Ci\^encia e a Tecnologia through national funds and by FEDER---Fundo Europeu de Desenvolvimento Regional through COMPETE2020 - Programa Operacional Competitividade e Internacionaliza\c{c}\~ao by these grants: UID/FIS/04434/2019; UIDB/04434/2020; UIDP/04434/2020; PTDC/FIS-AST/32113/2017 \& POCI-01-0145-FEDER-032113; PTDC/FIS-AST/28953/2017 \& POCI-01-0145-FEDER-028953; PTDC/FIS-AST/28987/2017 \& POCI-01-0145-FEDER-028987.

ACL is supported by an FCT fellowship (SFRH/BD/113746/2015), under the FCT Doctoral Program PhD::SPACE (PD/00040/2012).


This work was support by the Spanish Ministry of Science and Innovation (MICINN) under the 2013 Ram\'on y Cajal program RYC-2013-14875, the 2019 Juan de la Cierva program and the AYA2017-86389-P program.

SGS acknowledges support from the FCT through the Investigador FCT Contract No. CEECIND/00826/2018 and POPH/FSE (EC).


DE acknowledges support from from the European Research Council (ERC) under the European Union’s Horizon 2020 research and innovation programme (project {\sc Four Aces}; grant agreement No 724427).

\end{acknowledgements}

\bibliographystyle{aa_url}
\bibliography{Literature}{}

\end{document}

%% file: ListAuthors.tex
\author{
	Tobias M. Schmidt\inst{1}\and
	Paolo Molaro\inst{1,2}\and
	Michael T. Murphy\inst{3}\and
	Christophe Lovis\inst{4}\and
	Guido Cupani\inst{1,2}\and
	Stefano Cristiani\inst{1,2}\and
	Francesco A. Pepe\inst{4}\and
	Rafael Rebolo\inst{5,6}\and
	Nuno C. Santos\inst{7,8}\and
	Manuel Abreu\inst{9,10}\and
	Vardan Adibekyan\inst{7,11,8}\and
	Yann Alibert\inst{12}\and
	Matteo Aliverti\inst{13}\and
	Romain Allart\inst{4}\and
	Carlos Allende Prieto\inst{5,6}\and
	David Alves\inst{9,10}\and
	Veronica Baldini\inst{14}\and
	Christopher Broeg\inst{15}\and
	Alexandre Cabral\inst{9,10}\and
	Giorgio Calderone\inst{1}\and
	Roberto Cirami\inst{1}\and
	Jo\~{a}o Coelho\inst{9,10}\and
	Igor Coretti\inst{1}\and
	Valentina D'Odorico\inst{1,2}\and
	Paolo Di Marcantonio\inst{1}\and
	David Ehrenreich\inst{4}\and
	Pedro Figueira\inst{7,16}\and
	Matteo Genoni\inst{13}\and
	Ricardo G\'enova Santos\inst{5,6}\and
	Jonay I. Gonz\'alez Hern\'andez\inst{5,6}\and
	Florian Kerber\inst{17}\and
	Marco Landoni\inst{18,13}\and
	Ana C. O. Leite\inst{7,8}\and
	Jean-Louis Lizon\inst{17}\and
	Gaspare Lo Curto\inst{17}\and
	Antonio Manescau\inst{17}\and
	Carlos J.A.P. Martins\inst{7,11}\and
	Denis Meg\'evand\inst{4}\and
	Andrea Mehner\inst{16}\and
	Giuseppina Micela\inst{19}\and
	Andrea Modigliani\inst{17}\and
	Manuel Monteiro\inst{7}\and
	Mario J. P. F. G. Monteiro\inst{7,8}\and
	Eric Mueller\inst{17}\and
	Nelson J. Nunes\inst{9,10}\and
	Luca Oggioni\inst{13}\and
	Ant\'onio Oliveira\inst{9,10}\and
	Giorgio Pariani\inst{13}\and
	Luca Pasquini\inst{17}\and
	Edoardo Redaelli\inst{13}\and
	Marco Riva\inst{13}\and
	Pedro Santos\inst{9,10}\and
	Danuta Sosnowska\inst{4}\and
	S\'ergio G. Sousa\inst{7}\and
	Alessandro Sozzetti\inst{20}\and
	Alejandro Su\'arez Mascare\~no\inst{5,6}\and
	St\'ephane Udry\inst{4}\and
	Maria-Rosa Zapatero Osorio\inst{21}\and
	Filippo Zerbi\inst{13}
}

%% file: ListAffiliations.tex
\institute{
	Osservatorio Astronomico di Trieste, via G. B. Tiepolo 11, I-34143 Trieste, Italy, \email{tobias.schmidt@inaf.it}\and
	Institute for Fundamental Physics of the Universe, Via Beirut 2, I-34151 Miramare, Trieste, Italy\and
	Centre for Astrophysics and Supercomputing, Swinburne University of Technology, Hawthorn, Victoria 3122, Australia\and
	Observatoire Astronomique de l'Universit\'e de Gen\`eve, Chemin des Maillettes 51, Sauverny, CH-1290, Switzerland\and
	Instituto de Astrof\'{\i}sica de Canarias (IAC), Calle V\'{\i}a L\'actea s/n, E-38205 La Laguna, Tenerife, Spain\and
	Departamento de Astrof\'{\i}sica, Universidad de La Laguna (ULL), E-38206 La Laguna, Tenerife, Spain\and
	Instituto de Astrof\'{\i}sica e Ci\^encias do Espa\c co, CAUP, Universidade do Porto, Rua das Estrelas, 4150-762, Porto, Portugal\and
	Departamento de F\'{\i}sica e Astronomia, Faculdade de Ci\^encias, Universidade do Porto, Rua Campo Alegre, 4169-007, Porto, Portugal\and
	Instituto de Astrof\'isica e Ci\^encias do Espa\c{c}o, Faculdade de Ci\^encias da Universidade de Lisboa, Campo Grande, PT1749-016 Lisboa, Portugal\and
	Departamento de Física da Faculdade de Ciências da Universidade de Lisboa, Edifício C8, 1749-016 Lisboa, Portugal\and
	Centro de Astrof\'{\i}sica da Universidade do Porto, Rua das Estrelas, 4150-762 Porto, Portugal\and
	Physics Institute, University of Bern, Sidlerstrasse 5, 3012 Bern, Switzerland\and
	INAF - Osservatorio Astronomico di Brera, Via E. Bianchi 46, I-23807 Merate, Italy\and
	INAF - Istituto di Radio Astronomia, Via P. Gobetti, 101, 40129 Bologna\and
	Center for Space and Habitability, University of Bern, Gesellschaftsstrasse 6, 3012 Bern, Switzerland\and
	European Southern Observatory, Alonso de Co\'ordova 3107, Vitacura, Regio\'on Metropolitana, Chile\and
	European Southern Observatory, Karl-Schwarzschild-Strasse 2, 85748, Garching b. M\"unchen, Germany\and
	INAF - Istituto Nazionale di Astrofisica.  Osservatorio Astronomico di Cagliari.  Via della Scienza 5, 09047 Selargius (CA), Italy\and
	INAF - Osservatorio Astronomico di Palermo, Piazza del Parlamento 1, I-90134 Palermo, Italy\and
	INAF - Osservatorio Astrofisico di Torino, via Osservatorio 20, 10025 Pino Torinese, Italy\and
	Centro de Astrobiolog\'{\i}a (CSIC-INTA), Crta. Ajalvir km 4, E-28850 Torrej\'on de Ardoz, Madrid, Spain
}

%% file: Espresso_WavelengthCalibration_arxiv.bbl
\newcommand{\noop}[1]{}
\begin{thebibliography}{70}
\expandafter\ifx\csname natexlab\endcsname\relax\def\natexlab#1{#1}\fi

\bibitem[{{Astropy Collaboration} {et~al.}(2018){Astropy Collaboration},
  {Price-Whelan}, {Sip{\H{o}}cz}, {G{\"u}nther}, {Lim}, {Crawford}, {Conseil},
  {Shupe}, {Craig}, {Dencheva}, {Ginsburg}, {Vand erPlas}, {Bradley},
  {P{\'e}rez-Su{\'a}rez}, {de Val-Borro}, {Aldcroft}, {Cruz}, {Robitaille},
  {Tollerud}, {Ardelean}, {Babej}, {Bach}, {Bachetti}, {Bakanov}, {Bamford},
  {Barentsen}, {Barmby}, {Baumbach}, {Berry}, {Biscani}, {Boquien}, {Bostroem},
  {Bouma}, {Brammer}, {Bray}, {Breytenbach}, {Buddelmeijer}, {Burke},
  {Calderone}, {Cano Rodr{\'\i}guez}, {Cara}, {Cardoso}, {Cheedella}, {Copin},
  {Corrales}, {Crichton}, {D'Avella}, {Deil}, {Depagne}, {Dietrich}, {Donath},
  {Droettboom}, {Earl}, {Erben}, {Fabbro}, {Ferreira}, {Finethy}, {Fox},
  {Garrison}, {Gibbons}, {Goldstein}, {Gommers}, {Greco}, {Greenfield},
  {Groener}, {Grollier}, {Hagen}, {Hirst}, {Homeier}, {Horton}, {Hosseinzadeh},
  {Hu}, {Hunkeler}, {Ivezi{\'c}}, {Jain}, {Jenness}, {Kanarek}, {Kendrew},
  {Kern}, {Kerzendorf}, {Khvalko}, {King}, {Kirkby}, {Kulkarni}, {Kumar},
  {Lee}, {Lenz}, {Littlefair}, {Ma}, {Macleod}, {Mastropietro}, {McCully},
  {Montagnac}, {Morris}, {Mueller}, {Mumford}, {Muna}, {Murphy}, {Nelson},
  {Nguyen}, {Ninan}, {N{\"o}the}, {Ogaz}, {Oh}, {Parejko}, {Parley}, {Pascual},
  {Patil}, {Patil}, {Plunkett}, {Prochaska}, {Rastogi}, {Reddy Janga},
  {Sabater}, {Sakurikar}, {Seifert}, {Sherbert}, {Sherwood-Taylor}, {Shih},
  {Sick}, {Silbiger}, {Singanamalla}, {Singer}, {Sladen}, {Sooley},
  {Sornarajah}, {Streicher}, {Teuben}, {Thomas}, {Tremblay}, {Turner},
  {Terr{\'o}n}, {van Kerkwijk}, {de la Vega}, {Watkins}, {Weaver}, {Whitmore},
  {Woillez}, {Zabalza}, \& {Astropy Contributors}}]{Astropy2018}
{Astropy Collaboration}, {Price-Whelan}, A.~M., {Sip{\H{o}}cz}, B.~M., {et~al.}
  2018, \aj, 156, 123, \href{http://doi.org/10.3847/1538-3881/aabc4f}{doi},
  \href{https://ui.adsabs.harvard.edu/abs/2018AJ....156..123A}{ADS}

\bibitem[{{Astropy Collaboration} {et~al.}(2013){Astropy Collaboration},
  {Robitaille}, {Tollerud}, {Greenfield}, {Droettboom}, {Bray}, {Aldcroft},
  {Davis}, {Ginsburg}, {Price-Whelan}, {Kerzendorf}, {Conley}, {Crighton},
  {Barbary}, {Muna}, {Ferguson}, {Grollier}, {Parikh}, {Nair}, {Unther},
  {Deil}, {Woillez}, {Conseil}, {Kramer}, {Turner}, {Singer}, {Fox}, {Weaver},
  {Zabalza}, {Edwards}, {Azalee Bostroem}, {Burke}, {Casey}, {Crawford},
  {Dencheva}, {Ely}, {Jenness}, {Labrie}, {Lim}, {Pierfederici}, {Pontzen},
  {Ptak}, {Refsdal}, {Servillat}, \& {Streicher}}]{Astropy2013}
{Astropy Collaboration}, {Robitaille}, T.~P., {Tollerud}, E.~J., {et~al.} 2013,
  \aap, 558, A33, \href{http://doi.org/10.1051/0004-6361/201322068}{doi},
  \href{https://ui.adsabs.harvard.edu/abs/2013A&A...558A..33A}{ADS}

\bibitem[{{Baker} {et~al.}(2020){Baker}, {Blake}, \& {Reiners}}]{Baker2020}
{Baker}, A.~D., {Blake}, C.~H., \& {Reiners}, A. 2020, \apjs, 247, 24,
  \href{http://doi.org/10.3847/1538-4365/ab6a1c}{doi},
  \href{https://ui.adsabs.harvard.edu/abs/2020ApJS..247...24B}{ADS}

\bibitem[{{Bauer} {et~al.}(2015){Bauer}, {Zechmeister}, \&
  {Reiners}}]{Bauer2015}
{Bauer}, F.~F., {Zechmeister}, M., \& {Reiners}, A. 2015, \aap, 581, A117,
  \href{http://doi.org/10.1051/0004-6361/201526462}{doi},
  \href{https://ui.adsabs.harvard.edu/abs/2015A&A...581A.117B}{ADS}

\bibitem[{{Bolton} \& {Schlegel}(2010)}]{ABolton2010}
{Bolton}, A.~S. \& {Schlegel}, D.~J. 2010, \pasp, 122, 248,
  \href{http://doi.org/10.1086/651008}{doi},
  \href{https://ui.adsabs.harvard.edu/abs/2010PASP..122..248B}{ADS}

\bibitem[{{Bouchy} {et~al.}(2013){Bouchy}, {D{\'\i}az}, {H{\'e}brard},
  {Arnold}, {Boisse}, {Delfosse}, {Perruchot}, \& {Santerne}}]{Bouchy2013}
{Bouchy}, F., {D{\'\i}az}, R.~F., {H{\'e}brard}, G., {et~al.} 2013, \aap, 549,
  A49, \href{http://doi.org/10.1051/0004-6361/201219979}{doi},
  \href{https://ui.adsabs.harvard.edu/abs/2013A&A...549A..49B}{ADS}

\bibitem[{{Calderone} {et~al.}(2016){Calderone}, {Baldini}, {Cirami},
  {Coretti}, {Cristiani}, {Di Marcantonio}, {Landoni}, {M{\'e}gevand}, {Riva},
  \& {Santin}}]{Calderone2016}
{Calderone}, G., {Baldini}, V., {Cirami}, R., {et~al.} 2016, in Society of
  Photo-Optical Instrumentation Engineers (SPIE) Conference Series, Vol. 9913,
  \procspie, 99132K, \href{http://doi.org/10.1117/12.2231650}{doi},
  \href{https://ui.adsabs.harvard.edu/abs/2016SPIE.9913E..2KC}{ADS}

\bibitem[{{Cersullo} {et~al.}(2019){Cersullo}, {Coffinet}, {Chazelas}, {Lovis},
  \& {Pepe}}]{Cersullo2019}
{Cersullo}, F., {Coffinet}, A., {Chazelas}, B., {Lovis}, C., \& {Pepe}, F.
  2019, \aap, 624, A122,
  \href{http://doi.org/10.1051/0004-6361/201833852}{doi},
  \href{https://ui.adsabs.harvard.edu/abs/2019A&A...624A.122C}{ADS}

\bibitem[{{Chance} \& {Kurucz}(2010)}]{Chance2010}
{Chance}, K. \& {Kurucz}, R.~L. 2010, \jqsrt, 111, 1289,
  \href{http://doi.org/10.1016/j.jqsrt.2010.01.036}{doi},
  \href{https://ui.adsabs.harvard.edu/abs/2010JQSRT.111.1289C}{ADS}

\bibitem[{{Chand} {et~al.}(2004){Chand}, {Srianand}, {Petitjean}, \&
  {Aracil}}]{Chand2004}
{Chand}, H., {Srianand}, R., {Petitjean}, P., \& {Aracil}, B. 2004, \aap, 417,
  853, \href{http://doi.org/10.1051/0004-6361:20035701}{doi},
  \href{https://ui.adsabs.harvard.edu/abs/2004A&A...417..853C}{ADS}

\bibitem[{{Chazelas} {et~al.}(2012){Chazelas}, {Pepe}, \&
  {Wildi}}]{Chazelas2012}
{Chazelas}, B., {Pepe}, F., \& {Wildi}, F. 2012, in Society of Photo-Optical
  Instrumentation Engineers (SPIE) Conference Series, Vol. 8450, \procspie,
  845013, \href{http://doi.org/10.1117/12.926188}{doi},
  \href{https://ui.adsabs.harvard.edu/abs/2012SPIE.8450E..13C}{ADS}

\bibitem[{{Coffinet} {et~al.}(2019){Coffinet}, {Lovis}, {Dumusque}, \&
  {Pepe}}]{Coffinet2019}
{Coffinet}, A., {Lovis}, C., {Dumusque}, X., \& {Pepe}, F. 2019, \aap, 629,
  A27, \href{http://doi.org/10.1051/0004-6361/201833272}{doi},
  \href{https://ui.adsabs.harvard.edu/abs/2019A&A...629A..27C}{ADS}

\bibitem[{{DeGraffenreid} {et~al.}(2012){DeGraffenreid}, {Campbell}, \&
  {Sansonetti}}]{DeGraffenreid2012}
{DeGraffenreid}, W., {Campbell}, S.~C., \& {Sansonetti}, C.~J. 2012, Journal of
  the Optical Society of America B Optical Physics, 29, 1580,
  \href{http://doi.org/10.1364/JOSAB.29.001580}{doi},
  \href{https://ui.adsabs.harvard.edu/abs/2012JOSAB..29.1580D}{ADS}

\bibitem[{{Degraffenreid} \& {Sansonetti}(2002)}]{DeGraffenreid2002}
{Degraffenreid}, W. \& {Sansonetti}, C.~J. 2002, Journal of the Optical Society
  of America B Optical Physics, 19, 1711,
  \href{http://doi.org/10.1364/JOSAB.19.001711}{doi},
  \href{https://ui.adsabs.harvard.edu/abs/2002JOSAB..19.1711D}{ADS}

\bibitem[{{Dekker} {et~al.}(2000){Dekker}, {D'Odorico}, {Kaufer}, {Delabre}, \&
  {Kotzlowski}}]{Dekker2000}
{Dekker}, H., {D'Odorico}, S., {Kaufer}, A., {Delabre}, B., \& {Kotzlowski}, H.
  2000, in \procspie, Vol. 4008, Optical and IR Telescope Instrumentation and
  Detectors, ed. M.~{Iye} \& A.~F. {Moorwood}, 534--545,
  \href{http://doi.org/10.1117/12.395512}{doi},
  \href{http://adsabs.harvard.edu/abs/2000SPIE.4008..534D}{ADS}

\bibitem[{{Dzuba} {et~al.}(1999){Dzuba}, {Flambaum}, \& {Webb}}]{Dzuba1999a}
{Dzuba}, V.~A., {Flambaum}, V.~V., \& {Webb}, J.~K. 1999, \prl, 82, 888,
  \href{http://doi.org/10.1103/PhysRevLett.82.888}{doi},
  \href{https://ui.adsabs.harvard.edu/abs/1999PhRvL..82..888D}{ADS}

\bibitem[{{Evans} {et~al.}(2014){Evans}, {Murphy}, {Whitmore}, {Misawa},
  {Centurion}, {D'Odorico}, {Lopez}, {Martins}, {Molaro}, {Petitjean},
  {Rahmani}, {Srianand}, \& {Wendt}}]{Evans2014}
{Evans}, T.~M., {Murphy}, M.~T., {Whitmore}, J.~B., {et~al.} 2014, \mnras, 445,
  128, \href{http://doi.org/10.1093/mnras/stu1754}{doi},
  \href{https://ui.adsabs.harvard.edu/abs/2014MNRAS.445..128E}{ADS}

\bibitem[{{Griest} {et~al.}(2010){Griest}, {Whitmore}, {Wolfe}, {Prochaska},
  {Howk}, \& {Marcy}}]{Griest2010}
{Griest}, K., {Whitmore}, J.~B., {Wolfe}, A.~M., {et~al.} 2010, \apj, 708, 158,
  \href{http://doi.org/10.1088/0004-637X/708/1/158}{doi},
  \href{https://ui.adsabs.harvard.edu/abs/2010ApJ...708..158G}{ADS}

\bibitem[{{Horne}(1986)}]{Horne1986}
{Horne}, K. 1986, \pasp, 98, 609, \href{http://doi.org/10.1086/131801}{doi},
  \href{https://ui.adsabs.harvard.edu/abs/1986PASP...98..609H}{ADS}

\bibitem[{Hunter(2007)}]{Hunter2007}
Hunter, J.~D. 2007, Computing in Science \& Engineering, 9, 90,
  \href{http://doi.org/10.1109/MCSE.2007.55}{doi}

\bibitem[{{Hunter} \& {Ramsey}(1992)}]{Hunter1992}
{Hunter}, T.~R. \& {Ramsey}, L.~W. 1992, \pasp, 104, 1244,
  \href{http://doi.org/10.1086/133115}{doi},
  \href{https://ui.adsabs.harvard.edu/abs/1992PASP..104.1244H}{ADS}

\bibitem[{{King} {et~al.}(2012){King}, {Webb}, {Murphy}, {Flambaum},
  {Carswell}, {Bainbridge}, {Wilczynska}, \& {Koch}}]{King2012}
{King}, J.~A., {Webb}, J.~K., {Murphy}, M.~T., {et~al.} 2012, \mnras, 422,
  3370, \href{http://doi.org/10.1111/j.1365-2966.2012.20852.x}{doi},
  \href{https://ui.adsabs.harvard.edu/abs/2012MNRAS.422.3370K}{ADS}

\bibitem[{{Kotu{\v{s}}} {et~al.}(2017){Kotu{\v{s}}}, {Murphy}, \&
  {Carswell}}]{Kotus2017}
{Kotu{\v{s}}}, S.~M., {Murphy}, M.~T., \& {Carswell}, R.~F. 2017, \mnras, 464,
  3679, \href{http://doi.org/10.1093/mnras/stw2543}{doi},
  \href{https://ui.adsabs.harvard.edu/abs/2017MNRAS.464.3679K}{ADS}

\bibitem[{{Landoni} {et~al.}(2016){Landoni}, {Riva}, {Pepe}, {Aliverti},
  {Cabral}, {Calderone}, {Cirami}, {Cristiani}, {Di Marcantonio}, {Genoni},
  {M{\'e}gevand}, {Moschetti}, {Oggioni}, \& {Pariani}}]{Landoni2016}
{Landoni}, M., {Riva}, M., {Pepe}, F., {et~al.} 2016, in Society of
  Photo-Optical Instrumentation Engineers (SPIE) Conference Series, Vol. 9913,
  \procspie, 99133Q, \href{http://doi.org/10.1117/12.2231579}{doi},
  \href{https://ui.adsabs.harvard.edu/abs/2016SPIE.9913E..3QL}{ADS}

\bibitem[{{Levshakov} {et~al.}(2007){Levshakov}, {Molaro}, {Lopez},
  {D'Odorico}, {Centuri{\'o}n}, {Bonifacio}, {Agafonova}, \&
  {Reimers}}]{Levshakov2007}
{Levshakov}, S.~A., {Molaro}, P., {Lopez}, S., {et~al.} 2007, \aap, 466, 1077,
  \href{http://doi.org/10.1051/0004-6361:20066064}{doi},
  \href{https://ui.adsabs.harvard.edu/abs/2007A&A...466.1077L}{ADS}

\bibitem[{{Lovis et al.}(\noop{}2020, in prep.)}]{Lovis2020}
{Lovis et al.} \noop{}2020, in prep.

\bibitem[{{Mayor} {et~al.}(2003){Mayor}, {Pepe}, {Queloz}, {Bouchy},
  {Rupprecht}, {Lo Curto}, {Avila}, {Benz}, {Bertaux}, {Bonfils}, {Dall},
  {Dekker}, {Delabre}, {Eckert}, {Fleury}, {Gilliotte}, {Gojak}, {Guzman},
  {Kohler}, {Lizon}, {Longinotti}, {Lovis}, {Megevand}, {Pasquini}, {Reyes},
  {Sivan}, {Sosnowska}, {Soto}, {Udry}, {van Kesteren}, {Weber}, \&
  {Weilenmann}}]{Mayor2003}
{Mayor}, M., {Pepe}, F., {Queloz}, D., {et~al.} 2003, The Messenger, 114, 20,
  \href{https://ui.adsabs.harvard.edu/abs/2003Msngr.114...20M}{ADS}

\bibitem[{{M{\'e}gevand} {et~al.}(2014){M{\'e}gevand}, {Zerbi}, {Di
  Marcantonio}, {Cabral}, {Riva}, {Abreu}, {Pepe}, {Cristiani}, {Rebolo Lopez},
  {Santos}, {Dekker}, {Aliverti}, {Allende}, {Amate}, {Avila}, {Baldini},
  {Bandy}, {Bristow}, {Broeg}, {Cirami}, {Coelho}, {Conconi}, {Coretti},
  {Cupani}, {D'Odorico}, {De Caprio}, {Delabre}, {Dorn}, {Figueira}, {Fragoso},
  {Galeotta}, {Genolet}, {Gomes}, {Gonz{\'a}lez Hern{\'a}ndez}, {Hughes},
  {Iwert}, {Kerber}, {Land oni}, {Lizon}, {Lovis}, {Maire}, {Mannetta},
  {Martins}, {Molaro}, {Monteiro}, {Moschetti}, {Oliveira}, {Zapatero Osorio},
  {Poretti}, {Rasilla}, {Santana Tschudi}, {Santos}, {Sosnowska}, {Sousa},
  {Tenegi}, {Toso}, {Vanzella}, \& {Viel}}]{Megevand2014}
{M{\'e}gevand}, D., {Zerbi}, F.~M., {Di Marcantonio}, P., {et~al.} 2014, in
  Society of Photo-Optical Instrumentation Engineers (SPIE) Conference Series,
  Vol. 9147, \procspie, 91471H, \href{http://doi.org/10.1117/12.2055816}{doi},
  \href{https://ui.adsabs.harvard.edu/abs/2014SPIE.9147E..1HM}{ADS}

\bibitem[{{Milakovi{\'c}} {et~al.}(2020){Milakovi{\'c}}, {Pasquini}, {Webb}, \&
  {Lo Curto}}]{Milakovic2020a}
{Milakovi{\'c}}, D., {Pasquini}, L., {Webb}, J.~K., \& {Lo Curto}, G. 2020,
  \mnras, 493, 3997, \href{http://doi.org/10.1093/mnras/staa356}{doi},
  \href{https://ui.adsabs.harvard.edu/abs/2020MNRAS.493.3997M}{ADS}

\bibitem[{Milaković {et~al.}(2020)Milaković, Lee, Carswell, Webb, Molaro, \&
  Pasquini}]{Milakovic2020b}
Milaković, D., Lee, C.-C., Carswell, R.~F., {et~al.} 2020, \mnras, 500, 1,
  \href{http://doi.org/10.1093/mnras/staa3217}{doi}

\bibitem[{{Molaro}(2009)}]{Molaro2009}
{Molaro}, P. 2009, Astrophysics and Space Science Proceedings, 9, 389,
  \href{http://doi.org/10.1007/978-1-4020-9190-2_67}{doi},
  \href{https://ui.adsabs.harvard.edu/abs/2009ASSP....9..389M}{ADS}

\bibitem[{{Molaro} {et~al.}(2013{\natexlab{a}}){Molaro}, {Centuri{\'o}n},
  {Whitmore}, {Evans}, {Murphy}, {Agafonova}, {Bonifacio}, {D'Odorico},
  {Levshakov}, {Lopez}, {Martins}, {Petitjean}, {Rahmani}, {Reimers},
  {Srianand}, {Vladilo}, \& {Wendt}}]{Molaro2013a}
{Molaro}, P., {Centuri{\'o}n}, M., {Whitmore}, J.~B., {et~al.}
  2013{\natexlab{a}}, \aap, 555, A68,
  \href{http://doi.org/10.1051/0004-6361/201321351}{doi},
  \href{https://ui.adsabs.harvard.edu/abs/2013A&A...555A..68M}{ADS}

\bibitem[{{Molaro} {et~al.}(2013{\natexlab{b}}){Molaro}, {Esposito}, {Monai},
  {Lo Curto}, {Gonz{\'a}lez Hern{\'a}ndez}, {H{\"a}nsch}, {Holzwarth},
  {Manescau}, {Pasquini}, {Probst}, {Rebolo}, {Steinmetz}, {Udem}, \&
  {Wilken}}]{Molaro2013b}
{Molaro}, P., {Esposito}, M., {Monai}, S., {et~al.} 2013{\natexlab{b}}, \aap,
  560, A61, \href{http://doi.org/10.1051/0004-6361/201322324}{doi},
  \href{https://ui.adsabs.harvard.edu/abs/2013A&A...560A..61M}{ADS}

\bibitem[{{Molaro} {et~al.}(2008{\natexlab{a}}){Molaro}, {Levshakov}, {Monai},
  {Centuri{\'o}n}, {Bonifacio}, {D'Odorico}, \& {Monaco}}]{Molaro2008a}
{Molaro}, P., {Levshakov}, S.~A., {Monai}, S., {et~al.} 2008{\natexlab{a}},
  \aap, 481, 559, \href{http://doi.org/10.1051/0004-6361:20078864}{doi},
  \href{https://ui.adsabs.harvard.edu/abs/2008A&A...481..559M}{ADS}

\bibitem[{{Molaro} {et~al.}(2008{\natexlab{b}}){Molaro}, {Reimers},
  {Agafonova}, \& {Levshakov}}]{Molaro2008b}
{Molaro}, P., {Reimers}, D., {Agafonova}, I.~I., \& {Levshakov}, S.~A.
  2008{\natexlab{b}}, European Physical Journal Special Topics, 163, 173,
  \href{http://doi.org/10.1140/epjst/e2008-00818-4}{doi},
  \href{https://ui.adsabs.harvard.edu/abs/2008EPJST.163..173M}{ADS}

\bibitem[{{Murphy} \& {Berengut}(2014)}]{Murphy2014}
{Murphy}, M.~T. \& {Berengut}, J.~C. 2014, \mnras, 438, 388,
  \href{http://doi.org/10.1093/mnras/stt2204}{doi},
  \href{https://ui.adsabs.harvard.edu/abs/2014MNRAS.438..388M}{ADS}

\bibitem[{{Murphy} \& {Cooksey}(2017)}]{Murphy2017}
{Murphy}, M.~T. \& {Cooksey}, K.~L. 2017, \mnras, 471, 4930,
  \href{http://doi.org/10.1093/mnras/stx1949}{doi},
  \href{https://ui.adsabs.harvard.edu/abs/2017MNRAS.471.4930M}{ADS}

\bibitem[{{Murphy} {et~al.}(2012){Murphy}, {Locke}, {Light}, {Luiten}, \&
  {Lawrence}}]{Murphy2012}
{Murphy}, M.~T., {Locke}, C.~R., {Light}, P.~S., {Luiten}, A.~N., \&
  {Lawrence}, J.~S. 2012, \mnras, 422, 761,
  \href{http://doi.org/10.1111/j.1365-2966.2012.20656.x}{doi},
  \href{https://ui.adsabs.harvard.edu/abs/2012MNRAS.422..761M}{ADS}

\bibitem[{{Murphy} {et~al.}(2007){Murphy}, {Udem}, {Holzwarth}, {Sizmann},
  {Pasquini}, {Araujo-Hauck}, {Dekker}, {D'Odorico}, {Fischer}, {H{\"a}nsch},
  \& {Manescau}}]{Murphy2007}
{Murphy}, M.~T., {Udem}, T., {Holzwarth}, R., {et~al.} 2007, \mnras, 380, 839,
  \href{http://doi.org/10.1111/j.1365-2966.2007.12147.x}{doi},
  \href{https://ui.adsabs.harvard.edu/abs/2007MNRAS.380..839M}{ADS}

\bibitem[{{Murphy} {et~al.}(2003){Murphy}, {Webb}, \& {Flambaum}}]{Murphy2003}
{Murphy}, M.~T., {Webb}, J.~K., \& {Flambaum}, V.~V. 2003, \mnras, 345, 609,
  \href{http://doi.org/10.1046/j.1365-8711.2003.06970.x}{doi},
  \href{https://ui.adsabs.harvard.edu/abs/2003MNRAS.345..609M}{ADS}

\bibitem[{{Murphy} {et~al.}(2001){Murphy}, {Webb}, {Flambaum}, {Churchill}, \&
  {Prochaska}}]{Murphy2001b}
{Murphy}, M.~T., {Webb}, J.~K., {Flambaum}, V.~V., {Churchill}, C.~W., \&
  {Prochaska}, J.~X. 2001, \mnras, 327, 1223,
  \href{http://doi.org/10.1046/j.1365-8711.2001.04841.x}{doi},
  \href{https://ui.adsabs.harvard.edu/abs/2001MNRAS.327.1223M}{ADS}

\bibitem[{{Murphy et al.}(\noop{}2021, in prep.)}]{Murphy2020}
{Murphy et al.} \noop{}2021, in prep.

\bibitem[{{Nave} {et~al.}(2018){Nave}, {Kerber}, {Den Hartog}, \& {Lo
  Curto}}]{Nave2018}
{Nave}, G., {Kerber}, F., {Den Hartog}, E.~A., \& {Lo Curto}, G. 2018, in
  Society of Photo-Optical Instrumentation Engineers (SPIE) Conference Series,
  Vol. 10704, \procspie, 1070407,
  \href{http://doi.org/10.1117/12.2312286}{doi},
  \href{https://ui.adsabs.harvard.edu/abs/2018SPIE10704E..07N}{ADS}

\bibitem[{{Pepe} {et~al.}(2020){Pepe}, {Cristiani}, {Rebolo}, {Santos},
  {Dekker}, {Cabral}, {Di Marcantonio}, {Figueira}, {Lo Curto}, {Lovis},
  {Mayor}, {M{\'e}gevand}, {Molaro}, {Riva}, {Zapatero Osorio}, {Amate},
  {Manescau}, {Pasquini}, {Zerbi}, {Adibekyan}, {Abreu}, {Affolter}, {Alibert},
  {Aliverti}, {Allart}, {Allende Prieto}, {{\'A}lvarez}, {Alves}, {Avila},
  {Baldini}, {Bandy}, {Barros}, {Benz}, {Bianco}, {Borsa}, {Bourrier},
  {Bouchy}, {Broeg}, {Calderone}, {Cirami}, {Coelho}, {Conconi}, {Coretti},
  {Cumani}, {Cupani}, {D'Odorico}, {Damasso}, {Deiries}, {Delabre},
  {Demangeon}, {Dumusque}, {Ehrenreich}, {Faria}, {Fragoso}, {Genolet},
  {Genoni}, {G{\'e}nova Santos}, {Gonz{\'a}lez Hern{\'a}ndez}, {Hughes},
  {Iwert}, {Kerber}, {Knudstrup}, {Landoni}, {Lavie}, {Lillo-Box}, {Lizon},
  {Maire}, {Martins}, {Mehner}, {Micela}, {Modigliani}, {Monteiro}, {Monteiro},
  {Moschetti}, {Murphy}, {Nunes}, {Oggioni}, {Oliveira}, {Oshagh}, {Pall{\'e}},
  {Pariani}, {Poretti}, {Rasilla}, {Rebord{\~a}o}, {Redaelli}, {Santana
  Tschudi}, {Santin}, {Santos}, {S{\'e}gransan}, {Schmidt}, {Segovia},
  {Sosnowska}, {Sozzetti}, {Sousa}, {Span{\`o}}, {Su{\'a}rez Mascare{\~n}o},
  {Tabernero}, {Tenegi}, {Udry}, \& {Zanutta}}]{Pepe2020}
{Pepe}, F., {Cristiani}, S., {Rebolo}, R., {et~al.} 2020, arXiv e-prints,
  arXiv:2010.00316,
  \href{https://ui.adsabs.harvard.edu/abs/2020arXiv201000316P}{ADS}

\bibitem[{{Pepe} {et~al.}(2014){Pepe}, {Molaro}, {Cristiani}, {Rebolo},
  {Santos}, {Dekker}, {M{\'e}gevand}, {Zerbi}, {Cabral}, {Di Marcantonio},
  {Abreu}, {Affolter}, {Aliverti}, {Allende Prieto}, {Amate}, {Avila},
  {Baldini}, {Bristow}, {Broeg}, {Cirami}, {Coelho}, {Conconi}, {Coretti},
  {Cupani}, {D'Odorico}, {De Caprio}, {Delabre}, {Dorn}, {Figueira}, {Fragoso},
  {Galeotta}, {Genolet}, {Gomes}, {Gonz{\'a}lez Hern{\'a}ndez}, {Hughes},
  {Iwert}, {Kerber}, {Landoni}, {Lizon}, {Lovis}, {Maire}, {Mannetta},
  {Martins}, {Monteiro}, {Oliveira}, {Poretti}, {Rasilla}, {Riva}, {Santana
  Tschudi}, {Santos}, {Sosnowska}, {Sousa}, {Span{\'o}}, {Tenegi}, {Toso},
  {Vanzella}, {Viel}, \& {Zapatero Osorio}}]{Pepe2014}
{Pepe}, F., {Molaro}, P., {Cristiani}, S., {et~al.} 2014, arXiv e-prints,
  arXiv:1401.5918,
  \href{https://ui.adsabs.harvard.edu/abs/2014arXiv1401.5918P}{ADS}

\bibitem[{{Pepe} {et~al.}(2010){Pepe}, {Cristiani}, {Rebolo Lopez}, {Santos},
  {Amorim}, {Avila}, {Benz}, {Bonifacio}, {Cabral}, {Carvas}, {Cirami},
  {Coelho}, {Comari}, {Coretti}, {De Caprio}, {Dekker}, {Delabre}, {Di
  Marcantonio}, {D'Odorico}, {Fleury}, {Garc{\'\i}a}, {Herreros Linares},
  {Hughes}, {Iwert}, {Lima}, {Lizon}, {Lo Curto}, {Lovis}, {Manescau},
  {Martins}, {M{\'e}gevand}, {Moitinho}, {Molaro}, {Monteiro}, {Monteiro},
  {Pasquini}, {Mordasini}, {Queloz}, {Rasilla}, {Rebord{\~a}o}, {Santana
  Tschudi}, {Santin}, {Sosnowska}, {Span{\`o}}, {Tenegi}, {Udry}, {Vanzella},
  {Viel}, {Zapatero Osorio}, \& {Zerbi}}]{Pepe2010}
{Pepe}, F.~A., {Cristiani}, S., {Rebolo Lopez}, R., {et~al.} 2010, in Society
  of Photo-Optical Instrumentation Engineers (SPIE) Conference Series, Vol.
  7735, \procspie, 77350F, \href{http://doi.org/10.1117/12.857122}{doi},
  \href{https://ui.adsabs.harvard.edu/abs/2010SPIE.7735E..0FP}{ADS}

\bibitem[{{Probst} {et~al.}(2016){Probst}, {Lo Curto}, {{\'A}vila},
  {Brucalassi}, {Canto Martins}, {de Castro Le{\~a}o}, {Esposito},
  {Gonz{\'a}lez Hern{\'a}ndez}, {Grupp}, {H{\"a}nsch}, {Holzwarth},
  {Kellermann}, {Kerber}, {Mandel}, {Manescau}, {Pasquini}, {Pozna}, {Rebolo},
  {Renan de Medeiros}, {Stark}, {Steinmetz}, {Su{\'a}rez Mascare{\~n}o},
  {Udem}, {Urrutia}, \& {Wu}}]{Probst2016}
{Probst}, R.~A., {Lo Curto}, G., {{\'A}vila}, G., {et~al.} 2016, Society of
  Photo-Optical Instrumentation Engineers (SPIE) Conference Series, Vol. 9908,
  {Relative stability of two laser frequency combs for routine operation on
  HARPS and FOCES}, 990864, \href{http://doi.org/10.1117/12.2231434}{doi},
  \href{https://ui.adsabs.harvard.edu/abs/2016SPIE.9908E..64P}{ADS}

\bibitem[{{Probst} {et~al.}(2014){Probst}, {Lo Curto}, {Avila}, {Canto
  Martins}, {de Medeiros}, {Esposito}, {Gonz{\'a}lez Hern{\'a}ndez},
  {H{\"a}nsch}, {Holzwarth}, {Kerber}, {Le{\~a}o}, {Manescau}, {Pasquini},
  {Rebolo-L{\'o}pez}, {Steinmetz}, {Udem}, \& {Wu}}]{Probst2014}
{Probst}, R.~A., {Lo Curto}, G., {Avila}, G., {et~al.} 2014, Society of
  Photo-Optical Instrumentation Engineers (SPIE) Conference Series, Vol. 9147,
  {A laser frequency comb featuring sub-cm/s precision for routine operation on
  HARPS}, 91471C, \href{http://doi.org/10.1117/12.2055784}{doi},
  \href{https://ui.adsabs.harvard.edu/abs/2014SPIE.9147E..1CP}{ADS}

\bibitem[{{Rahmani} {et~al.}(2013){Rahmani}, {Wendt}, {Srianand}, {Noterdaeme},
  {Petitjean}, {Molaro}, {Whitmore}, {Murphy}, {Centurion}, {Fathivavsari},
  {D'Odorico}, {Evans}, {Levshakov}, {Lopez}, {Martins}, {Reimers}, \&
  {Vladilo}}]{Rahmani2013}
{Rahmani}, H., {Wendt}, M., {Srianand}, R., {et~al.} 2013, \mnras, 435, 861,
  \href{http://doi.org/10.1093/mnras/stt1356}{doi},
  \href{https://ui.adsabs.harvard.edu/abs/2013MNRAS.435..861R}{ADS}

\bibitem[{{Redman} {et~al.}(2014){Redman}, {Nave}, \&
  {Sansonetti}}]{Redman2014}
{Redman}, S.~L., {Nave}, G., \& {Sansonetti}, C.~J. 2014, \apjs, 211, 4,
  \href{http://doi.org/10.1088/0067-0049/211/1/4}{doi},
  \href{https://ui.adsabs.harvard.edu/abs/2014ApJS..211....4R}{ADS}

\bibitem[{{Reiners} {et~al.}(2016){Reiners}, {Mrotzek}, {Lemke}, {Hinrichs}, \&
  {Reinsch}}]{Reiners2016}
{Reiners}, A., {Mrotzek}, N., {Lemke}, U., {Hinrichs}, J., \& {Reinsch}, K.
  2016, \aap, 587, A65, \href{http://doi.org/10.1051/0004-6361/201527530}{doi},
  \href{https://ui.adsabs.harvard.edu/abs/2016A&A...587A..65R}{ADS}

\bibitem[{{Riva} {et~al.}(2014){Riva}, {Aliverti}, {Moschetti}, {Land oni},
  {Dell'Agostino}, {Pepe}, {M{\'e}gevand}, {Zerbi}, {Cristiani}, \&
  {Cabral}}]{Riva2014b}
{Riva}, M., {Aliverti}, M., {Moschetti}, M., {et~al.} 2014, in Society of
  Photo-Optical Instrumentation Engineers (SPIE) Conference Series, Vol. 9147,
  \procspie, 91477G, \href{http://doi.org/10.1117/12.2056499}{doi},
  \href{https://ui.adsabs.harvard.edu/abs/2014SPIE.9147E..7GR}{ADS}

\bibitem[{{Rosenband}(2008)}]{Rosenband2008}
{Rosenband}, T. 2008, in APS Meeting Abstracts, Vol.~39, APS Division of
  Atomic, Molecular and Optical Physics Meeting Abstracts, OPJ.20,
  \href{https://ui.adsabs.harvard.edu/abs/2008APS..DMP.J2002R}{ADS}

\bibitem[{{Sansonetti} \& {Weber}(1984)}]{Sansonetti1984}
{Sansonetti}, C.~J. \& {Weber}, K.~H. 1984, Journal of the Optical Society of
  America B Optical Physics, 1, 361,
  \href{http://doi.org/10.1364/JOSAB.1.000361}{doi},
  \href{https://ui.adsabs.harvard.edu/abs/1984JOSAB...1..361S}{ADS}

\bibitem[{{Savitzky} \& {Golay}(1964)}]{SavitzkyGolay1964}
{Savitzky}, A. \& {Golay}, M.~J.~E. 1964, Analytical Chemistry, 36, 1627,
  \href{https://ui.adsabs.harvard.edu/abs/1964AnaCh..36.1627S}{ADS}

\bibitem[{{Steinmetz} {et~al.}(2008){Steinmetz}, {Wilken}, {Araujo-Hauck},
  {Holzwarth}, {H{\"a}nsch}, {Pasquini}, {Manescau}, {D'Odorico}, {Murphy},
  {Kentischer}, {Schmidt}, \& {Udem}}]{Steinmetz2008}
{Steinmetz}, T., {Wilken}, T., {Araujo-Hauck}, C., {et~al.} 2008, Science, 321,
  1335, \href{http://doi.org/10.1126/science.1161030}{doi},
  \href{https://ui.adsabs.harvard.edu/abs/2008Sci...321.1335S}{ADS}

\bibitem[{{Uzan}(2011)}]{Uzan2011}
{Uzan}, J.-P. 2011, Living Reviews in Relativity, 14, 2,
  \href{http://doi.org/10.12942/lrr-2011-2}{doi},
  \href{https://ui.adsabs.harvard.edu/abs/2011LRR....14....2U}{ADS}

\bibitem[{{Vogt} {et~al.}(2019){Vogt}, {Kerber}, {Mehner}, {Yu}, {Pfrommer},
  {Lo Curto}, {Figueira}, {Parraguez}, {Pepe}, {M{\'e}gevand}, {Riva}, {Di
  Marcantonio}, {Lovis}, {Amate}, {Molaro}, {Cabral}, \& {Osorio}}]{Vogt2019}
{Vogt}, F. P.~A., {Kerber}, F., {Mehner}, A., {et~al.} 2019, \prl, 123, 061101,
  \href{http://doi.org/10.1103/PhysRevLett.123.061101}{doi},
  \href{https://ui.adsabs.harvard.edu/abs/2019PhRvL.123f1101V}{ADS}

\bibitem[{{Vogt} {et~al.}(1994){Vogt}, {Allen}, {Bigelow}, {Bresee}, {Brown},
  {Cantrall}, {Conrad}, {Couture}, {Delaney}, {Epps}, {Hilyard}, {Hilyard},
  {Horn}, {Jern}, {Kanto}, {Keane}, {Kibrick}, {Lewis}, {Osborne},
  {Pardeilhan}, {Pfister}, {Ricketts}, {Robinson}, {Stover}, {Tucker}, {Ward},
  \& {Wei}}]{Vogt1994}
{Vogt}, S.~S., {Allen}, S.~L., {Bigelow}, B.~C., {et~al.} 1994, in \procspie,
  Vol. 2198, Instrumentation in Astronomy VIII, ed. D.~L. {Crawford} \& E.~R.
  {Craine}, 362, \href{http://doi.org/10.1117/12.176725}{doi},
  \href{http://adsabs.harvard.edu/abs/1994SPIE.2198..362V}{ADS}

\bibitem[{{Webb} {et~al.}(1999){Webb}, {Flambaum}, {Churchill}, {Drinkwater},
  \& {Barrow}}]{Webb1999}
{Webb}, J.~K., {Flambaum}, V.~V., {Churchill}, C.~W., {Drinkwater}, M.~J., \&
  {Barrow}, J.~D. 1999, \prl, 82, 884,
  \href{http://doi.org/10.1103/PhysRevLett.82.884}{doi},
  \href{https://ui.adsabs.harvard.edu/abs/1999PhRvL..82..884W}{ADS}

\bibitem[{{Whitmore} \& {Murphy}(2015)}]{Whitmore2015}
{Whitmore}, J.~B. \& {Murphy}, M.~T. 2015, \mnras, 447, 446,
  \href{http://doi.org/10.1093/mnras/stu2420}{doi},
  \href{https://ui.adsabs.harvard.edu/abs/2015MNRAS.447..446W}{ADS}

\bibitem[{{Whitmore} {et~al.}(2010){Whitmore}, {Murphy}, \&
  {Griest}}]{Whitmore2010}
{Whitmore}, J.~B., {Murphy}, M.~T., \& {Griest}, K. 2010, \apj, 723, 89,
  \href{http://doi.org/10.1088/0004-637X/723/1/89}{doi},
  \href{https://ui.adsabs.harvard.edu/abs/2010ApJ...723...89W}{ADS}

\bibitem[{{Wildi} {et~al.}(2012){Wildi}, {Chazelas}, \& {Pepe}}]{Wildi2012}
{Wildi}, F., {Chazelas}, B., \& {Pepe}, F. 2012, Society of Photo-Optical
  Instrumentation Engineers (SPIE) Conference Series, Vol. 8446, {A passive
  cost-effective solution for the high accuracy wavelength calibration of
  radial velocity spectrographs}, 84468E,
  \href{http://doi.org/10.1117/12.926841}{doi},
  \href{https://ui.adsabs.harvard.edu/abs/2012SPIE.8446E..8EW}{ADS}

\bibitem[{{Wildi} {et~al.}(2010){Wildi}, {Pepe}, {Chazelas}, {Lo Curto}, \&
  {Lovis}}]{Wildi2010}
{Wildi}, F., {Pepe}, F., {Chazelas}, B., {Lo Curto}, G., \& {Lovis}, C. 2010,
  Society of Photo-Optical Instrumentation Engineers (SPIE) Conference Series,
  Vol. 7735, {A Fabry-Perot calibrator of the HARPS radial velocity
  spectrograph: performance report}, 77354X,
  \href{http://doi.org/10.1117/12.857951}{doi},
  \href{https://ui.adsabs.harvard.edu/abs/2010SPIE.7735E..4XW}{ADS}

\bibitem[{{Wildi} {et~al.}(2011){Wildi}, {Pepe}, {Chazelas}, {Lo Curto}, \&
  {Lovis}}]{Wildi2011}
{Wildi}, F., {Pepe}, F., {Chazelas}, B., {Lo Curto}, G., \& {Lovis}, C. 2011,
  Society of Photo-Optical Instrumentation Engineers (SPIE) Conference Series,
  Vol. 8151, {The performance of the new Fabry-Perot calibration system of the
  radial velocity spectrograph HARPS}, 81511F,
  \href{http://doi.org/10.1117/12.901550}{doi},
  \href{https://ui.adsabs.harvard.edu/abs/2011SPIE.8151E..1FW}{ADS}

\bibitem[{{Wilken} {et~al.}(2012){Wilken}, {Curto}, {Probst}, {Steinmetz},
  {Manescau}, {Pasquini}, {Gonz{\'a}lez Hern{\'a}ndez}, {Rebolo}, {H{\"a}nsch},
  {Udem}, \& {Holzwarth}}]{Wilken2012}
{Wilken}, T., {Curto}, G.~L., {Probst}, R.~A., {et~al.} 2012, \nat, 485, 611,
  \href{http://doi.org/10.1038/nature11092}{doi},
  \href{https://ui.adsabs.harvard.edu/abs/2012Natur.485..611W}{ADS}

\bibitem[{{Wilken} {et~al.}(2010{\natexlab{a}}){Wilken}, {Lovis}, {Manescau},
  {Steinmetz}, {Pasquini}, {Lo Curto}, {H{\"a}nsch}, {Holzwarth}, \&
  {Udem}}]{Wilken2010a}
{Wilken}, T., {Lovis}, C., {Manescau}, A., {et~al.} 2010{\natexlab{a}}, \mnras,
  405, L16, \href{http://doi.org/10.1111/j.1745-3933.2010.00850.x}{doi},
  \href{https://ui.adsabs.harvard.edu/abs/2010MNRAS.405L..16W}{ADS}

\bibitem[{{Wilken} {et~al.}(2010{\natexlab{b}}){Wilken}, {Lovis}, {Manescau},
  {Steinmetz}, {Pasquini}, {Lo Curto}, {H{\"a}nsch}, {Holzwarth}, \&
  {Udem}}]{Wilken2010b}
{Wilken}, T., {Lovis}, C., {Manescau}, A., {et~al.} 2010{\natexlab{b}}, in
  Society of Photo-Optical Instrumentation Engineers (SPIE) Conference Series,
  Vol. 7735, \procspie, 77350T, \href{http://doi.org/10.1117/12.857248}{doi},
  \href{https://ui.adsabs.harvard.edu/abs/2010SPIE.7735E..0TW}{ADS}

\bibitem[{{Wolfe} {et~al.}(1976){Wolfe}, {Brown}, \& {Roberts}}]{Wolfe1976}
{Wolfe}, A.~M., {Brown}, R.~L., \& {Roberts}, M.~S. 1976, \prl, 37, 179,
  \href{http://doi.org/10.1103/PhysRevLett.37.179}{doi},
  \href{https://ui.adsabs.harvard.edu/abs/1976PhRvL..37..179W}{ADS}

\bibitem[{{Zechmeister} {et~al.}(2014){Zechmeister}, {Anglada-Escud{\'e}}, \&
  {Reiners}}]{Zechmeister2014}
{Zechmeister}, M., {Anglada-Escud{\'e}}, G., \& {Reiners}, A. 2014, \aap, 561,
  A59, \href{http://doi.org/10.1051/0004-6361/201322746}{doi},
  \href{https://ui.adsabs.harvard.edu/abs/2014A&A...561A..59Z}{ADS}

\end{thebibliography}
